\documentclass[fleqn,usenatbib]{mnras}
\usepackage{newtxtext,newtxmath}
\usepackage[T1]{fontenc}
\usepackage{ae,aecompl}

\usepackage{graphicx}	
\usepackage{amsmath}	
\usepackage[pdftex,dvipsnames]{xcolor}  
\usepackage{soul} 
\usepackage{xargs} 
\usepackage{siunitx}
\usepackage{longtable}
\usepackage{booktabs}
\usepackage{pdflscape}

\usepackage[colorinlistoftodos]{todonotes}
\newcommandx{\tobedone}[2][1=]{\todo[linecolor=red,backgroundcolor=red!25,bordercolor=red,inline,#1]{#2}}
\newcommandx{\changed}[2][1=]{\todo[linecolor=blue,backgroundcolor=blue!25,bordercolor=blue,inline,#1]{#2}\noindent}
\newcommandx{\asa}[2][1=]{\todo[linecolor=BurntOrange,backgroundcolor=BurntOrange!25,bordercolor=BurntOrange,inline,#1]{#2}\noindent}
\newcommandx{\jcw}[2][1=]{\todo[linecolor=SpringGreen,backgroundcolor=SpringGreen!40,bordercolor=SpringGreen,inline,#1]{#2}\noindent}
\newcommandx{\soko}[2][1=]{\todo[linecolor= Aquamarine,backgroundcolor= Aquamarine!40,bordercolor= Aquamarine,inline,#1]{#2}\noindent}
\newcommandx{\comment}[2][1=]{\todo[linecolor=red,backgroundcolor=red!25,bordercolor=red,inline,#1]{#2}}

\newcommandx{\mf}[2][1=]{\todo[linecolor=BurntOrange,backgroundcolor=blue!25,bordercolor=BurntOrange,inline,#1]{#2}\noindent}

\newcommandx{\afr}[2][1=]{\todo[linecolor=Purple,backgroundcolor=cyan!25,bordercolor=Blue,inline,#1]{#2}\noindent}

\newcommand{\vsini}{V\,sin\,$i$}
\newcommand{\chisq}{$\chi^2$}
\newcommand{\SM}{\texttt{STAR-MELT}}

\title[STAR-MELT]{The STAR-MELT Python package\thanks{\url{https://github.com/justyncw/STAR_MELT}} for emission line analysis of YSOs\thanks{Based on data obtained with ESO programmes 106.20Z8.002 and 106.20Z8.009.}}

\author[J. Campbell-White et al.]{Justyn Campbell-White,$^{1}$\thanks{E-mail: jcampbellwhite001@dundee.ac.uk}
Aurora Sicilia-Aguilar,$^{1}$
Carlo F. Manara,$^{2}$
\newauthor Soko Matsumura,$^{1}$
Min Fang$^{3}$,
Antonio Frasca,$^{4}$
and Veronica Roccatagliata$^{5,6,7}$
\\
$^{1}$SUPA, School of Science and Engineering, University of Dundee, Nethergate, Dundee DD1 4HN, U.K.\\
$^{2}$European Southern Observatory, Karl-Schwarzschild-Strasse 2, 85748 Garching bei M\"unchen, Germany\\
$^{3}$Purple Mountain Observatory, Chinese Academy of Sciences, 10 Yuanhua Road, Nanjing 210023, China\\
$^{4}$INAF - Osservatorio Astrofisico di Catania, via S. Sofia, 78, 95123 Catania\\
$^{5}$Dipartimento di Fisica “Enrico Fermi”, Universita’ di Pisa, Largo Pontecorvo 3, 56127 Pisa, Italy\\
$^{6}$INFN, Sezione di Pisa, Largo Bruno Pontecorvo 3, 56127 Pisa, Italy\\
$^{7}$INAF-Osservatorio Astrofisico di Arcetri, Largo E. Fermi 5, 50125 Firenze, Italy
}

\date{Accepted XXX. Received YYY; in original form ZZZ}

\pubyear{2021}

\begin{document}
\label{firstpage}
\pagerange{\pageref{firstpage}--\pageref{lastpage}}
\maketitle

\begin{abstract}
We introduce the \SM\, Python package that we developed to facilitate the analysis of time-resolved emission line spectroscopy of young stellar objects. \SM\ automatically extracts, identifies and fits emission lines. We summarise our analysis methods that utilises the time domain of high-resolution stellar spectra to investigate variability in the line profiles and corresponding emitting regions. This allows us to probe the innermost disc and accretion structures of YSOs. Local temperatures and densities can be determined using Boltzmann statistics, the Saha equation, and the Sobolev large velocity gradient approximation. \SM\ allows for new results to be obtained from archival data, as well as facilitating timely analysis of new data as it is obtained. We present the results of applying \SM\ to three YSOs, using spectra from UVES, XSHOOTER, FEROS, HARPS, and  ESPaDOnS. We demonstrate what can be achieved for data with disparate time sampling, for stars with different inclinations and variability types. For EX Lupi, we confirm the presence of a localised and stable stellar-surface  hot spot associated with the footprint of the accretion column. For GQ Lupi A, we find that the maximum infall rate from an accretion column is correlated with lines produced in the lowest temperatures. For CVSO109 we investigate the rapid temporal variability of a redshifted emission wing, indicative of rotating and infalling material in the inner disc. Our results show that \SM\ is a useful tool for such analysis, as well as other applications for emission lines. 
\end{abstract}




\section{Introduction}
\label{sec:intro}
Low- and intermediate-mass young stellar objects (YSOs) acquire the majority of their mass in the protostellar phase, but accretion continues during the pre-main-sequence (PMS) phase via an accretion disc over a few million years. Accretion has a large impact on the future of the stars \citep{hartmann_accretion_2016} and the formation of planetary systems \citep{1974MNRAS.168..603L,2011ARA&A..49...67W,2016JGRE..121.1962M}. The accretion process governs the transport of matter from the disc to the star and angular momentum from the star to the disc; and therefore affects disc stability, rotation, future stellar activity and potential habitability of planets. This is important for Classical T-Tauri (CTT) stars, with late-type stars  observed to remain active during a substantial part of their lives \citep{gallet_tidal_2017}. It is also important in Herbig Ae/Be stars, which have masses between CTTs and massive young stellar objects. Such intermediate-mass stars may undergo different accretion mechanisms; from the magnetically channeled accretion of CTTs, to a direct disc-to star accretion through a boundary layer in more massive stars \citep{mendigutia_accretion_2011,vioque_gaia_2018}. Disc locking and angular momentum removal by winds/outflows \citep{shu_magnetocentrifugally_1994,herbst_stellar_2002} regulate stellar rotation, but the efficiency of these processes is highly dependent on the accretion rate and the stellar magnetic field \citep{matt_magnetic_2012,cauley_testing_2012}. 

Accretion and stellar activity also make the detection of young newly-formed planets challenging \citep{johns-krull_candidate_2016}, since these processes mimic and mask both photometric and spectroscopic planet signatures \citep{crockett_search_2012,kospal_radial_2014}. Furthermore, accretion regulates the disc masses and thus affects the migration speeds of both low-mass, non-gap-opening planets \citep[e.g.][]{Tanaka02,Paardekooper11} as well as massive, gap-opening planets \citep[e.g.][]{Kanagawa18a}.  An accretion associated feature such as a magnetospheric cavity may also act as a stopping mechanism of migrating planets \citep{Lin96}. 

It is therefore clear that a thorough understanding of the varying accretion and activity processes for young stars is fundamentally important when studying the evolution of main sequence stars and planetary systems. The main challenge is actually observing the sub-AU scales of accretion and the innermost disc. Whilst the most nearby PMS stars and discs can be studied via interferometry down to a few AU \citep[e.g.][]{2016ApJ...820L..40A,2018ApJ...869L..41A,2019A&A...632A..53G}, the spatial resolution and sensitivity is still limited, and such accuracy cannot yet be achieved for the overall population of PMS stars that are in general fainter. The innermost part of the disc contains very little gas that is generally warm and thus accounts for low emission contribution, even from facilities such as ALMA \citep{sicilia-aguilar_rosetta_2016}. Sub-mm interferometry does also not trace the highly energetic processes related to accretion. 

However, the spectra of young stars are known to show many emission lines since the class was defined \citep{joy45}. The wealth of metallic emission lines observed allows us to probe the accretion structures, winds, and innermost region of the accretion disc.
Investigations of densities, temperatures and spatial locations of material along the line-of-sight using emission lines have been conducted for a small number of PMS stars \citep[e.g.][]{beristain_permitted_1998,beristain_iron_2000,sicilia-aguilar_optical_2012,petrov_doppler_2014,2017A&A...606A..48A}. Furthermore, time-resolved data for many emission lines dramatically improve these results, revealing how far the structures are from the star and their corresponding latitudes \citep[e.g.][]{sicilia-aguilar_accretion_2015,2018A&A...614A.108S,sicilia-aguilar_reading_2020,mcginnis_magnetic_2020}, down to scales not possible with direct imaging techniques. Tracking atomic gas parcels in the innermost disc on short time-scales is also possible - both in emission or absorption \citep[e.g.][]{mora_dynamics_2004}. Observations conducted over a stellar period \citep[typically $\sim$ 7 days for CTTs,][]{bouvier_coyotes_1995} reveal rotational modulations in the emission lines. Furthermore, archival time-resolved data over several years can reveal the long-term variability of accretion structures and activity spots.

High-resolution data of PMS stars are already available from numerous archival programs aiming at spectroscopic planet detection around young stars. These, together with new data expected in the coming years from dedicated surveys, \citep[such as PENELLOPE,][and ODYSSEUS, Espaillat et al. in prep]{2021arXiv210312446M}, provide a unique possibility for a comprehensive view of their accretion processes and inner discs. We hence require a means of accurately and precisely analysing such a large quantity of data in a timely manner. In this paper we introduce the \SM\, (\texttt{ST}ellar \texttt{A}cc\texttt{R}etion \texttt{M}apping with \texttt{E}mission \texttt{L}ine \texttt{T}omography) Python package that we have developed to carry out these detailed investigations. The \SM\, package is a set of modules and functions to carry out the extraction of the accretion spectrum from the high resolution data, and we foresee many further applications beyond this. We are able to automate the process of identifying, matching and individually fitting emission lines, allowing for quick access to results from a large number of stars, for a diverse range of spectroscopic instruments. 

This work is organised as follows: An overview of our emission line  analyses is given in section\,\ref{sec:tomography}. The \SM\ package description is given in section\,\ref{sec:star_melt}, followed by example applications of \SM\  for three YSOs with different properties, as well as different amounts of data with various sampling cadences, and the corresponding discussion in section\,\ref{sec:discussion}. Our summaries, conclusions and outlook for the coming years of new data are given in section \ref{sec:conc}.

\section{Emission line tomography: Investigating the inner disc and accretion structures}
\label{sec:tomography}

Observational signatures of accretion in PMS stars are excess emission from x-ray to near infrared (IR). The majority of this emission from the ultra-violet (UV) regime to lower energies predominantly comprises re-processed x-ray shock emission, which originate from the accretion columns and boundary layers \citep{hartmann_accretion_2016}.  In addition to this excess continuum emission, further key indicators of accretion are the permitted emission lines of e.g. H$\alpha$, H$\beta$, Ca II and He I, which have been used to estimate accretion rates \citep[e.g.][]{fang_star_2009,herczeg_measuring_2009,alcala_x-shooter_2014}. PMS stars also possess many metallic neutral and ionised lines, such as Fe I/II, Ca I/II, Ti I/II, Si II \citep{hamann_emission-line_1992,hamann92-HAeBe,hamann_emission-line_1994,2013ApJ...778...71G,alcala_x-shooter_2014}. Because these lines are so persistent, they can be used to map the innermost accretion disc and accretion funnels, and to detect activity-related spots via emission line tomography \citep{sicilia-aguilar_optical_2012,fang_gw_2014,sicilia-aguilar_accretion_2015,sicilia-aguilar_2014-2017_2017,sicilia-aguilar_reading_2020}.

Although strong lines are most common in strongly-accreting systems \citep{hamann_emission-line_1992,hamann92-HAeBe,beristain_permitted_1998}, current instrumentation (such as ESPRESSO, FEROS, HARPS, XSHOOTER, ESPaDOnS) 
detect He I, Ca II, Fe I/II, Si II, Ti II lines (and more) in the vast majority of young stars \citep{alcala_x-shooter_2014},
also revealing line velocity changes when comparing spectra taken in different epochs. Extracting the line velocities and profile details, is not immediate. In particular, the velocity modulations observed in narrow lines associated to locations very close to the stellar photosphere \citep{dodin12} will be of the order of the projected rotational velocity (\vsini) based on the system inclination \citep{petrov_facing_2014,kospal_radial_2014}, although other velocity signatures could be due to mass motion of the structures and shocks. Since lines from different species may originate in slightly different places \citep{dupree_tw_2012}, the extraction and analysis needs to address each line individually, fitting the continuum with great care to avoid introducing noise in the result \citep{sicilia-aguilar_accretion_2015}. Moreover, for some lines the underlying photospheric spectrum cannot be neglected and must be subtracted to get the correct shape and velocity of the emission line profile. To obtain further information regarding the physical conditions of the emitting gas (density, temperature), the lines need to be also properly identified and well-fitted. 

The emission line tomography technique utilises the time domain to look for distortions in the stellar emission line profiles. It is similar to the application of Doppler tomography, a visualisation technique where the phase and velocity of the line profiles are considered; e.g. applied to the accretion disc of cataclysmic variable stars \citep{1988MNRAS.235..269M}, or for chromospheric mapping of stellar surface features from UV emission lines \citep{1989A&A...215...79N,1999A&A...350..571B} and H$\alpha$ plages \citep{2008A&A...481..229F}. These previous studies have applied the technique to more evolved stars, however, the distortions due to accretion have similar spectral signatures. For PMS stars, these temporal distortions can be attributed to rotating spots, winds, accretion funnels, and inner disc structures. Together with approximations of physical properties, including temperatures and densities of the emitting material, we use these Doppler signatures of metallic emission lines to obtain tomographic information regarding the accretion structures, activity (hot) spots and the innermost hot atomic gas in the disc. The metallic lines observed in high resolution spectra of young stars span a large range of excitation potentials and transition probabilities. This allows us to probe various depths, densities and temperatures over a few stellar radii along the accretion columns, shocks, and post-shock regions, and to detect winds and hot spots.

\subsection{Physical approximations from line ratios}

In addition to investigating the line profiles and temporal variations to make tomographic inferences about the inner-disc and accretion in PMS stars, physical approximations can be drawn from the data by comparing the observed line ratios to those of theoretical calculations. At present, \SM\ includes two simple methods that allow to compare the relative line intensities for ionised and neutral species using the Saha equations \citep{mihalas_stellar_1978,hubeny14}, as well as to constrain the temperature and column density-velocity gradient for broad lines using the Sobolev approximation for lines originating in the same upper level \citep{beristain_permitted_1998}. Both methods have been applied in the past to different observations of young stars \citep[e.g. DR Tau, EX Lupi and Z CMa;][]{beristain_permitted_1998,sicilia-aguilar_optical_2012,sicilia-aguilar_reading_2020}. Whilst these routines are automated within \SM, caution needs to be exercised for individual objects; such as ensuring lines are selected as likely originating from similar locations (i.e. similar line profiles), or from similar wavelength regions to avoid extinction influences. Furthermore, deviations from  Local Thermodynamic Equilibrium (LTE) including non-thermal ionisation, pumping of the upper levels of some lines by higher-energy transitions, radiative transfer effects including optical depth and self-absorption, and saturation are among some of the processes that can be encountered in YSOs and that may limit the use of these approximations for certain targets and/or lines.

The Saha equation \ref{eq:saha} \citep{mihalas_stellar_1978} allows to determine the relative populations of ionised and neutral species, 
\begin{equation}
\label{eq:saha}
\frac{n_{j+1} n_{e}}{n_{j}}=\left(\frac{2 \pi m_{e} k T}{h^{2}}\right)^{3 / 2} \frac{2 U_{j+1}(T)}{U_{j}(T)} e^{-\chi_{1} / k T},
\end{equation}
where $n_{j+1}$ and $n_j$ are the densities of the two species, $n_e$ is the electron density, $T$ is the temperature, $m_e$ is the electron mass, $k$ and $h$ are  Boltzmann and Planck's constants, respectively, $\chi_{1}$ is the ionisation potential, and $U_x(T)$ is the partition function for species $x$. The equation assumes that the species are in LTE, so that the population between different levels can be estimated using the Boltzmann statistics, 
\begin{equation}
\label{eq:boltz}
\frac{n_{i}}{n_{0}}=\frac{g_{i}}{g_{0}} e^{-E_i / k T}.
\end{equation}
Here, the population of an excited level $i$ is given with respect to the ground level, as a function of the temperature, the energy of the upper level $E_i$, and $g_x$ is the statistical weight of level $x$. The relative intensities of the lines will depend exclusively of their atomic parameters.

Because in general the spectra are not flux-calibrated, the Saha calculation is performed in \SM\ as done in \citet{sicilia-aguilar_reading_2020}, using the relative intensities observed between selected lines to compare to the relative intensities derived from Saha models spanning a range of temperatures and electron densities, and estimating as best fit the one that results in a lower standard deviation between the values associated to all the lines. The uncertainty is derived including the regions where the standard deviation value is up to 3 times the minimum value, and the contribution of individual lines is weighted according to the quality of the atomic parameters as listed in the National Institute of Standards and Technology (NIST) database\footnote{ \url{https://physics.nist.gov/PhysRefData/ASD/lines_form.html }}. 

The Sobolev approximation, when applied to lines arising from the same upper level, allows us to overcome the radiative transfer effects that can change the relative populations of the various atomic levels \citep{beristain_permitted_1998}. Its main limitation is that it should only be applied to broad lines, since it uses the Large Velocity Gradient (LVG) approximation to treat the escape probabilities. Considering two lines from the same upper level, the ratio of the weakest to the strongest line ($r_{w/s}$) can be written as
\begin{equation}
\label{eq:sob_ratio}
r_{w / s}=\frac{A_{k i}^{w} \lambda_{s}}{A_{k j}^{s} \lambda_{w}} \frac{1-e^{-\tau_{w}}}{\tau_{w}} \frac{\tau_{s}}{1-e^{-\tau_{s}}},
\end{equation}
where $A_{ki}^x$ is the transition probability, $\lambda_x$ is the wavelength, and $\tau_x$ is the opacity for line $x$. The LVG approximation allows to rewrite the opacities as
\begin{equation}
\label{eq:tau}
\tau=\frac{h c}{4 \pi} \frac{n_{i} B_{i k}-n_{j} B_{k i}}{d v / d l} \approx \frac{h c}{4 \pi} B_{i k} g_{i} \frac{e^{-E_{i} / k T}}{U(T)} \frac{n d l}{d v}
\end{equation}
where the previously used symbols retain their meanings, and $c$ is the speed of light, $B_{xy}$ are the Einstein's coefficients, and we have introduced $\mathcal{N}=n dl/dv$, the modified column density or density-velocity gradient.

This approximation allows us to estimate the best-fitting density-velocity gradient and temperature for every pair of lines from the same upper level. The calculation resulting from a single pair of lines will be degenerated regarding both variables, but we can use all pairs of lines to derive further constraints. The main theoretical limitations of this method are those imposed by LTE and the LVG approximation, but one needs to consider further experimental constraints, in particular, in case of lines with large uncertainties in their atomic parameters and with similar strengths.

In general, none of these methods should be used to compare lines that have different profile types and/or velocities, since this is indication that the lines have different origins. Comparing lines in very different parts of the spectra can also affect the results if calibration varies across the wavelength, or due to extinction. In future work we will see if including an extinction law to different wavelengths will allow this method to function across larger ranges.

\section{The STAR-MELT Python Package}
\label{sec:star_melt}


\begin{figure}
	\centering
	\includegraphics[width=\columnwidth]{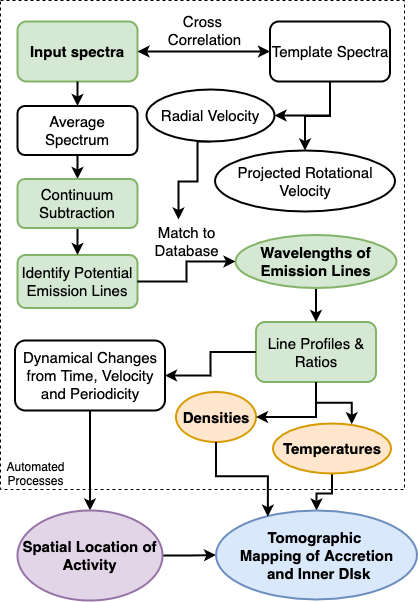}
	\caption{\SM\ flow chart, outlining the key steps in the package workflow. Steps within the outer dashed box are carried out automatically, subject to selection of emission lines for profiles and subsequent steps.}
	\label{fig:flow_chart}	
\end{figure}

We have designed the \SM\ Python package to automatically carry out as many as possible of the steps required to map the accretion and inner disc of PMS stars, whilst facilitating user interaction as needed. Broadly, this covers reading in the data for a given target star and spectroscope, determining the radial and rotational velocities using a template/standard star, identifying and matching the atomic emission lines to a reference database, and various visualisation and analysis functions (Fig.\,\ref{fig:flow_chart}). There are also several convenience functions to handle reading in data, customising output plots and exporting results. Further interactivity with \SM\, is achieved via the use of Jupyter Notebooks. This allows for dropdown widgets and sliders, such that one could run the \SM\, Jupyter notebook on their own data without the need of editing any code\footnote{A tutorial Jupyter Notebook for \SM\ is provided within the \href{https://github.com/justyncw/STAR_MELT}{GitHub repository}.}. The \SM\, package features multiple modules and functions to also allow for a fully scripted automation for obtaining summary statistics of the specified data, such as number of emission lines identified/fit (optionally for specific elements) for all target stars across all observations. \SM\ uses various \textsc{astropy} \citep{astropy_collaboration_astropy_2013,astropy_collaboration_astropy_2018} functions, in addition to further package dependencies that will be referenced in their corresponding descriptions in this section.

The relevant data are extracted directly from the reduced \texttt{FITS} spectra\footnote{See the online documentation for the up-to-date list of direct \texttt{FITS} instrument compatibility.}. The \SM\ analysis functions are designed to be compatible with a simple wavelength, flux and (optional) error data frame, for a given (modified-) Julian date observation. These can be user provided as a csv or text file from any instrument source; with a simple routine that will create the required data frame. However, there is further information that is extracted from the \texttt{FITS} headers that will aid in the analysis flow and it is recommended to use the integrated \SM\ read-in functions. The mean and median flux values for each set of observations are determined and also stored in the data frame of observations. There is a further option here to normalise/scale the flux data frame, by dividing each flux value by the median flux across the observation. This is useful for comparing data from different instruments.

Some spectroscopes have gaps in the wavelength coverage, which are automatically removed for each instrument where applicable. As are regions of the wavelength range badly affected by telluric absorptions, such as 6872 - 6920 \,\AA\, and 7592 - 7690\,\AA\ \citep{1964ApOpt...3.1401C}. Further wavelength exclusions can be specified by the user for all observations. Similarly, one can also choose to consider a smaller wavelength range than that automatically obtained from the \texttt{FITS} data. The user can also specify at this stage which of the observations (dates) they wish to analyse. This is particularly useful for large amounts of archival data, covering many years. 

 \begin{figure*}
	\centering
	\includegraphics[width=\textwidth]{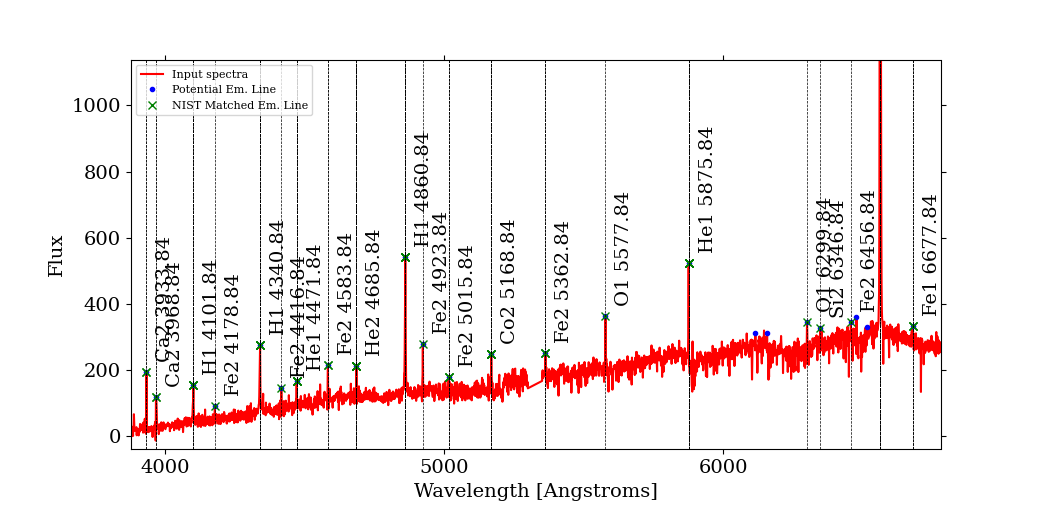}
	\caption{Emission lines identified from the median spectra of EX Lupi HARPS observations. The S/N threshold at which to identify potential lines defaults to 3$\sigma$ and can be specified by the user. Potential lines are indicated by the blue circles, with the green crosses showing those that have been successfully matched to the NIST database, within the velocity and wavelength tolerance. For the line labelling, if a line has a multiple match to the database, only the closest match will be displayed in the plot (all matches are given in the resulting list and will be filtered/selected in the following steps).}
	\label{fig:lines_id}
\end{figure*}

\subsection{Radial and Rotational Velocities}

Radial and projected rotational velocities (RV and \vsini, respectively) of the target star are calculated using the standard cross correlation function (CCF) with a template star of similar spectral type\footnote{Using the \textsc{PyAstronomy} \url{https://github.com/sczesla/PyAstronomy} functions \textsc{crosscorrRV}, \textsc{rotBroad} and \textsc{fastRotBroad} \citep{czesla_pya_2019}.}. Table\,\ref{tab:standard_stars} shows the list of standard stars provided in the \SM\, package library at the time of publication. 

To automate this process for any target star, a SIMBAD query is performed\footnote{Using \textsc{astroquery} \citep[][]{ginsburg_astroquery_2019}} on the name (identifier) of the target star to pull the most recent spectral type. \SM\ then selects the standard star with the closest spectral type from the \SM\ library. Since the RV of the target PMS stars may be variable, unavailable, or not well defined, we do not pull that information from the SIMBAD database; instead, the user can calculate it within \SM\ or supply it manually. The RV of the standard star, which is well defined, is queried from SIMBAD to allow for the calculations. Obtaining the RVs for the standard stars from SIMBAD allows for further standard stars to be added in future releases, or for the user to provide their own standard stars. \SM\ can also compute the CCFs using modelled stellar atmosphere spectra as the template. This is useful for checking that the value obtained from the standard star is correct, and for where the standard star spectral type does not well sample the target star.

Whilst the user can specify a specific observation for which to calculate the RV and \vsini, for the purposes of emission line identification, we use the median spectra across all observation dates considered. This avoids the introduction of uncertainties for each observation, leaving instead only a potential systematic offset. Providing the wavelength calibration of the reduced data is accurate, errors of a few km/s in the RV calculation would not have a large impact on matching the emission lines to the reference database. To improve the velocity accuracy for FEROS, we do, however, apply a manual barycentric correction to velocities calculated from the FEROS pipeline \citep{2013A&A...556A...3M}. Note that for cases where we intend to study the relative radial velocity variations for the lines, we recommend to derive the stellar radial velocity for one spectrum (e.g. the average or highest S/N spectrum) and use it for the rest. Therefore, the uncertainties in the line radial velocity would depend only on the line fit and on the accuracy of the wavelength solution provided by the telescope.

Our analysis primarily uses the \vsini\ for specifying the velocity range in which to match the emission lines (as discussed in Sec.\,\ref{sec:tomography}). Hence, for fast rotators where the \vsini\, calculation fails due to rapidly increasing CCFs, we assume a value of 50 km/s and set the corresponding range for velocity windows.

At present, \SM\ does not include any veiling correction nor estimate of the spectral type. In particular, veiling is not addressed because we intend to use \SM\ to study the prevalence of line-dependent veiling \citep{dodin12}, which is known to affect spectral types, radial, and rotational velocity estimates and can have particularly dramatic effects in radial velocity estimates for stars with many lines \citep{kospal_radial_2014,sicilia-aguilar_accretion_2015}, but this needs to be kept in mind when comparing with other results and line ratios from different spectral ranges. However, \SM\ is compatible with output from the ROTFIT reduction software \citep{2003A&A...405..149F,2006A&A...454..301F,2013MNRAS.434.1422M}, which we use in this work to check that our results are not affected by photospheric absorption in the lines.


\subsection{Automatic Emission Line Identification}

With the RV of the target star determined, we are now able to automatically identify and match the emission lines to our reference database (Fig.\,\ref{fig:lines_id}). The database provided with the \SM\ package is compiled from the NIST database\footnote{\url{http://physics.nist.gov/PhysRefData/ASD/lines form.html}} \citep{kelleher_new_1999,ralchenko_new_2005,a_kramida_nist_2020}. All neutral and singly ionised atomic transitions (for lines with transition probabilities) from elements H through Zn are included for the air wavelength range of 3000-20000\,\AA. This includes information about the wavelength, excitation potential (or energy of the upper level, E$_{k}$ in eV), multiplicity, relative intensity, and transition probabilities (or Einstein coefficient for spontaneous emission, A$_{ki}$), each of which are required in our analysis. Constraints on the transition probabilities and energies of lines can be applied before or after the matching process. By default, the wavelengths in air are used for the matching. The vacuum wavelengths are also included in the database and can be optionally used in the matching process\footnote{A further vacuum wavelength range of 2-50\,\AA\ is also included for X-ray data, which is not discussed in this paper but will be expanded upon in future work.}. During the matching process, we also indicate previously observed lines from the spectra of EX Lupi, ASASSN-13db and ZCMa \citep{sicilia-aguilar_optical_2012,sicilia-aguilar_accretion_2015,sicilia-aguilar_2014-2017_2017,sicilia-aguilar_reading_2020}, which can help the user to identify common lines in case of doubt.

Emission lines are automatically extracted by roughly subtracting the continuum contribution from the median-scaled flux values, using sigma-clipping, and then applying a \textsc{SciPy} Savitzky-Golay (S-G) filter \citep{1964AnaCh..36.1627S}.\footnote{\SM\ can, in principle, also detect absorption lines in the same manner, by using the lower threshold instead of the upper. Since this was not a main intent when designing the package and carrying out our analysis, the absorption line detection requires further testing.} The filter is a standard interpretive polynomial smoothing filter, we used a coefficient length of 31 with a default 3rd order polynomial; both of which can be modified by the user. These default values were a good balance between smoothing out the narrow lines completely, and only applying a minimal local shift around broad lines (see Sec. \ref{sec:discussion} for further details on the adjustments made for different types of line identification). This continuum subtraction process works well for stars with spectral types B through M. The peaks of the emission lines are then identified via another stage of sigma-clipping at a user-specified threshold level. 

The wavelength positions of these identified lines are then shifted by the stellar RV contribution and matched to the reference line list, with a default absolute tolerance of 0.2\,\AA. A further constraint is then applied to check that the shift introduced by the RV is less than the dispersion resulting from the \vsini. This absolute and velocity tolerance can also be set by the user for broader or highly shifted lines. 
 
The positional list of matched lines are saved as a data frame, along with the corresponding atomic properties from NIST. This can be downloaded and stored locally. It can also be re-read by the package so that future analysis can refer back to the already determined list of lines. If an identified line has more than one possible match in the database, this is indicated by the `multi' column in the data frame; allowing the user to further investigate which of the lines are the most likely matches (see notes to Fig.\,\ref{fig:lines_id}). The steps thus far in the \SM\, package are computationally inexpensive, with the largest increase depending on the number of observations. It is possible that some lines are misidentified due to noise in the spectra. Further refinement to the matched lines can be completed by fitting simple models to the line profiles. 

 The model line fitting used in \SM\ employs a goodness of fit measure to remove badly fitted (and potentially misidentified) lines, hence producing a final list of refined matched lines. Again, all such information can be downloaded at any stage for future reference. The line fitting has to be carried out on a one-by-one basis, taking some time to fit all identified lines (tests result in $\sim$20 minutes for $\sim$500 lines, using a computer with a 2.3 GHz Quad-Core Intel Core i5 processor and 16 GB 2133 MHz LPDDR3 RAM). Therefore, if the user does not wish to carry this out for all of the lines, selections can be made from the list of matched lines to decide which ones to fit, for example, all Fe lines with an excitation energy less than a given value. With the selections made for which lines to fit, the user then specifies for which observation dates, or average spectra, they would like to fit the lines for. If an overview of the lines is required, the average spectra yield much better signal/noise ratio than the individual lines. If temporal investigations is preferred, all observation dates of interest can be selected and fit.

\subsection{Line Fitting}

To perform the line fitting, \SM\ uses the Python package \textsc{lmfit}\footnote{\url{https://lmfit.github.io/lmfit-py/index.html}} \citep{newville_lmfit_2014}, which carries out non-linear least-squares model fitting. The advantage of using the \textsc{lmfit} package is that parameter attributes (such as the central position, or $\sigma$ values) for the variables used in the model fitting can be either be automatically obtained and set from the observational data itself; or manually constrained by the user. In the case of the latter, the constraints can be set interactively within the Jupyter notebook, and do not require the editing of any code. Furthermore, all best-fit model parameters, along with the model data, are easily obtained from the model class. Since the line profiles may be very complicated, \SM\ is designed to provide simple, yet accurate, model fits that allow for detailed comparisons of parameters.

Up to four Gaussian components can be fitted for each emission line. The default fitting option is to adopt a model comprising one Gaussian plus a linear component to approximate the emission line plus the continuum, respectively. For lines that possess better signal/noise (S/N) and/or more complexity, up to two more positive Gaussian components can be included. A second positive Gaussian can identify narrow components within broad components of the lines (e.g. figures within CVSO109 discussion, Sec.\,\ref{cvso-sect}). A negative Gaussian can also be combined with any of these positive Gaussians, which can model absorption components from winds/outflows/infall (e.g. figures within GQ Lupi A discussion, Sec.\,\ref{gqlupi-sect}). In addition to the component parameters, the module also returns calculated parameters such as the integrated flux and line asymmetry. The option to fit further line profiles, such as Lorentzian and Voigt models, will be incorporated in future development of the package. A minimum goodness-of-fit (the reduced \chisq statistic) can be specified to only return those fits meeting the criteria, and thus provide the user with a set of well-fitted, identified emission lines. This helps to remove spurious lines or those that are not present on certain dates.

With the selections made for which lines to be fitted specified (such as a given set of elements with a maximum excitation potential), the fitting process described above is then fully automated within \SM. The list of matched emission lines is used to subset the main flux data frame around the given wavelength range for each identified line. This function is also a convenience function that can be called to produce a plot of the lines, with either wavelength or velocity as the dependent variable. For the line fitting, the velocity dispersion around the specified line position is required. The velocity range to consider can be specified from the \vsini\, of the star, or be manually set by the user. When fitting multiple lines, this is carried out iteratively for each identified line from the previous \SM\, functions.

Whilst line profiles can be explored directly from the reduced observational data, which require more individual, bespoke analysis and discussion, the parameters from the model-fits such as the central positions can be used to generate periodograms to probe overall variations for the temporal variability of the emission lines. The integrated flux values from the model-fits can be used to calculate observed line ratios, which can be compared to their theoretical counterparts, as described in Sec.\,\ref{sec:tomography}. Further discussion and examples are given in our analysis of three YSOs in Sec.\,\ref{sec:discussion}. 

\begin{figure}
    \centering
    \includegraphics[width=\columnwidth]{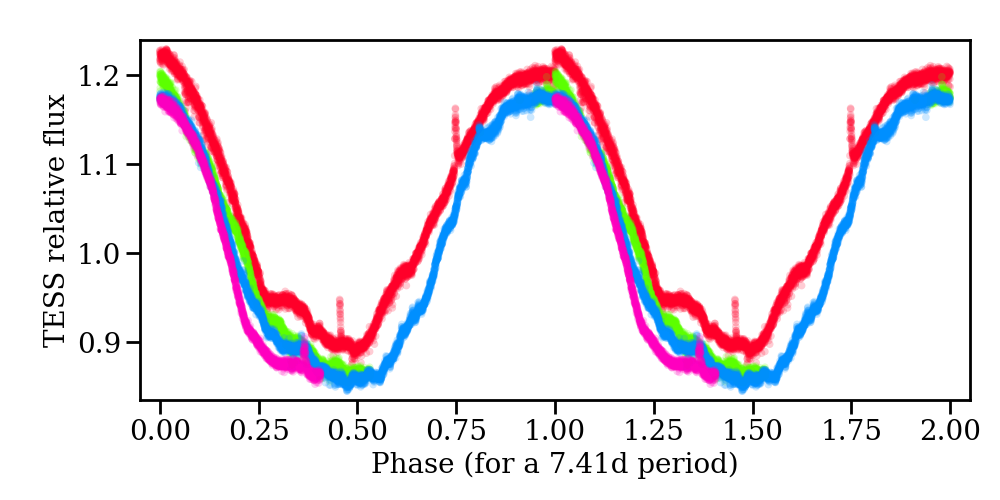}
    \caption{\textit{TESS} data for EX Lupi, wrapped according to the 7.41d period. The data are colour-coded according to different period cycles.}
    \label{exlupi-tess}
\end{figure}

\begin{figure*}
    \centering
    \begin{tabular}{cc}
    \includegraphics[width=7cm]{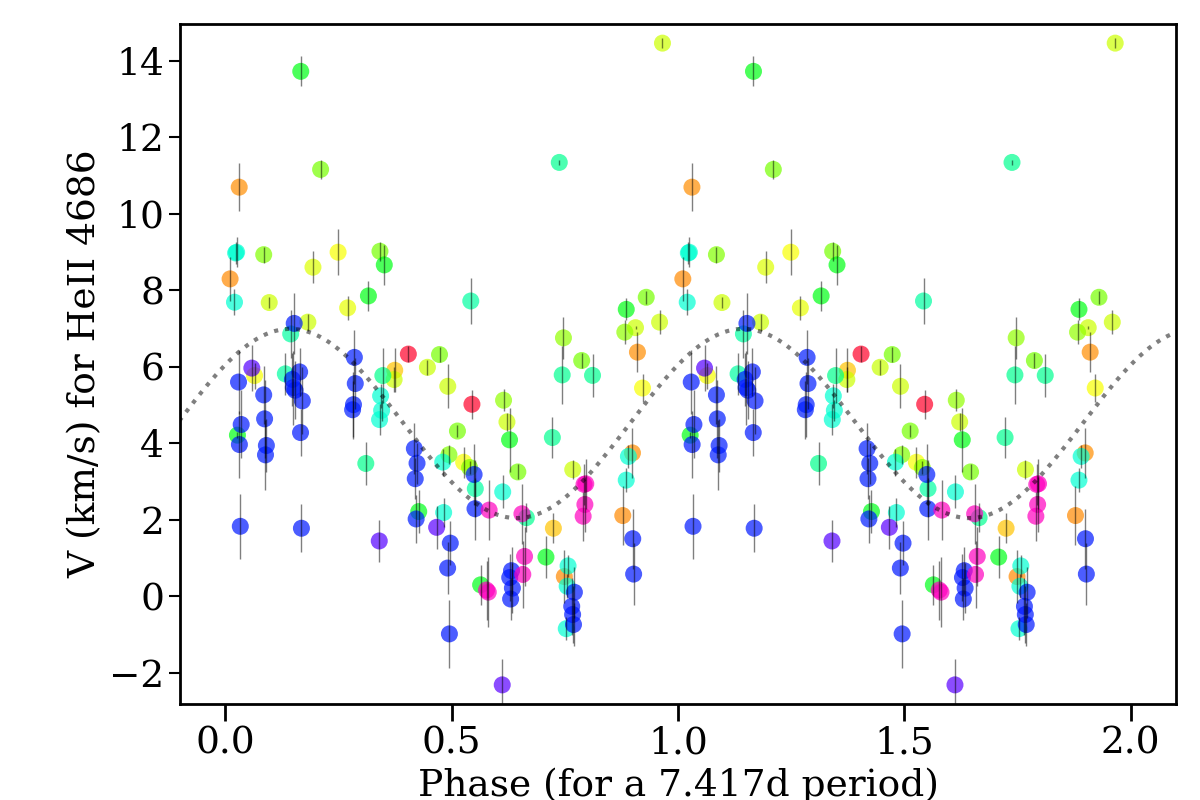} &
    \includegraphics[width=7cm]{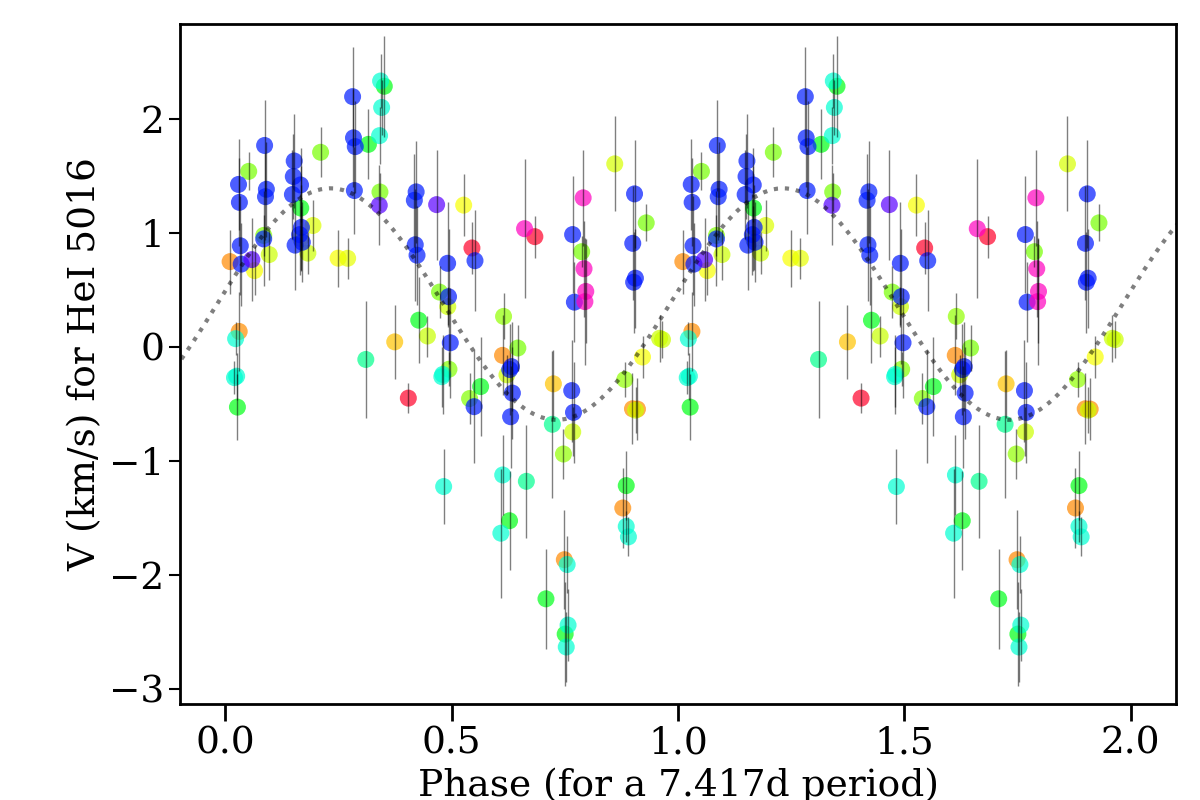}\\
    \includegraphics[width=7cm]{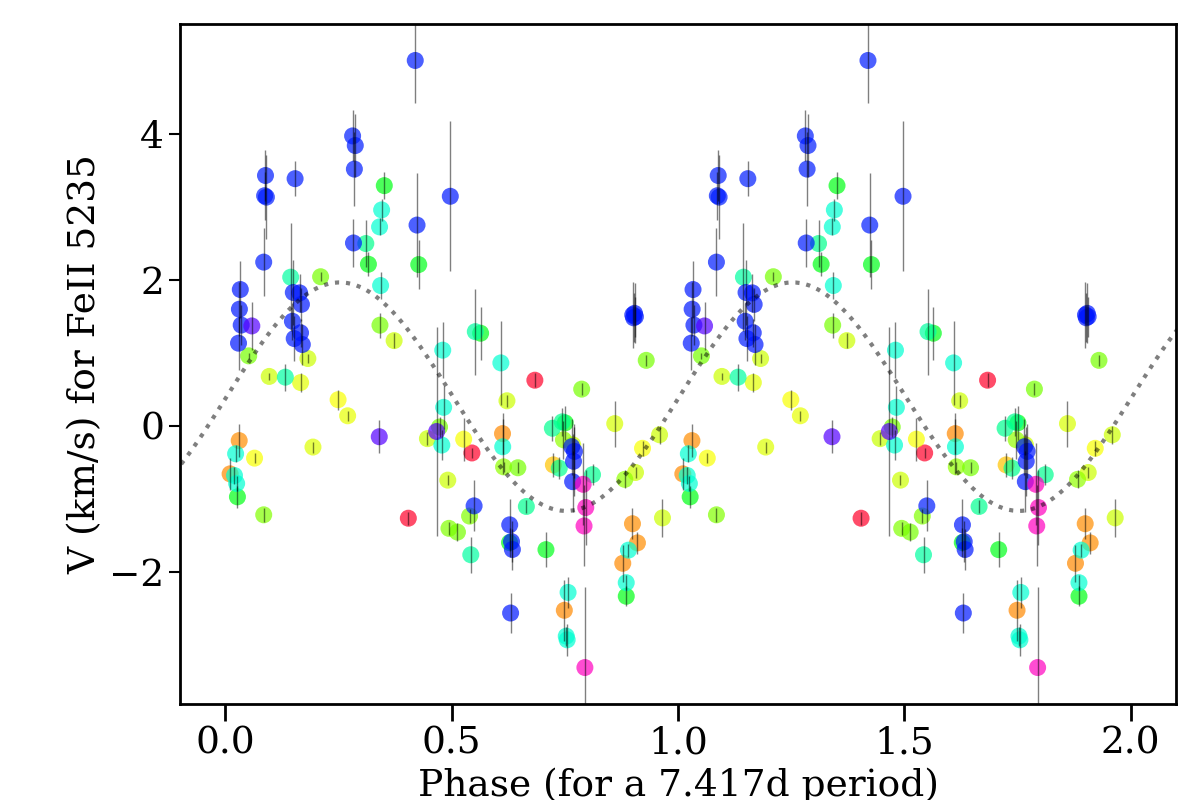} &
    \includegraphics[width=7cm]{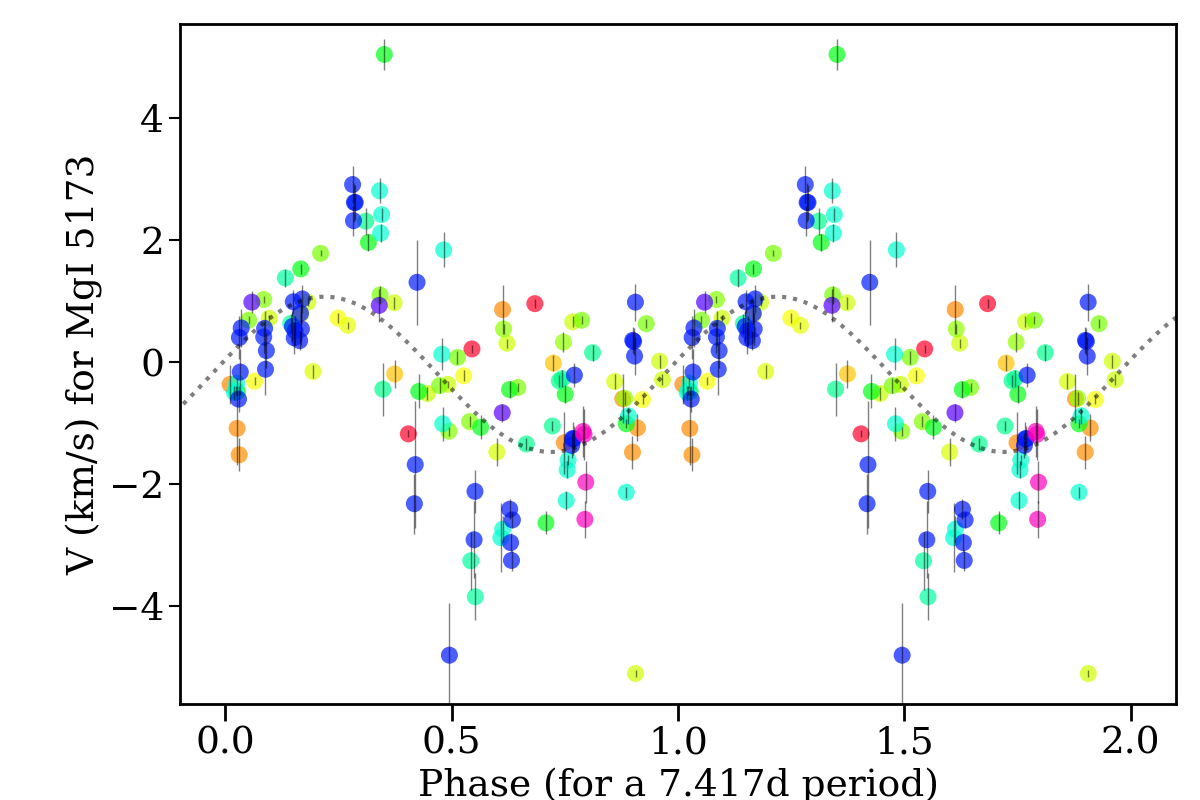} \\
    \includegraphics[width=7cm]{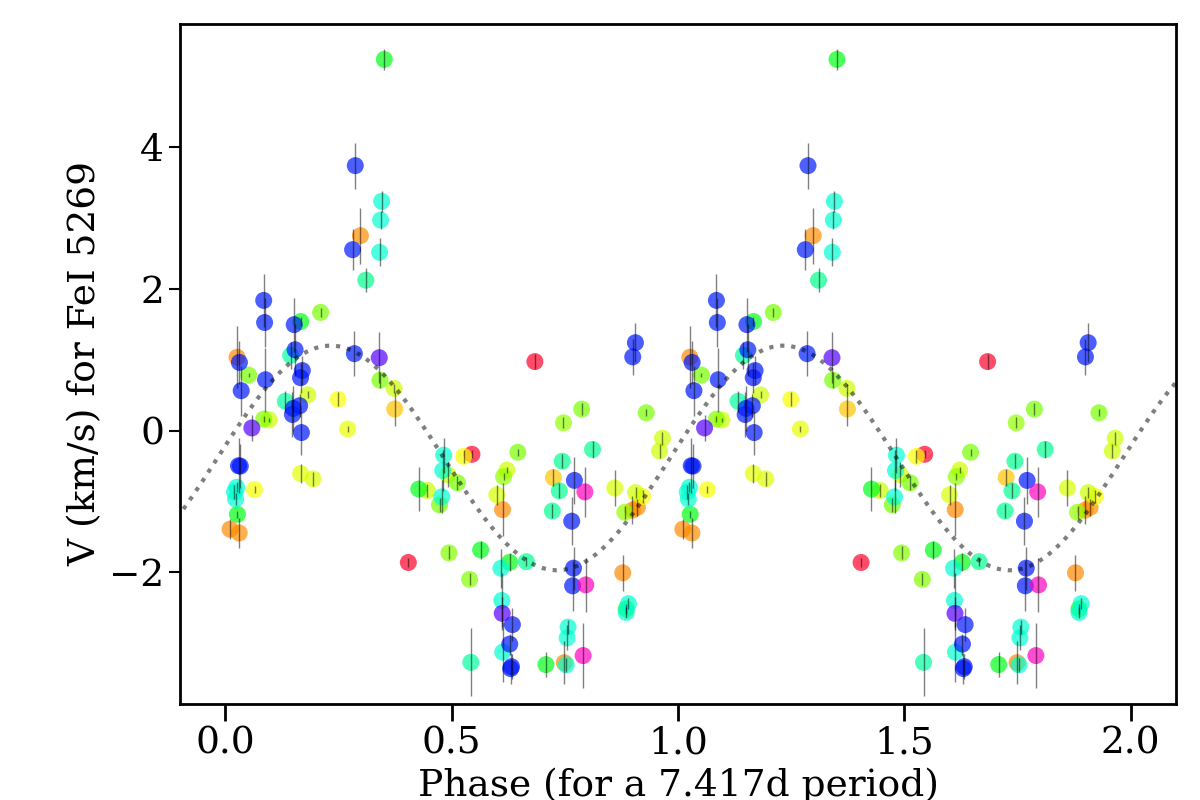}&
    \includegraphics[width=7cm]{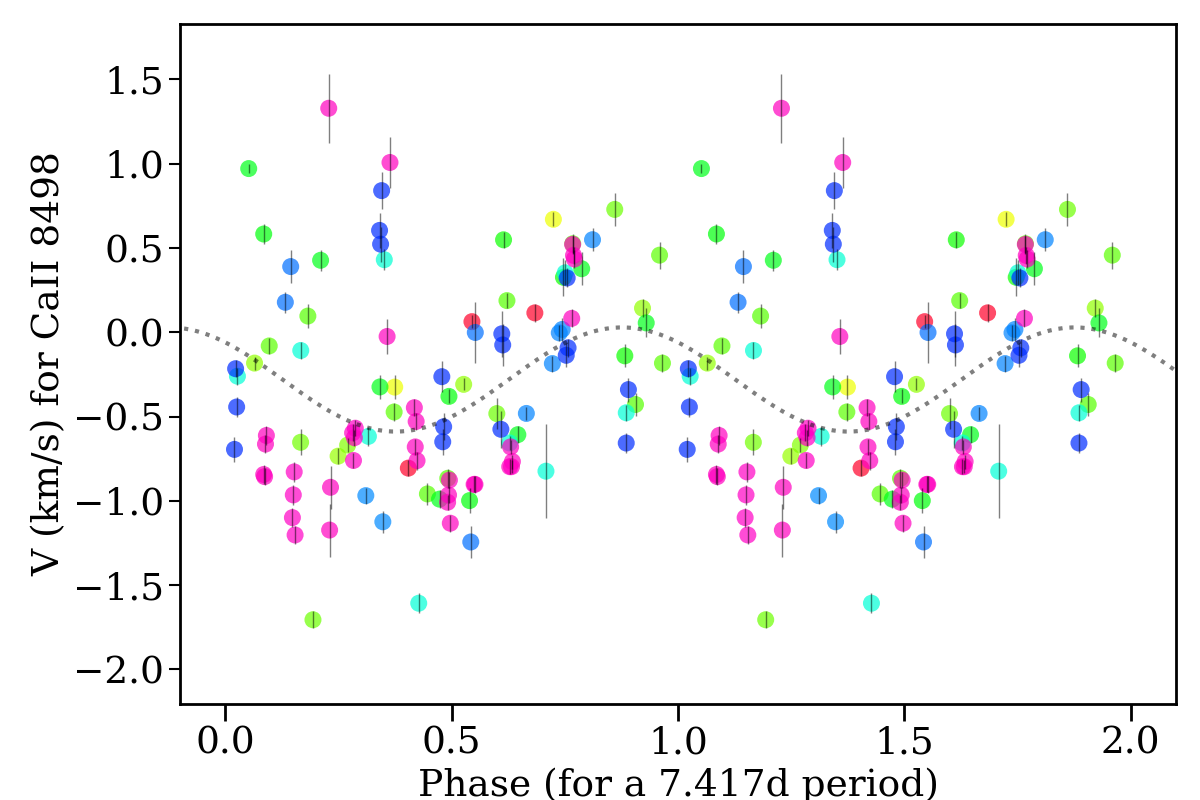}
       \end{tabular}
    \caption{Some examples of periodicity observed in the velocities of the lines extracted with \SM\ for EX Lupi, all of them wrapped with the 7.417d period obtained for the He II 4686\,\AA\ line. The colour scale is set to reflect the number of periods elapsed since the first observation, and changes from red to purple from the first to the last MJD displayed. The dotted line is the best fit for the expected modulation (equation \ref{rota-eq}). The phase for the He II 4686\AA\ and Ca II 8498\,\AA\ lines are notably different from the rest.}
    \label{exlupi-line-period}
\end{figure*}

In terms of the \SM\ work-flow, we have now obtained the necessary analytical information from the observational data to fully investigate the emission lines across our observations. Although there are various options the user may wish to specify for their data, all such options have default values. Thus, in principle, everything up to this stage can be fully automated by \SM, with good compatibility with the different instruments. The physical approximations described Sec.\,\ref{sec:tomography}, and the following discussion section pertain to stellar accretion investigations. However, it is clear that up to here, \SM\ has numerous potential applications to aid in identification and fitting of all kinds of astronomical emission lines. The \SM\ package is available to the community on \href{{https://github.com/justyncw/STAR_MELT}}{GitHub}\footnote{\url{https://github.com/justyncw/STAR_MELT}}, along with an example Jupyter notebook that features all steps outlined in this Section and example data from this work. The \SM\ notebook can be run in an online executable environment via the \href{https://mybinder.org/v2/gh/justyncw/STAR_MELT/HEAD?filepath=STAR_MELT_example_notebook.ipynb}{binder platform}\footnote{\url{https://mybinder.org/v2/gh/justyncw/STAR_MELT/HEAD}}. We also welcome collaborative projects using the package.

\begin{figure}
    \centering
    \includegraphics[width=8cm]{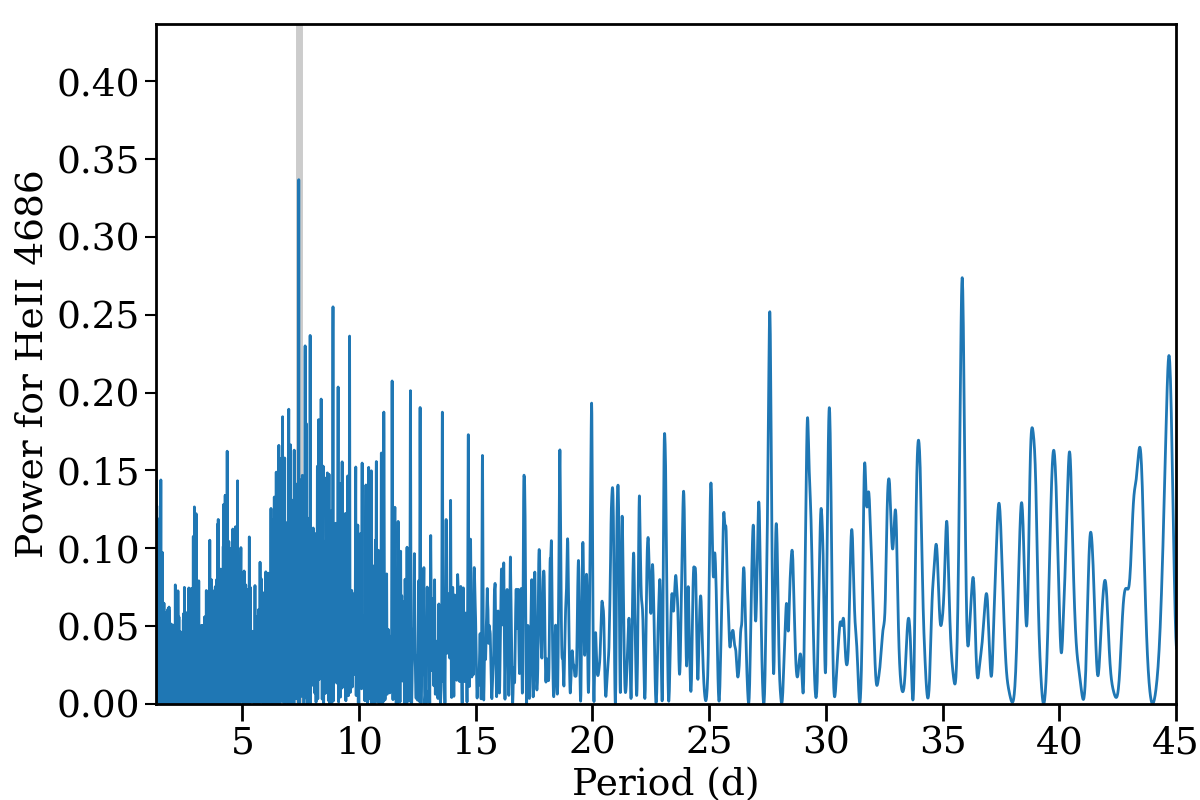}
    \caption{Lomb-Scargle periodogram revealing the 7.417d period from the radial velocity of the He II 4686\,\AA\ line as observed for EX Lupi.}
    \label{exlupi-gls}
\end{figure}


\section{STAR-MELT application \& discussion}
\label{sec:discussion}

The application of \SM\ is demonstrated by analysing three YSOs: EX Lupi, GQ Lupi A and CVSO109. We have studied EX Lupi in previous work \citep{sicilia-aguilar_optical_2012,kospal_radial_2014,sicilia-aguilar_accretion_2015}, and we detail here the comparisons of the results achieved using \SM; along with investigations of the YY Ori-type profiles in GQ Lupi A and variability analysis of CVSO109 from the ongoing PENELLOPE survey \citep{2021arXiv210312446M}.

\subsection{EX Lupi}

EX Lupi is a M0 star \citep{herbig77,sipos09,alcala17} and the prototype of EXor variables; a class of low-mass young stars with repetitive accretion-powered outbursts that recur within timescales of months to years \citep{herbig01,herbig08}. Although very large outburst where the accretion rate increases by 2-3 orders of magnitude or more are rare and have been documented only twice in the 1950's and in 2008 \citep{herbig77,jones08}, smaller accretion variations are observed on a regular basis \citep{lehmann95,herbig07,sicilia-aguilar_accretion_2015} and we are considering them as part of the quiescence stage.

EX Lupi spectra contain a wealth of emission lines in both outburst and quiescence, with complex profiles that include broad and narrow, strongly variable components \citep{aspin10,sicilia-aguilar_optical_2012,sicilia-aguilar_accretion_2015}. The broad components are more evident in outburst and at times when the accretion rate is high. The narrow lines are present at about all times (with some variations in intensity and S/N) and  originate in the accretion post-shock region at the footprint of the accretion column over the stellar surface. The lines display rotational modulation in quiescence, which is a signature of the accretion structures being remarkably stable over time \citep{sicilia-aguilar_accretion_2015}.

Because of the number of lines and their stability, EX Lupi is an ideal target to test the capabilities and functionality of \SM. We used the data available for EX Lupi during its quiescence period, which contains the epochs that have been studied in more detail regarding line velocities and periodicities \citep{sicilia-aguilar_accretion_2015}, plus some more recent data. These include the data obtained with FEROS, HARPS and ESPaDOnS (see Appendix\,\ref{app:inst} for instrument descriptions). We use the data from 2007-2012 that had been previously analysed \citep[excluding the data from the 2008 outburst;][]{kospal_radial_2014,sicilia-aguilar_accretion_2015}, together with archival observations obtained between 2013-2019, with a total of  152 observations (see Tab.\,\ref{tab:obs_log}). Although we exclude the data obtained during the large 2008 outburst, the data cover epochs where the accretion varies between a factor of few to one order of magnitude. For the analysis, regions strongly contaminated by telluric features were excluded, which affects the 5500-5700, 6760-6970, 8150-8390, and 9100-9250\,\AA\ ranges, primarily. The radial velocity of EX Lupi (around 0 km/s) is recovered (-0.05$\pm$0.33 km/s) by \SM. The \vsini\ derived has an upper limit of 5km/s, with the main limitation being the slow rotation of the target compared to the \SM\ template. This value is consistent with the previously determined \vsini\ of $\sim$ 4.4$\pm$2 km/s \citep{sipos09}, which we adopt in the following analysis.

Careful continuum sampling was found to be important for the extraction of the broadest lines (such as the H Balmer series or the Ca II IR triplet). In general, averaging over a window of 2\,\AA\ and fitting the continuum with a 3-degree polynomial produced the best result overall to extract the metallic lines. The continuum becomes uncertain in the proximity of very broad lines, so that for stars with very different types of line profiles, it is more efficient to run \SM\ twice and extract and fit separately the broad and the narrow lines. Our analysis here is based on the narrow metallic and He lines, which have FWHM under 50 km/s.

Because the lines are in general broad, we experimented with various criteria for automated line finding. The best criterion to identify a maximum number of lines while minimising the contamination by features that are not real was found to be between 2 and 2.5$\sigma$, where $\sigma$ corresponds to the standard deviation per pixel of the continuum. 
The main issues observed in the automated line identification occurred in the regions with telluric features, as well as for lines that appear both as a narrow emission and a photospheric absorption feature. Those include a large number of the Fe I lines \citep[see][]{sicilia-aguilar_accretion_2015}, and a detailed photospheric fit with a template with a similar spectral type will be needed to be able to extract such lines, to be implemented in the future. Non-existing lines identified automatically in regions where the continuum is uncertain due to tellurics, bad pixels, or nearby strong lines can be easily filtered out by fitting the lines. The best fitting window depends on the line properties, and thus for instance the broader He I and He II lines were best fitted using a window of $\pm$50 km/s, while narrower Fe I and Fe II lines can best fitted with a $\pm25$ km/s window.

All the strong lines flagged in \citet{sicilia-aguilar_accretion_2015} could be recovered and fitted (see Tab.\,\ref{tab:lines}). One of the main potential issues of the automated continuum fit could be to introduce noise in the velocity of the lines that invalidates the periodicity studies \citep{kospal_radial_2014}. Reproducing the periodic velocity signatures observed in \citet{sicilia-aguilar_accretion_2015}, who reported a rotational period of 7.41\,d, is thus a good test for the robustness of \SM. This period is now also confirmed by \textit{TESS} data (see Fig.\,\ref{exlupi-tess}) obtained via the Barbara A. Mikulski Archive for Space Telescopes (MAST\footnote{\url{https://mast.stsci.edu}}). The periodicity observed in the lines is clearly recovered (see Figs.\,\ref{exlupi-line-period} and \ref{exlupi-gls}), as well as the offset in the phase between the very energetic He II line and the Fe lines. As previously found, the modulation is weaker in very strong lines that are expected to be produced over larger and more distributed regions around the stellar surface, but has improved in significance when extracted with \SM. For instance, the modulation for the Ca II IR lines, despite having a very small amplitude, produces cleaner results when extracted more consistently with \SM; and suggests a difference in phase with respect to the rest. This could indicate varying temperatures on the hot spot(s) over the stellar surface, and will be studied in detail subsequent work. Shorter periods and worse-defined rotational modulation can be also a signature of more than one spot (or a more extended region in longitude) producing line emission. In general, we find that the \SM\ extraction is more self-consistent and the possibility of choosing a fitting window appropriate for each line improves the results even for the weaker lines that have lower peak-to-peak amplitude in their velocity modulations.

\begin{figure*}
  \includegraphics[width=14cm]{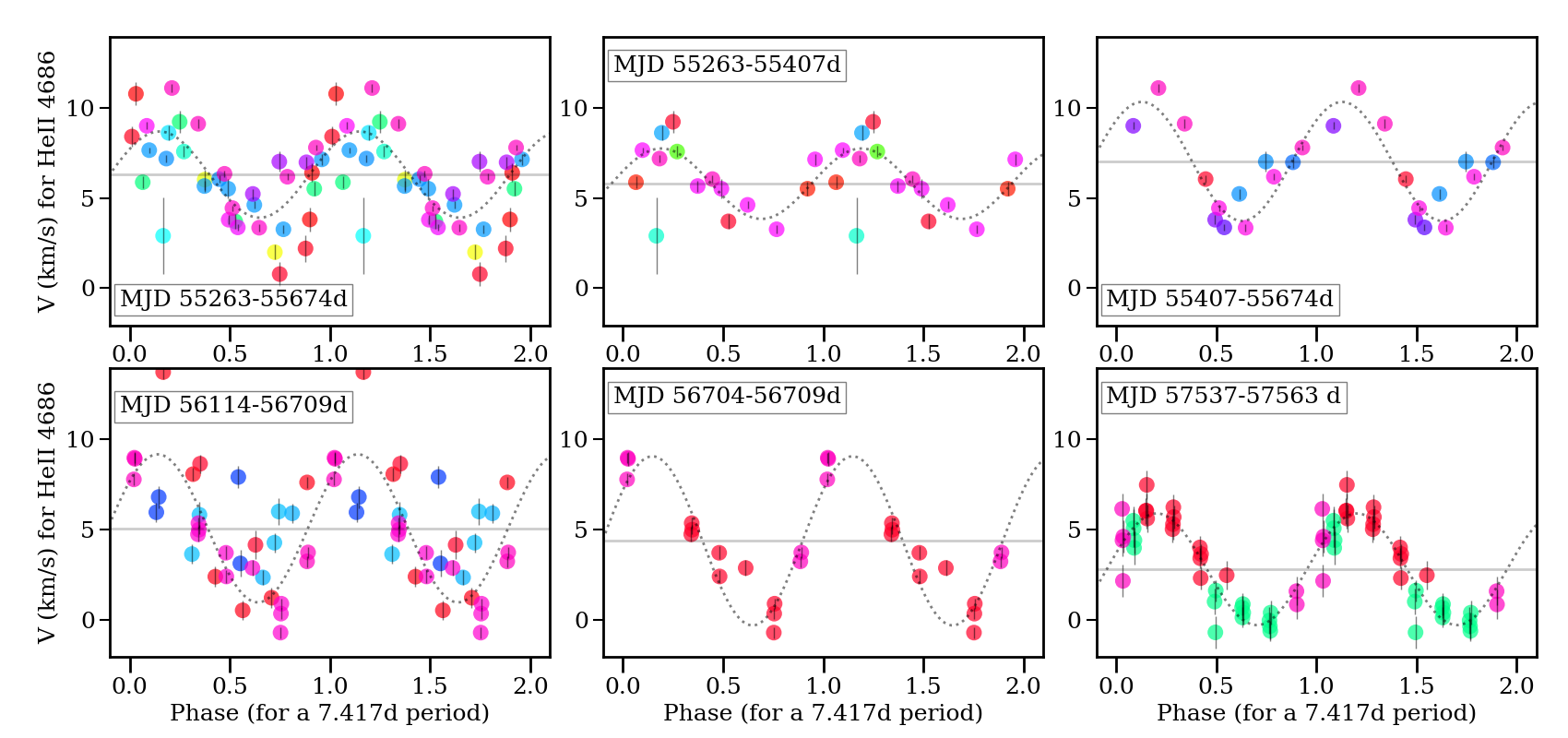} 
    \caption{Example of longer timescale variations observed in the He II line  for EX Lupi, for different timescales and epochs. Only data with goodness-of-fit$<$0.1 are plotted. The colours indicate the epoch at which the data were obtained, as a continuous scale from red (the initial MJD) to purple (the final MJD). The dotted line is the best fit for the expected modulation (equation \ref{rota-eq}.). The horizontal line shows the offset or central position of the modulation ($V_0'$).} 
    \label{exlupi-rvtime}
\end{figure*}

As in \citet{sicilia-aguilar_accretion_2015}, the amplitudes of the observed modulations can be related to the latitude of the spots that originates them. We observe the same trend, with the He II line displaying the largest amplitude, Fe II lines having amplitudes around 1-2 km/s, and CaII IR lines having often a nearly-negligible modulation. \SM\ can more efficiently extract some of the lower energy lines such as Fe I and Mg I, revealing now significant modulations. This, together with the increased number of datapoints, allows us to estimate the rotational period from many other lines, confirming that the large majority of lines are consistent with a rotational modulation with period $\sim$7.4d (see Table \ref{exlup-rvline}), with the exceptions found for lines with few points and/or low amplitude.
In general, the modulation can be written as
\begin{equation}
    RV_{line} = V_0' + vsini\, cos \theta_s\, sin \phi, \label{rota-eq}
\end{equation}
where $V_0'$ is an offset velocity after correcting for the stellar radial velocity (which may depend on any residual motion such as infall, on optical depth effects, and may also have a non-negligible systematic uncertainty for some lines\footnote{Different references provided by NIST often give line wavelengths with systematic differences of 1-2 km/s, so that line-to-line zero-point variations below this level have to be treated with extreme caution. Note that such an offset is systematic, so relative radial velocity variations are not affected. }), \vsini\ is the projected rotational velocity, $\theta_s$ is the latitude of the spot, and $\phi$ is the phase. Results for some lines are given in Tab.\,\ref{exlup-rvline}.  Therefore, for lines originating on a spot at or very near the stellar surface, the maximum modulation observed will be \vsini\ \citep{bouvier_magnetospheric_2007}. This is consistent with our observations, confirming the origin in the post-shock region at the footprint of the accretion column and very close to the stellar surface. Lines originating in a raised structure could have modulations larger than \vsini.

\begin{figure}
    \centering
    \includegraphics[width=8cm]{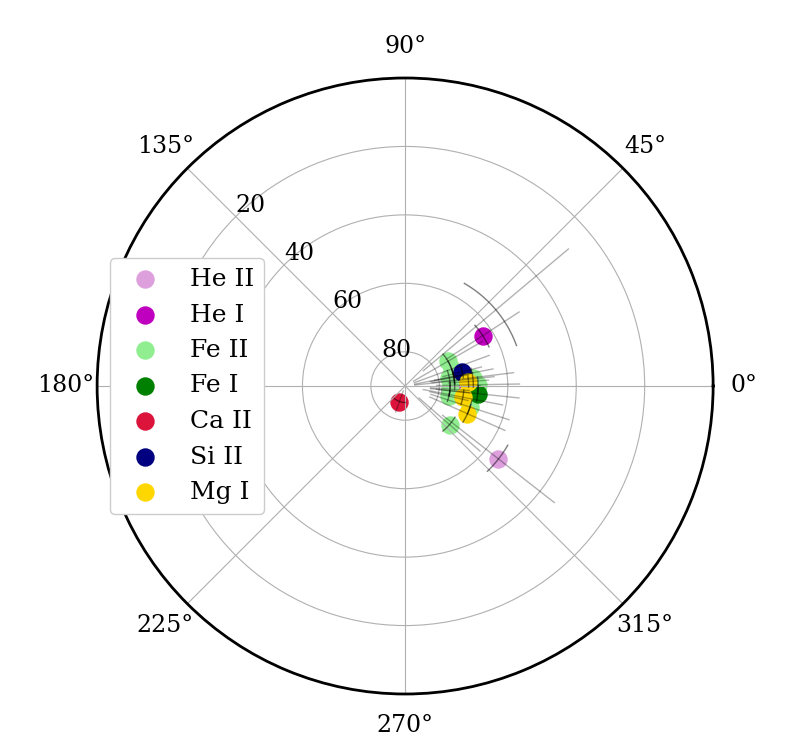}
    \caption{Location of the line-emitting region for different species for EX Lupi, assuming that it is located on the stellar surface.The lines are labeled according to their species, and the data is shown in polar coordinates, with the stellar pole closer to our line-of-sight in the centre. The relative phase of the lines is converted to longitude over the stellar surface.}
    \label{fig:exlupi-polar-species}
\end{figure}

\begin{figure*}
    \centering
    \begin{tabular}{cc}
    \includegraphics[height=7cm]{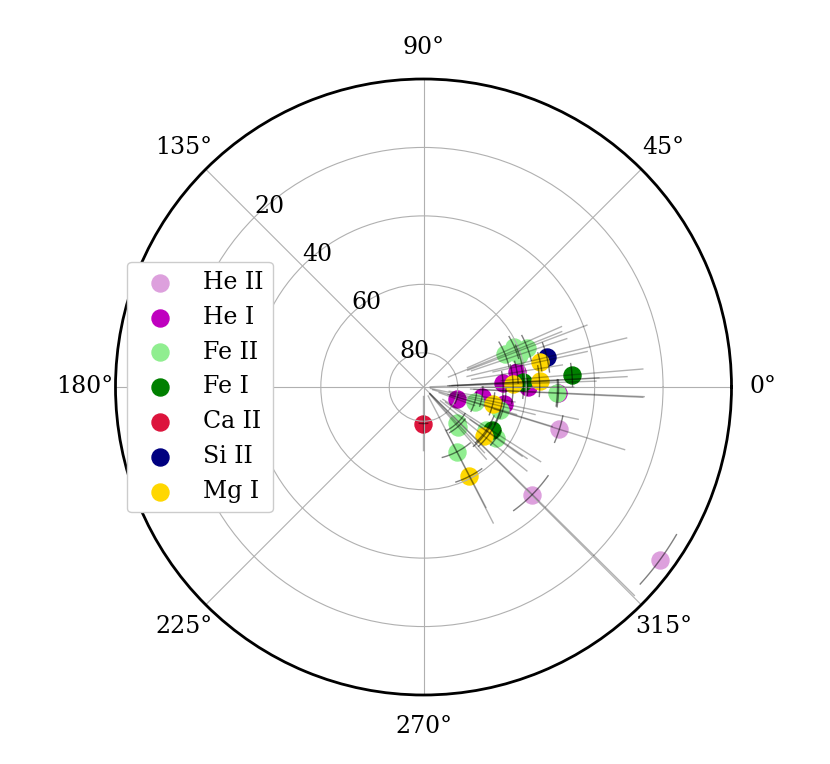} &
    \includegraphics[height=6.7cm]{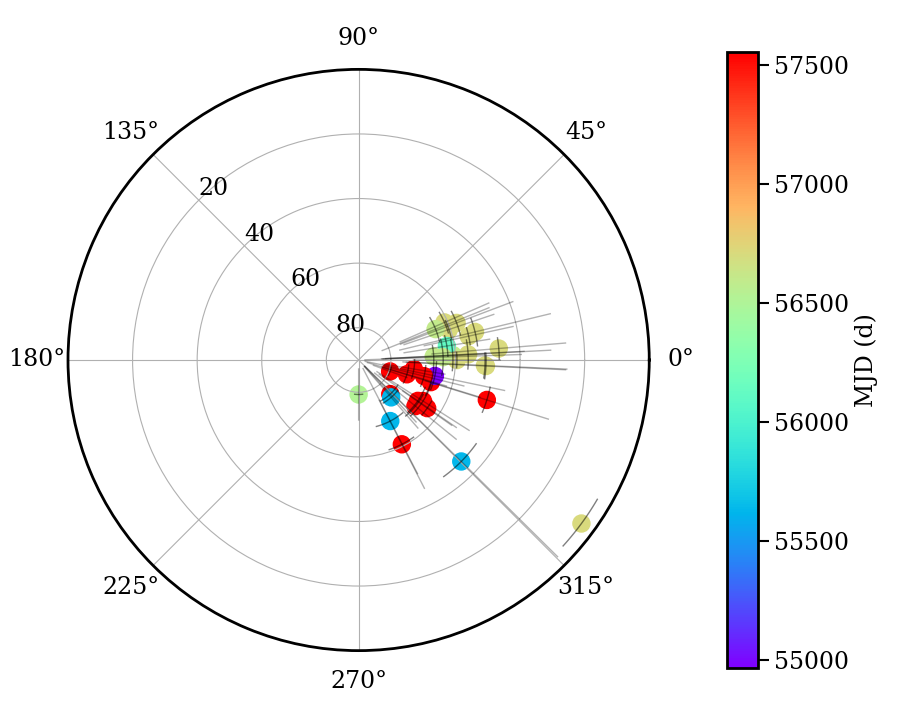}\\
    \end{tabular}
    \caption{Location of the line emission in EX Lupi, assuming that the spot is located on the stellar surface. The location is derived by fitting the line radial velocity for various time intervals that contain a significant number of points. The left panel displays the location of lines from various species. The right panel shows the variation according to the central MJD of the corresponding interval.}
    \label{fig:exlup-polar-time}
\end{figure*}

 The amplitude and phase of the line radial velocity modulation thus depends on the location of the spot. In addition, a double or multiple surface spot would tend to produce a shorter period \citep[e.g. as observed in RX J1604.3-2130A;][]{sicilia20-j1604}, and in general, emission originating from more than one structure (e.g. over the surface and/or along an extended accretion column) could dilute the rotational modulation. For lines with rotational modulation and radial velocity amplitudes below \vsini\, we can therefore assume that the spot is very close to the stellar surface and thus derive its latitude and relative phase. The results of this exercise are listed in Table \ref{exlup-rvline} and shown in Figure \ref{fig:exlupi-polar-species}. \SM\ allows us to extract lines with higher accuracy. The amplitudes of the modulations for all lines are higher as extracted with \SM\ compared to previous, more manual extraction \citep{sicilia-aguilar_accretion_2015}, and the uncertainties in the phase are lower. The improved line fits constrain the location of metallic lines to a small spot on the stellar surface, which is visible at all times. The main uncertainty in the latitude does not come from the method, but reflects the fact that the line amplitudes vary in time, which is an indication of a possible change in latitude over the stellar surface, and/or vertically. The uncertainty in \vsini\ would produce a systematic offset.

The differences in phase and amplitude between the He II line and the rest \citep{sicilia-aguilar_accretion_2015} are also confirmed, although the new data also reveal that the amplitude of the He II line is very variable over this longer period of time (see Figure \ref{exlupi-line-period}).  Such a phase difference may indicate a temperature structure either along the stellar surface or on the vertical line, with line emission from material at slightly different latitudes or heights over a vertically-extended column. The lines are also weaker during periods of lower accretion, which causes some of the lines to not be detected across all epochs, although the periodicity remains the same. 

There is evidence for longer-term variations in offset velocity, amplitude, and phase  (see Fig.\,\ref{exlupi-rvtime}). 
To disentangle the effects of time variations from variations in spot properties over the stellar surface, we fitted Eq. \ref{rota-eq} to the radial velocity of the strongest lines in different MJD intervals (see Figure \ref{fig:exlup-polar-time} and Table \ref{tab:exlupi-polar}). The time intervals were set by the requirement of having at least 10 datapoints for the fit to be significant. This means that they include several periods and have a duration of weeks to nearly a year, which may dilute any variations occurring on shorter timescales. The conclusion is that, although there are clear latitude and longitude variations in the location of the line-emitting region, this location remains small  (spot-like) and within the same quadrant during the nearly 7 years for which the time analysis is possible. Therefore, the filling factors of the line-emitting region remains always small, which could also explain the very high temperatures observed despite the low quiescence accretion rate.

\begin{figure}
    \centering
    \includegraphics[width=\columnwidth]{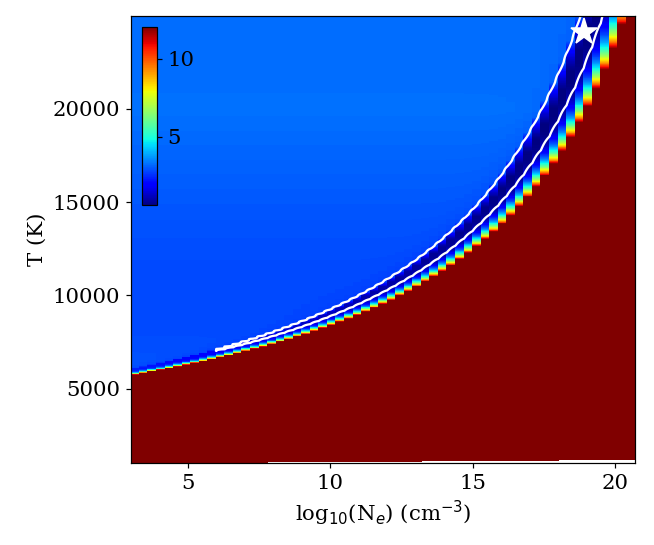}
    \caption{Results of the Saha equation applied to the 7 He lines identified in EX Lupi by \SM. The colour scheme represents the relative variations in the line intensity ratio for each combination of temperature and electron density (such that 0.1 would correspond to 10\% average deviations). The white contour marks the regions for which the relative variations of the line ratio are up to 3 times the minimum value observed, which is considered as our best-fit region, and a white star mark the best fit.}
    \label{exlupi-sahaHe}
\end{figure}

\begin{figure*}
    \centering
    \begin{tabular}{ccc}
    \includegraphics[width=5.5cm]{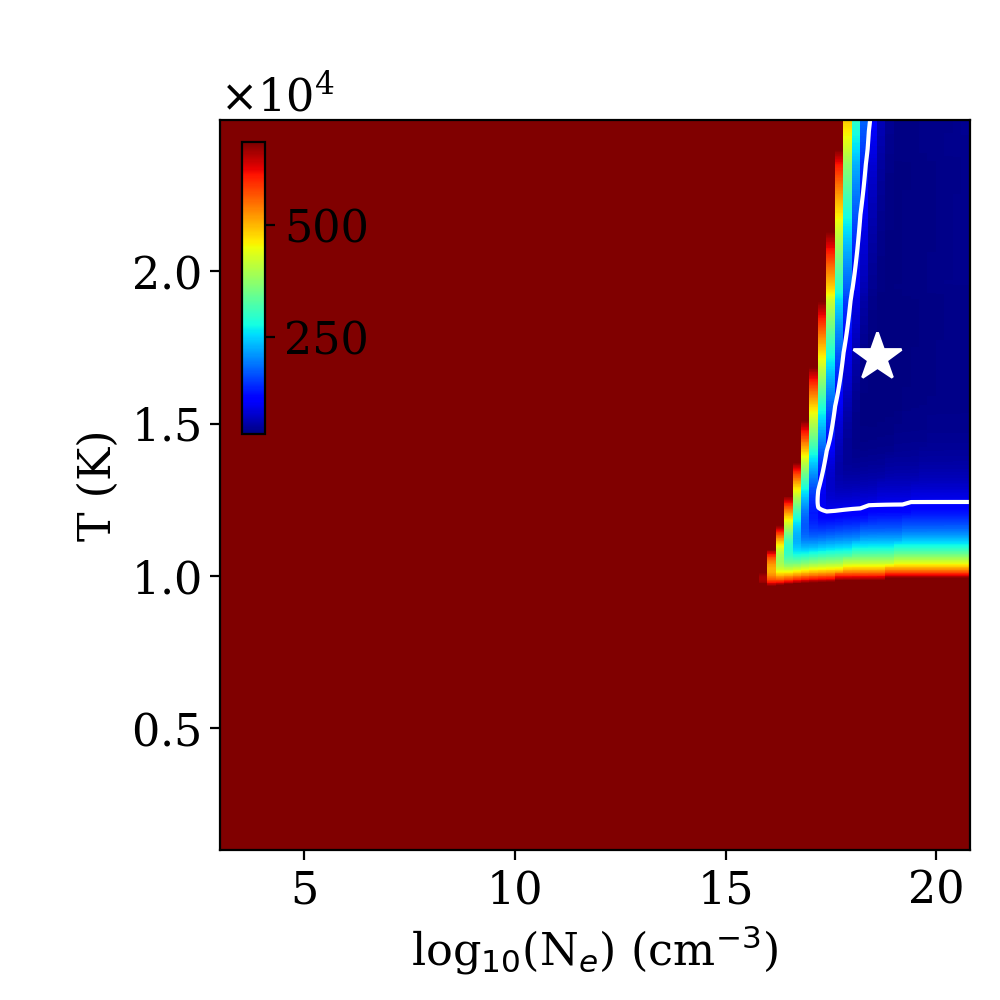} &
    \includegraphics[width=5.5cm]{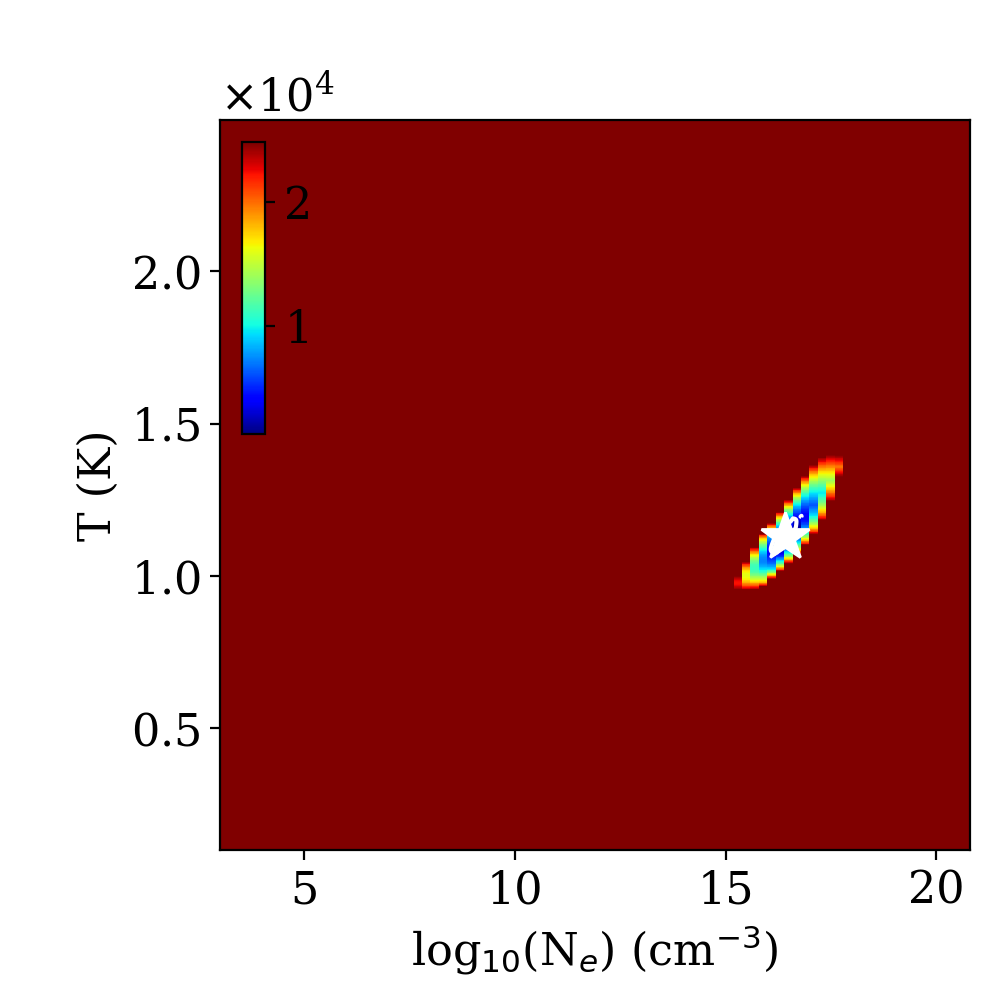} &
    \includegraphics[width=5.1cm]{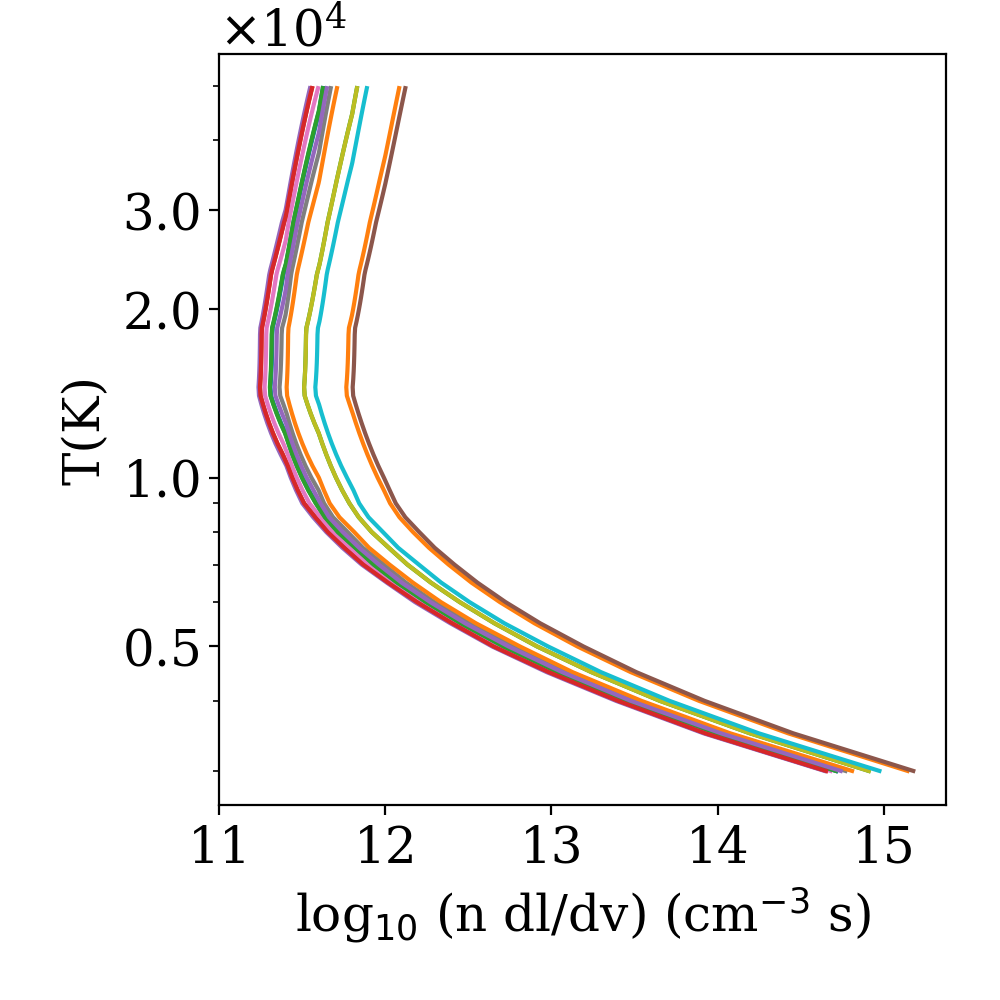} \\
    \end{tabular}
    \caption{Results of the Saha equation and Sobolev LVG approximation analysis for the FeI/FeII lines identified in EX Lupi. From left to right: Saha results of fitting all 18 lines selected in velocity, Saha results of fitting 5 lines with high quality data and similar wavelengths, Sobolev results for the Fe II pair at 5316.6 and 5534.84\,\AA\ (each curve represents the theoretical values for the observed line ratio on a given observation date).}
    \label{exlupi-sahaFe}
\end{figure*}

The second result we tested with \SM\ is the temperature ranges observed in EX Lupi. Two main sets of lines can be used for this, including the He I/He II and the Fe I/Fe II lines. The results for He are consistent with the high temperatures (over 20,000 K) estimated in \citet{sicilia-aguilar_accretion_2015} (see Fig.\,\ref{exlupi-sahaHe}).  The very high temperatures and densities \citep[e.g. compared to][]{hartmann_accretion_2016} indicate, as had been pointed out before, contributions of non-thermal ionisation. This is one of the main limitations of the calculation, together with the fact that the line profiles differ slightly between ions and neutrals, so that as before, the results are dominated by the single He II line and probably suggest a temperature higher than the regions where much of the more extended wings of He I are produced. There are a total of 20 Fe I/II lines that are identified and strong enough to produce reliable fits. Among those, two of them show very large velocity offsets ($>$3 km/s in absolute value) that suggests that they may correspond to different species and were thus excluded from the analysis. The remaining 18 produce a significantly lower temperature compared to He, as expected. Further selection based on data quality and on similar wavelengths (4900-5400\,\AA) to minimise possible effects of variations of the underlying continuum tighten further the result (see Fig.\,\ref{exlupi-sahaFe}, left and middle). One of the main limitations of the Fe I/Fe II result is that at present, many of the weaker Fe I lines are not well-fitted due to the presence of photospheric absorption, which will be addressed in future versions of \SM. 

We also show the results from the Sobolev LVG approximation for a pair of Fe II lines that originate from the same upper energy level (Fig.\,\ref{exlupi-sahaFe}, right). These lines are relatively broad, hence we can assume that the LVG approximation holds true. Both lines were fitted across each individual spectra, using the default goodness-of-fit criteria. The resulting theoretical values for temperature and density-velocity gradient at the corresponding observed line ratios are shown. There is a shift in degenerate values across the observation dates, which may correspond to changes in the accretion, i.e. temperature changes. For the temperature to be consistent with the Saha results (T $\sim$ 1 -- 1.5$\times 10^4$ K), this suggests a density-velocity gradient of $\sim$ 1 $\times 10^{12}$ cm$^3$ s. The fact that we do not observe other Fe lines with low A$_{ki}$ \citep[e.g. like those observed in ZCMa,][]{sicilia-aguilar_reading_2020} also concurs with a lower density.  

The temperatures derived from the Fe lines are consistent with the higher end of what is expected in accretion shocks \citep[$\sim$10,000 K, ][]{gullbring98,lima10,dodin12}, also noting that the temperature is dependent on the size of the spot \citep{hartmann_accretion_2016}. Since the high densities and temperatures strongly suggest the possibility of non-LTE effects, the Sobolev LVG approximation would be more suitable in this case, since it does not depend on how the original level was populated. Nevertheless, as already pointed out by \citep{sicilia-aguilar_accretion_2015}, some of the Fe lines are close to saturation, which would also impose a limitation.

The analysis of EX Lupi allows us to test the validity of the line finding algorithm, as well as to define the best strategy to analyse further sources and improve the line-finding strategy for lines with different properties. In addition to reproducing the previously inferred results, the new line velocity analysis of the extended dataset now available for EX Lupi, which covers twelve years of data from 2007 to 2019, with relatively dense sampling from 2009 to 2019, confirms the presence of a very well-localised and stable accretion-related spot on or very near the stellar surface proposed by \citet{sicilia-aguilar_accretion_2015}. The  spot is relatively stable for even longer periods of time, despite the variations in the line strengths and accretion rate during those epochs.  The confirmation of the rotational period of 7.417~d results in a corotation radius of 0.063~au, or between 8.5-9\,R$_*$ \citep[considering the stellar radius estimate of 1.5-1.6 R$_\odot$,][]{sipos09}. This value is larger than the nominal 5 R$_*$ typically assumed in accretion models \citep{gullbring98}, and it also means that the magnetospheric cavity in the inner disk of EX Lupi is larger than those typically assumed, being nearly as large as systems accreting in a propeller regime \citep[e.g. LkCa 15][]{donati_magnetic_2019}. The line data can be saved to use them with a more complex (e.g. radiative transfer) models or fits.

\begin{figure}
    \centering
    \includegraphics[width=\columnwidth]{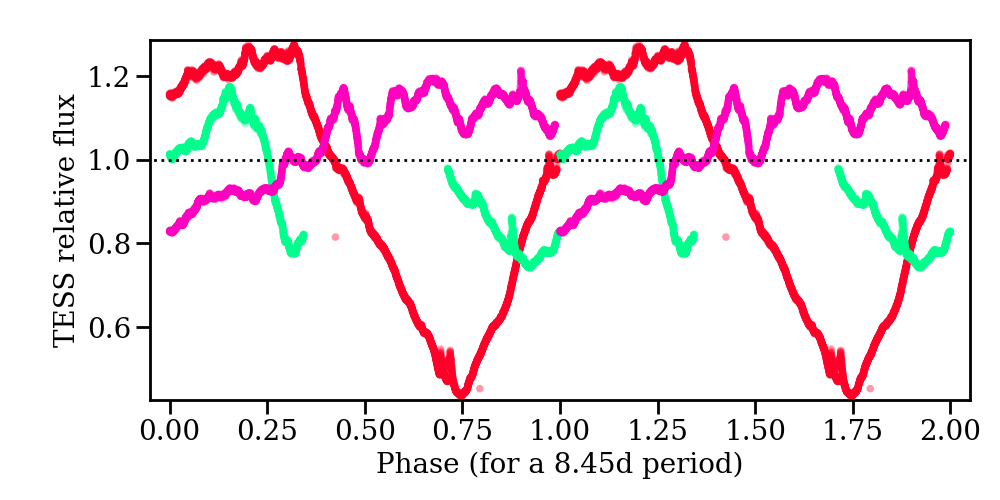} \\
    \includegraphics[width=\columnwidth]{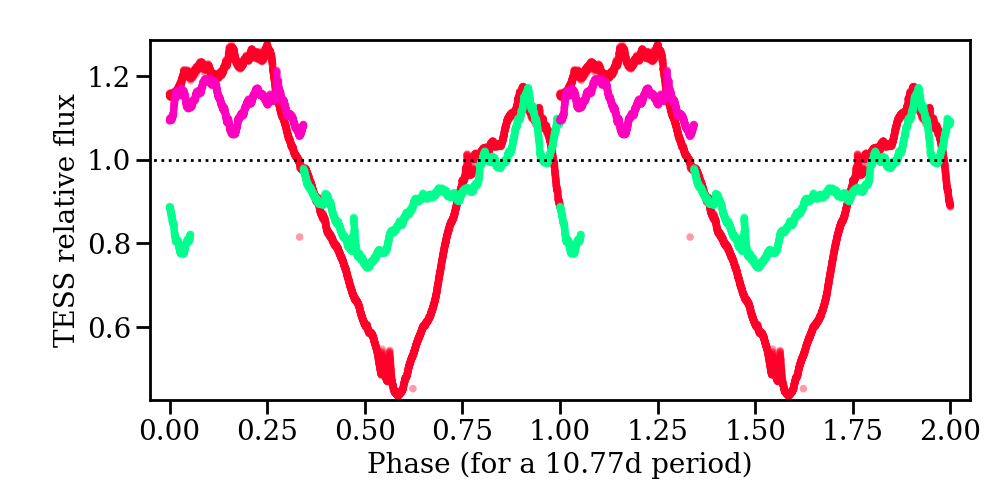} \\
    \caption{\textit{TESS} normalised flux for GQ Lupi A, wrapped according to a 8.45d period \citep{broeg07} (top) and according to a 10.77d period (bottom). The data are colour-coded according to different period cycles.}
    \label{gqlupi-tess}
\end{figure}

\subsection{GQ Lupi A \label{gqlupi-sect}}

GQ Lupi is a young K7 star \citep{herbig77,donati12} with a moderate accretion rate \citep[\.{M}=10$^{-9}$M$_\odot$/yr,][]{frasca_x-shooter_2017}, mostly known for its wide-orbit companions \citep{neuhauser05,alcala_2mass_2020}, although for this part of the analysis we concentrate on the primary alone. Since the GQ Lupi b is about 6 mag fainter than GQ Lupi A \citep{neuhauser05}, the photometry and spectroscopy are all dominated by the primary. The ESO archive contains a total of 76 spectra observed with HARPS (19 spectra), FEROS (39 spectra), and XSHOOTER (5 spectra each for the UVB and VIS arms, 8 spectra for the NIR arm) between 2005 and 2016, which make it an ideal target to test the capabilities of \SM\ for analysing and comparing multi-instrument, multi-epoch data. GQ Lupi A is known to have a rotational period of 8.5d and an inclination of 27 deg, as well as potential hot spots that trigger the observed photometric variability \citep{broeg07,donati12}. It also has a very strong and variable large-scale magnetic field \citep{donati12}.

\begin{figure}
    \centering
    \includegraphics[width=8cm]{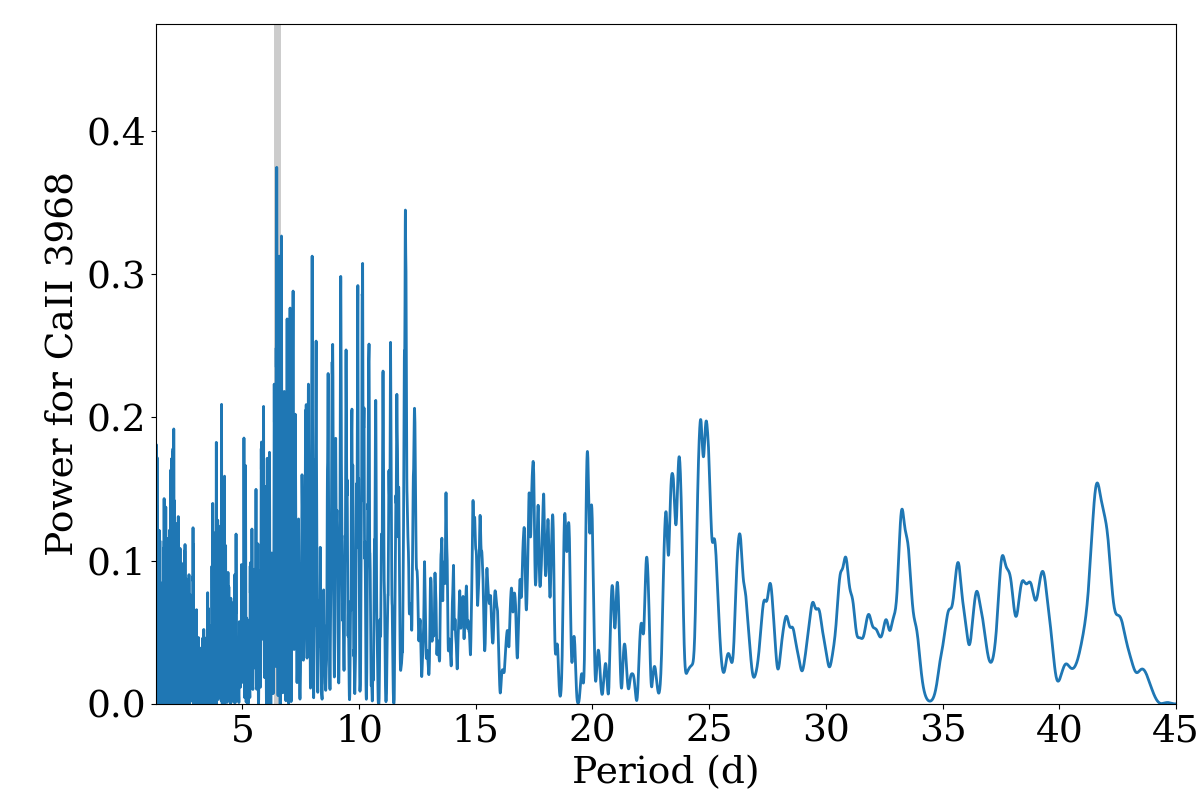}
    \caption{Example of the Lomb-Scargle periodogram for the Ca II 3968\AA\ line in GQ Lupi. Note that although there is a peak at 6.47\,d, the strength of the peak results in a false-alarm probability of 2\% for a white-noise model, and there are several other peaks similar in height.}
    \label{fig:GQLupi-GLS}
\end{figure}

\begin{figure}
    \centering
    \includegraphics[width=8.cm]{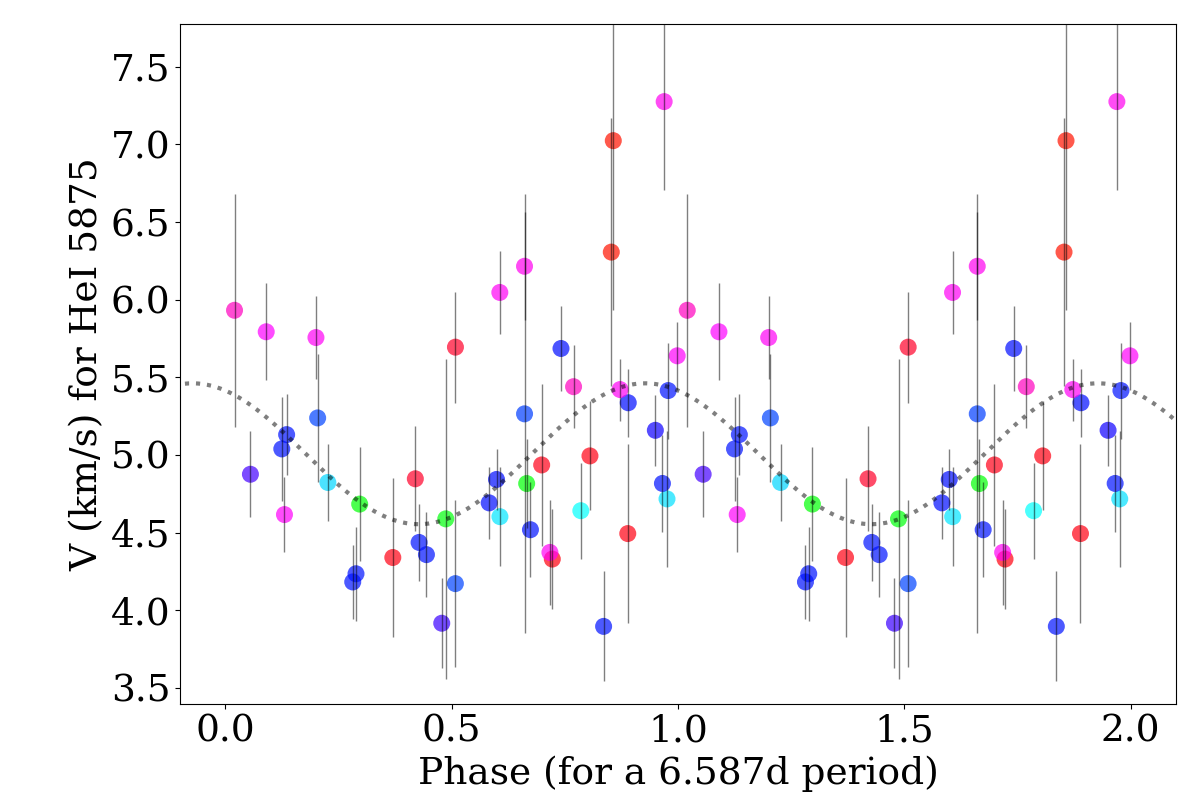}
    \includegraphics[width=8.cm]{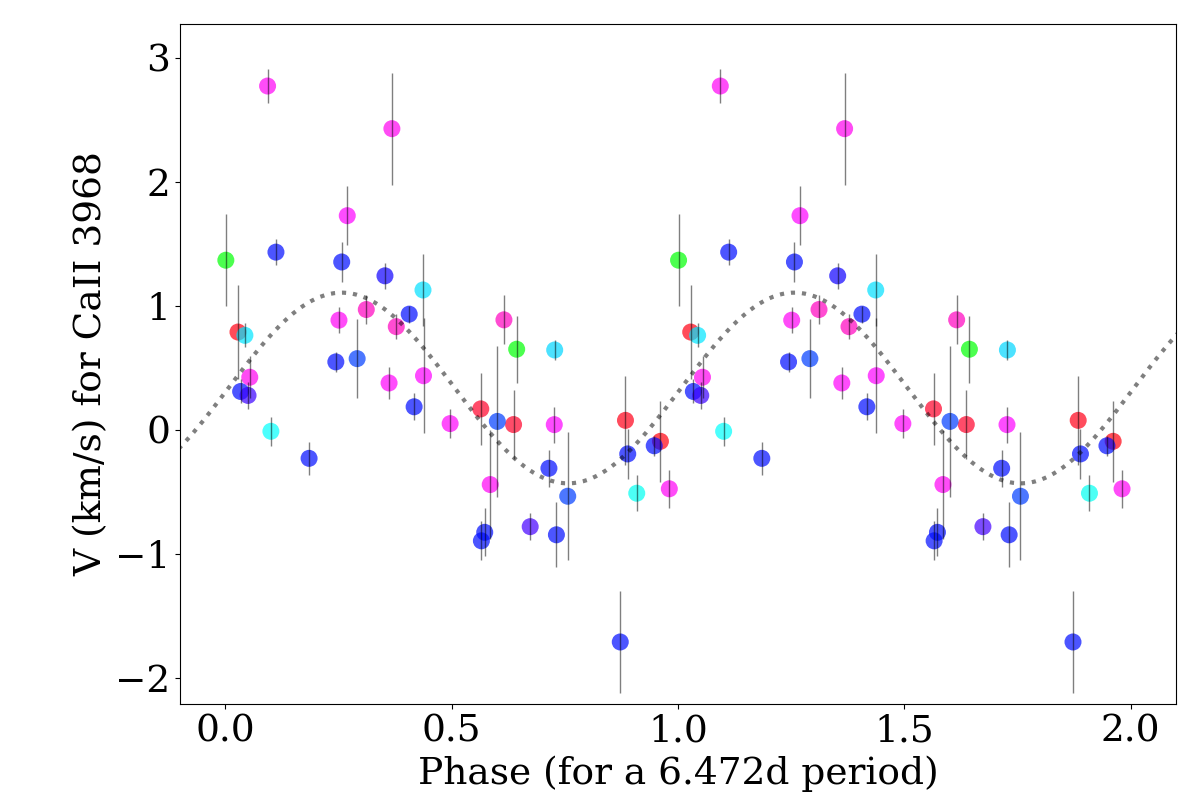}
    \caption{Tentative periods around 6-7 d observed for the line radial velocities in GQ Lupi A for He I 5875\,\AA\ (wrapped for a 6.587\,d period) and Ca II 3934\,\AA\ (wrapped for a 6.472\,d period). The colours indicate different period cycles,  with the colour scale changing from red to purple from the first to the last MJD epoch.}
    \label{GQlup-modulation}
\end{figure}

\begin{figure}
    \centering
    \includegraphics[width=0.9\columnwidth]{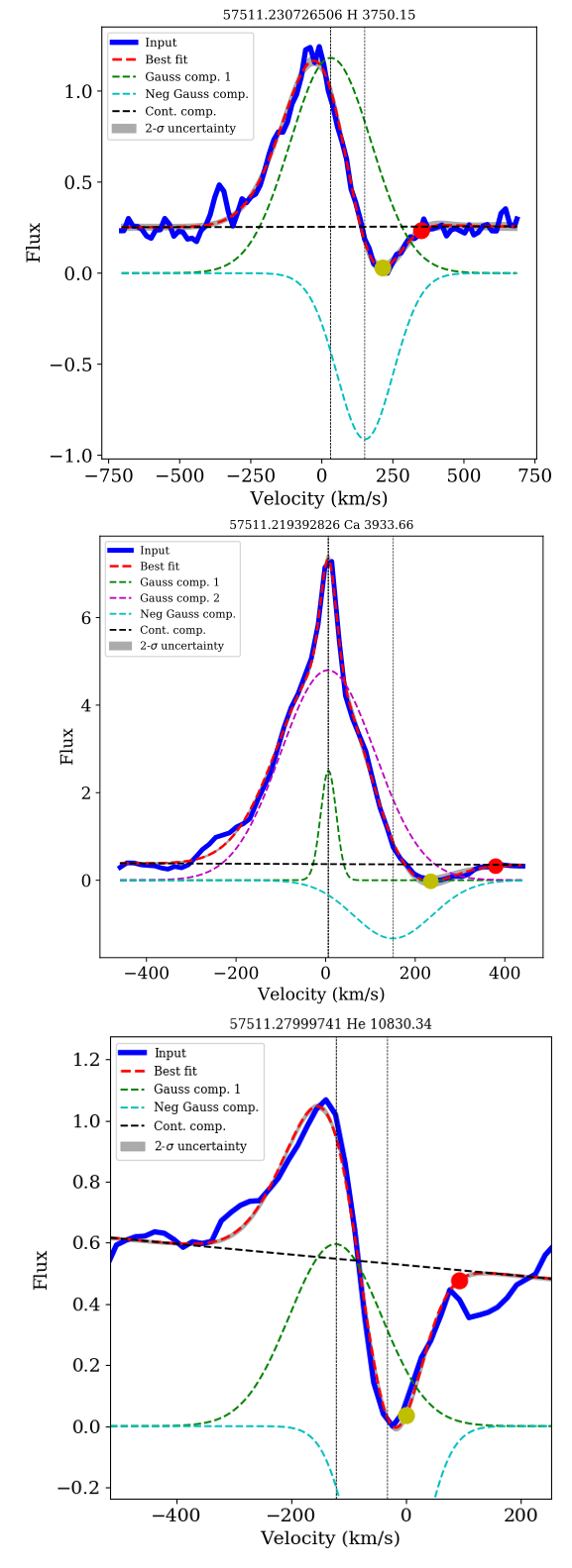} 
    \caption{Example model fits of the XSHOOTER GQ Lupi A spectra for the H I 3750\,\AA\ (top) Ca II K 3934\,\AA\ (middle) and He I 10830\,\AA\ (bottom) lines. Each line displays a redshifted absorption component that is modelled by the negative Gaussian component. The positions of the maximum redshifted absorption velocity are indicated by the red circles, and the maximum absorption is marked by the yellow circles. The dotted vertical lines indicate the centre of the Gaussian components.}
    \label{fig:GQLup_abs-fig}
\end{figure}

The \textit{TESS} data on GQ Lupi are inconclusive regarding periodicity (see Fig.\,\ref{gqlupi-tess}).  \textit{TESS} reveals a large dip, similar to the variability pointed out by \citet{broeg07}, but this dip is not observed on subsequent periods despite all observations being taken within a 25\,d interval. This, together with the shape of the \textit{TESS} lightcurve,
suggests that the main cause for variations could be extinction events, which may be linked to the stellar rotation as it has been observed in other systems \citep{sicilia20-j1604}. The best-fitting period for the \textit{TESS} data is 10.77d, although the entire dataset covers less than 3 such periods so the result has a low significance.  A secondary peak appears at  7.01d, but neither peak nor the $\sim$8.5\,d rotational period are conclusive.

A combination of the FEROS and HARPS data was used to estimate the rotational and radial velocity of the source, resulting in \vsini = 5.35$\pm$0.74 km/s and RV=$-$3.07$\pm$0.05 km/s. Because most of the lines are relatively broad, fitting the continuum works best if we use a large under-sampling window and an order 3 polynomial. 
For the lower resolution XSHOOTER data, especially in the UV region, which is very rich in emission lines, a large under-sampling window (8\AA) is needed, and the best fit is obtained with an order 1 polynomial. Automated identification by \SM\ finds 9 lines in the combined HARPS and FEROS data, compared to 14 that are found in the same range of the XSHOOTER data (see Tab.\,\ref{tab:lines}). Both tend to miss H$\alpha$, H$\beta$, and H$\gamma$ because they are very broad and have deep central absorptions. The main difference lies in lines that are very weak and that become diluted by the higher resolution of FEROS and HARPS (such as the He II line at 4686\,\AA, or the He I line at 5015\,\AA).
XSHOOTER reveals 41 lines, most of which are in the UV arm of the spectrum. The broad He I 10830\,\AA\ line is again a challenge for the automated detection due to its deep absorption. 
The conclusion is that, to extract such broad lines or lines that have central absorption components, the user may need to relax the typical matching range. In general, the \vsini\ is a good criterion to identify lines that are centrally peaked and not significantly red- or blue-shifted, but for stars like GQ Lupi, that show a variety of narrow and broad lines profiles, the best strategy is to extract them separately.

Looking at the emission lines extracted for the high-resolution FEROS and HARPS data, we detect a potential  6-7\,d radial velocity modulation (Fig. \ref{fig:GQLupi-GLS}). The modulation is most evident in the He I  and Ca II lines. Other lines (including Fe I and He II) reveal low-significance (few per cent false-alarm probability for white-noise) modulations between 6-7d (see Fig.\,\ref{GQlup-modulation}), which is suggestive of a quasi-period shorter than the nominal rotation period of 8.45\,d.   As observed for EX Lupi, the variations in the radial velocity of the lines do not exceed \vsini\, but the modulation is by far less clean. One of the reasons is that there are fewer datapoints and they are spread over a rather long timescale. 
Some lines, such as the He I 5875\,\AA, display redshifted profiles that may be an indication of wind absorption, so the interpretation of the periodicity is complicated, since it could include elements from the stellar rotation but also variability associated with the wind.
In some cases, marginally-significant peaks arise at periods $\sim$4d, which could be caused if spots are found on the two sides of the star. Not all the lines show a sinusoidal profile, and the fact that some of the lines are not observed all the time may indicate that the line-emitting region is very rapidly changing or not always visible. This is all in agreement with the rapidly-changing complex signatures observed in the TESS lightcurve as well as with the rapid evolution of the spots and magnetic field structure \citep{donati12}. Rapid changes can explain why the modulation is, at best, quasi-periodic, and the false-alarm probability is never as low as observed for EX Lupi.

Although the radial velocity modulations of the lines are inconclusive in this case regarding the structure of the accretion structures, examination of the line profiles reveal that many of the lines display inverse PCygni or YY Ori-type profiles, with redshifted absorptions that go below the continuum (see Fig.\,\ref{fig:GQLup_abs-fig}).  Similar profiles had been observed at certain epochs in the past \citep{batalha01}, although in our case they appear to be always present in the XSHOOTER epochs. Redshifted absorptions are found in the higher Balmer lines, Ca II lines, and He I IR triplet lines. Many other lines (such as the Ca II IR triplet) have asymmetric profiles suggestive of redshifted absorption, although the absorption is not as marked and does not go below the continuum. Since this type of profile requires the accretion column to be observed along the line-of-sight, it is in agreement with the interpretation of the photometric variability as resulting from occultations by the disc material. Interestingly, because the inclination of the star has been estimated to be around 27 deg \citep{broeg07,donati12}, it would also indicate a misalignment between the inner and the outer disc. In fact, the outer disc has a significantly different inclination \citep[60 deg;][]{macgregor17}, which together with the system being a multiple stellar system, could lead to a complex and warped structure.

For the YY Ori-type profiles, the H I and He I profiles were fitted using \SM\ with a single positive and a negative Gaussian component, as well as the continuum linear component (Fig.\,\ref{fig:GQLup_abs-fig}). The Ca II-K line displays a defined narrow component in addition to the broad emission component and the redshifted absorption and was hence fitted with two positive Gaussians, the negative absorption Gaussian and the linear continuum component. The maximum redshifted absorption velocity, $V_{red}$, is calculated from the best fit model as the velocity at which the absorption correspond to 10\% of the maximum absorption  \citep{sicilia-aguilar_2014-2017_2017}. Table\,\ref{tab:GQLup_vred} shows the observed $V_{red}$ values obtained from this absorption profile fitting. Across all individual observations where this feature was well fitted, the mean values of $V_{red}$ are 102$\pm$12, 313$\pm$27, and 390$\pm$17 km/s for the He I, H I, and Ca II, respectively. We also tested the photospheric removed XSHOOTER spectra from ROTFIT \citep{frasca_x-shooter_2017}, to ensure that these absorption features were not distorted by the stellar photosphere. In each case, concurrent values of $V_{red}$ were obtained from the ROTFIT reduced spectra. 

We find that $V_{red}$ shows a significant anti-correlation with the total excitation energy of the transitions (Ek for the neutral lines and Ek plus ionisation potential energy for the Ca II lines). The highest infall velocities are reached by the lines produced in lower temperature regions, the same result we observed in ASASSN-13db \citep{sicilia-aguilar_2014-2017_2017}, and opposite to the behaviour reported by \citet{bertout82} and \citet{petrov_doppler_2014}, for SCrA and CoD-35$^{\circ}$10525. We note that we only use three species, whereas previous work has used many more metallic lines. It could therefore be that we are detecting an abundance difference between the H and He and the Ca, however, there is still difference in $V_{red}$ values between the H and He. These lines also show variation in the $V_{red}$ positions across the different observation dates, but with no clear correlation, nor enough data points to infer any periodicity. 

\begin{table}
\centering
  \caption{GQ Lupi maximum redshifted absorption velocity, $V_{red}$, for the YY Ori-type line profiles \label{tab:GQLup_vred}}
  \begin{tabular}{lccccc}
    \hline
    Species/$\lambda$  & Excitation Energy & Mean Vred & Infall Radius  \\
    (\AA) & (Ev) & (km/s) & (R$_{*}$) \\
    \hline
CaII 3934 & 9.26 & 390$\pm$17 &  5.4 \\

HI 3722 & 13.53 & 340$\pm$12 & 2.6 \\
HI 3750 & 13.50 & 341$\pm$24 & 2.6 \\
HI 3771 & 13.49 & 290$\pm$7  & 1.8 \\
HI 3798 & 13.46 & 326$\pm$12 & 2.3 \\
HI 3835 & 13.43 & 290$\pm$18 & 1.8 \\
HI 3889 & 13.39 & 293$\pm$6  & 1.8 \\
HeI 10830 & 20.96 & 102$\pm$12 & 1.1  \\
\hline 
\end{tabular} 
\end{table} 

Table \ref{tab:GQLup_vred} also shows the corresponding infall radii, calculated using the stellar parameters of M$_*$ = 1.05 M$_\odot$, R$_*$ = 1.7 R$_\odot$ and $i$ = 27 deg \citep{donati12} and assuming the material is in free-fall to the star. We assume that the infall is directed along the pole of the star and hence adjust the projected $V_{red}$ by the inclination angle to achieve infall radii in this plane. These values do not account for possible complex geometries in the accretion column, but give an approximate value for the minimum infall radii. These values for the H and He are less than the nominal 5 R$_*$ typically assumed for accretion, the Ca II is in good agreement.

\begin{figure}
    \centering
    \includegraphics[width=\columnwidth]{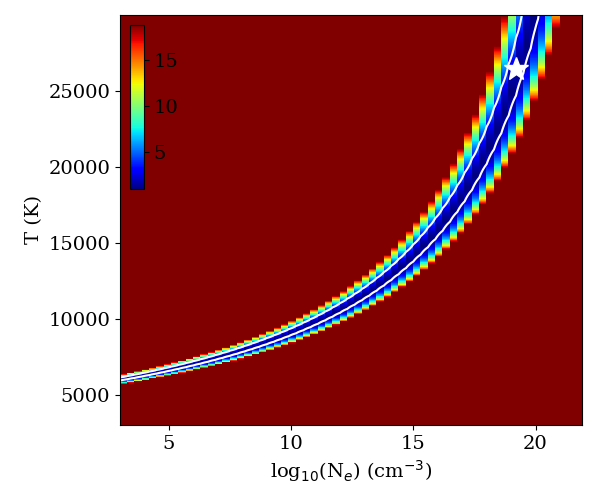}
    \caption{Saha results from the He I 4026, 4471, 5876 and He II 4686 lines in GQ Lupi, identified with \SM.}
    \label{fig:GQLup_saha}
\end{figure}

\begin{figure}
    \centering
    \includegraphics[width=\columnwidth]{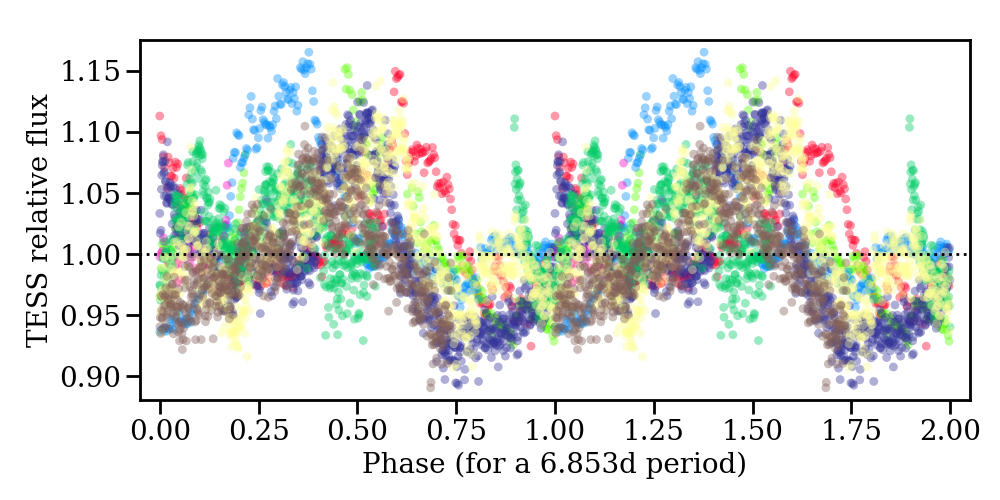}
    \caption{\textit{TESS} data for CVSO109, wrapped according to a 6.853d period. The data are colour-coded according to the different period cycles to better display the variations.}
    \label{fig:CVSO109_tess}
\end{figure}

\begin{figure*}
	\centering
	\includegraphics[width=\columnwidth]{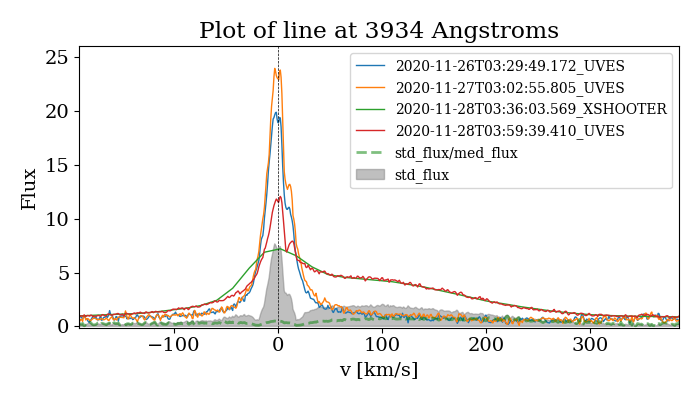}
	\includegraphics[width=\columnwidth]{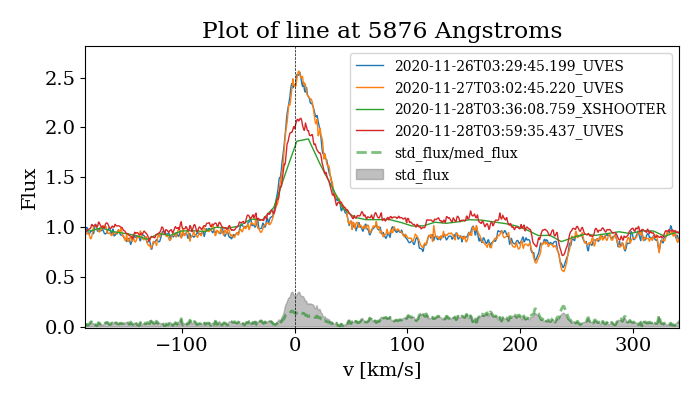}
	\includegraphics[width=\columnwidth]{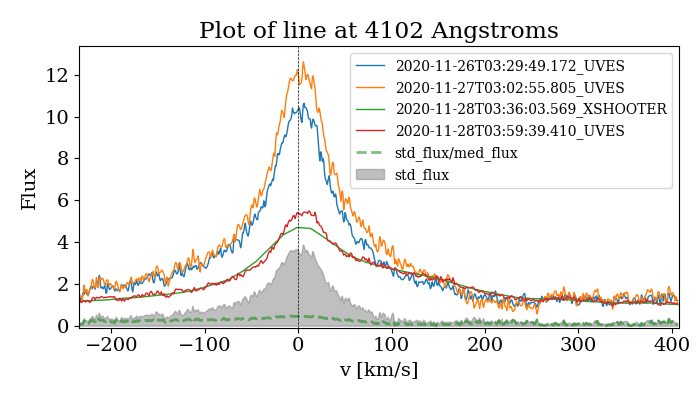}
	\includegraphics[width=\columnwidth]{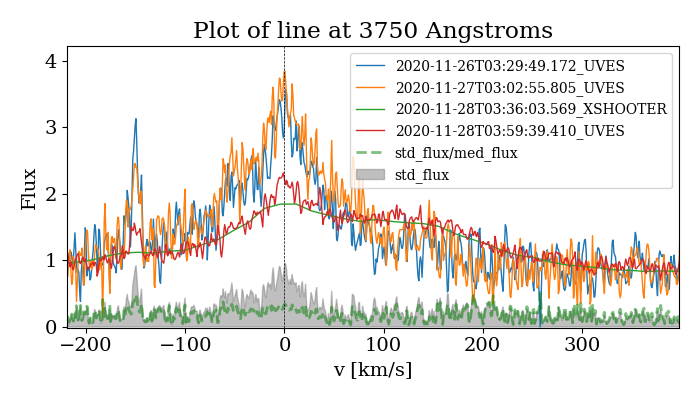}
	\caption{Variability plots of the redshifted emission wing in CVSO109 across the four epochs. Top left: Ca II K 3934\,\AA\ (model fit shown in Fig\,\ref{fig:CVSO_H_fit}, top row),  top right: He I 5876\,\AA\, bottom left: H I 4102\,\AA\ (model fit shown in Fig\,\ref{fig:CVSO_H_fit}, bottom row), bottom right: H I 3750\,\AA. }
	\label{fig:cvso_bump_comp}
\end{figure*}

\begin{figure*}
	\begin{tabular}{cccc}
	Ca II 3934 UVES 1 & Ca II 3934 UVES 2 & Ca II 3934 XSHOOTER & Ca II 3934 UVES 3 \\
	\includegraphics[width=0.23\textwidth]{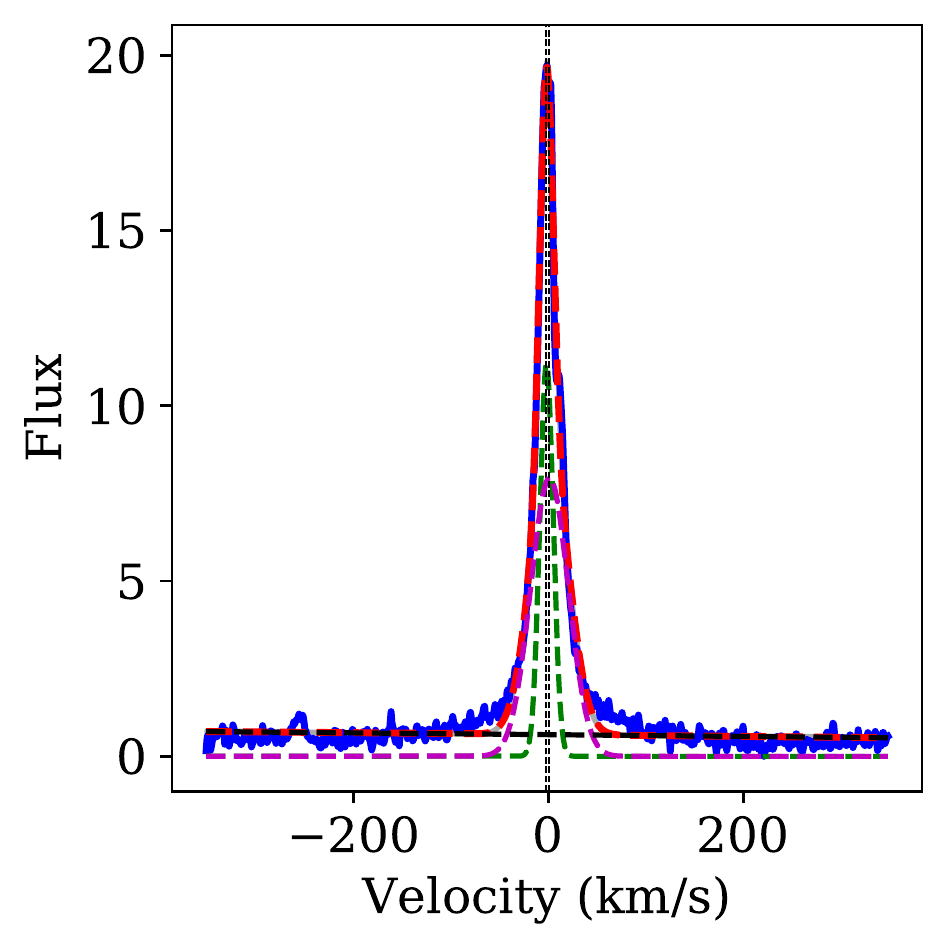} &
	\includegraphics[width=0.23\textwidth]{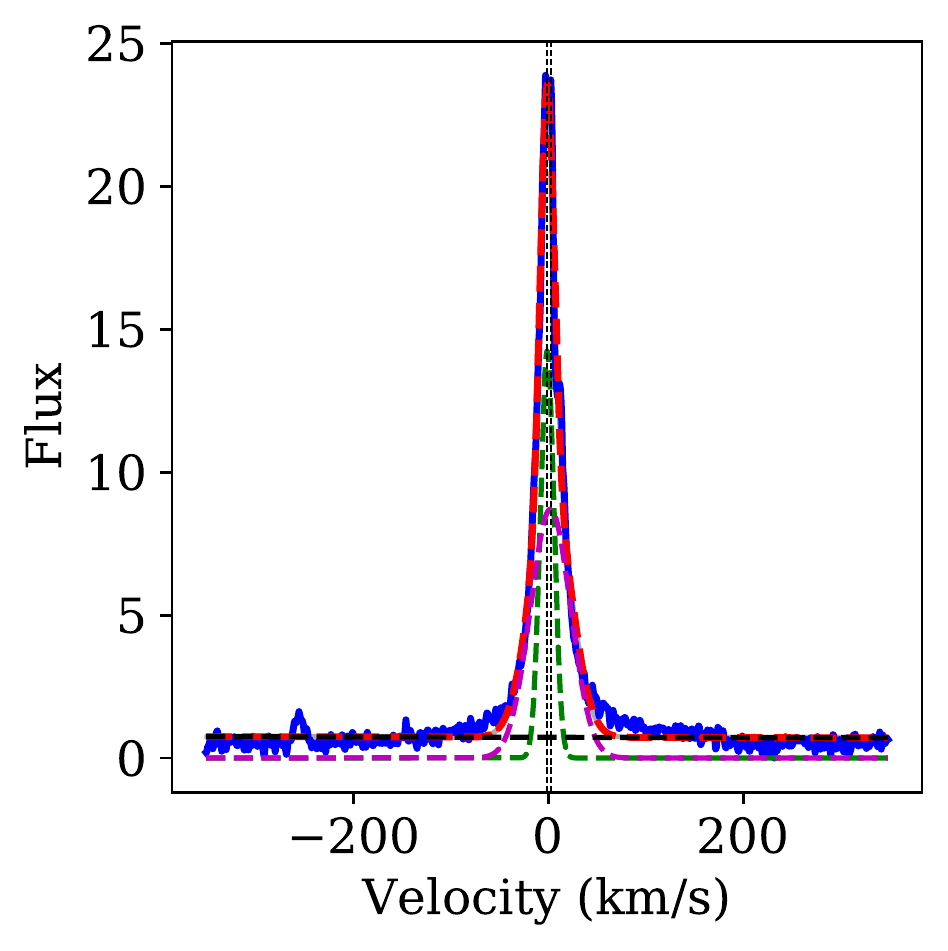} &
	\includegraphics[width=0.23\textwidth]{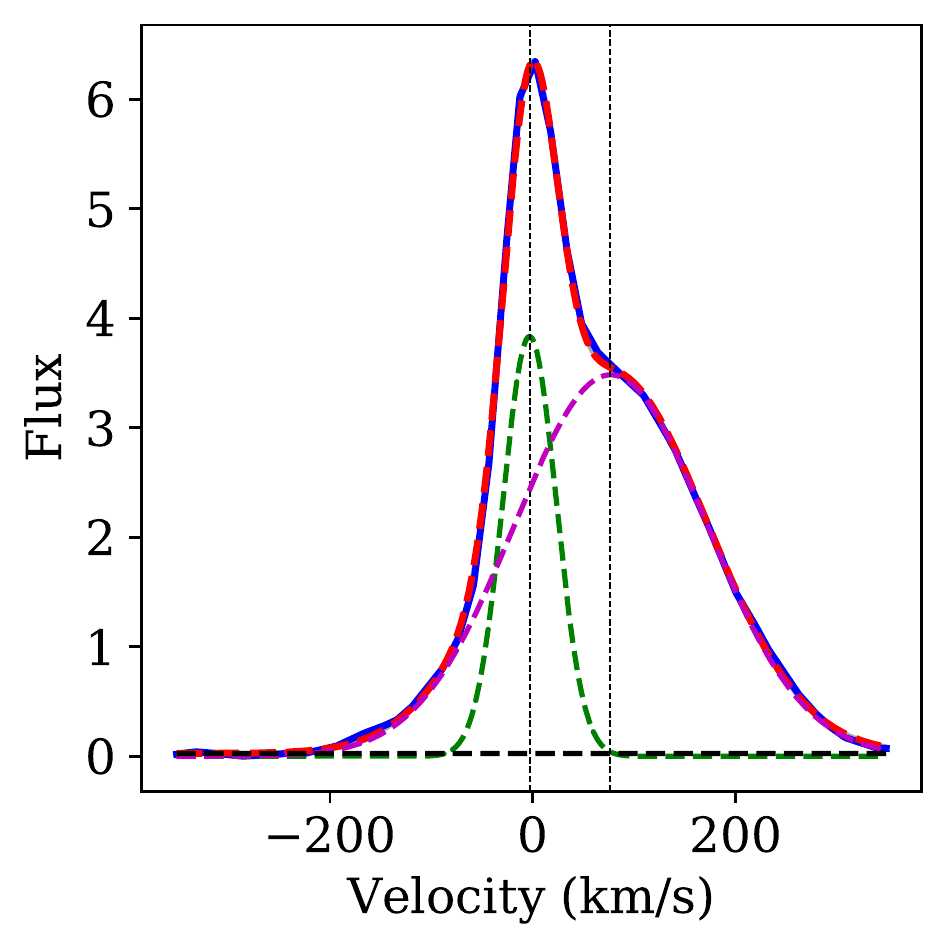} &
	\includegraphics[width=0.23\textwidth]{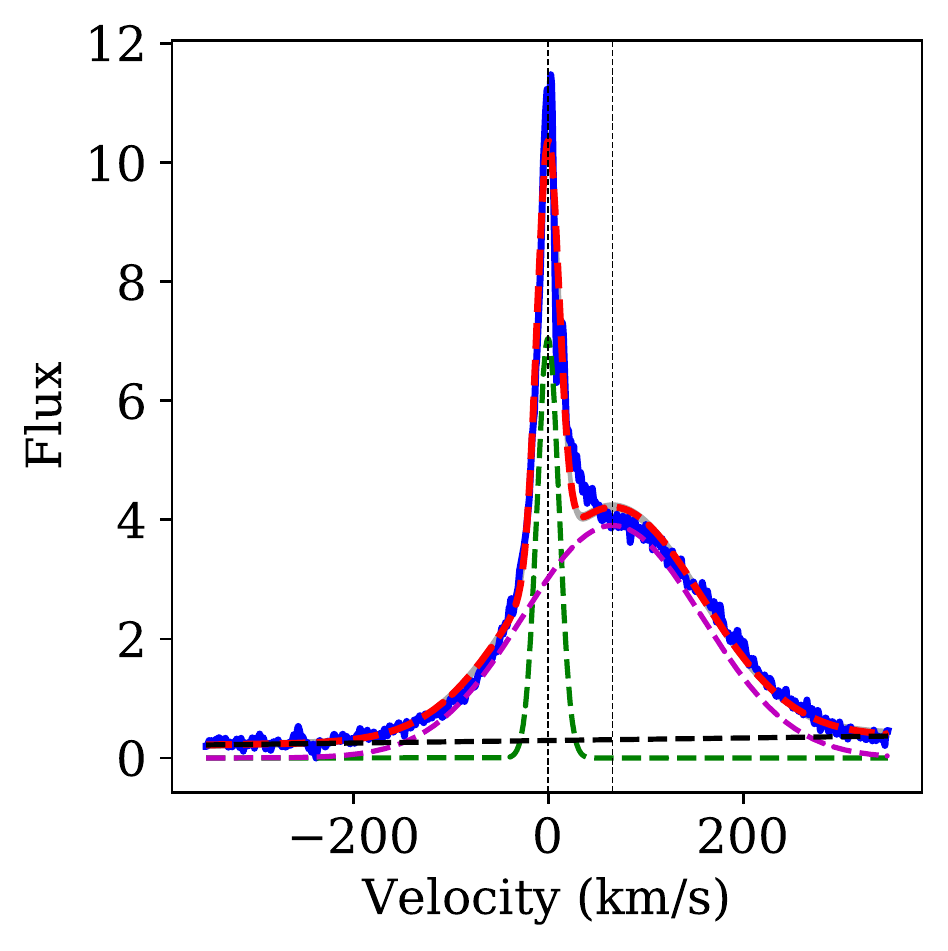} \\
	&&&\\
	H I 4102 UVES 1 & H I 4102 UVES 2 & H I 4102 XSHOOTER & H I 4102 UVES 3 \\
	\includegraphics[width=0.23\textwidth]{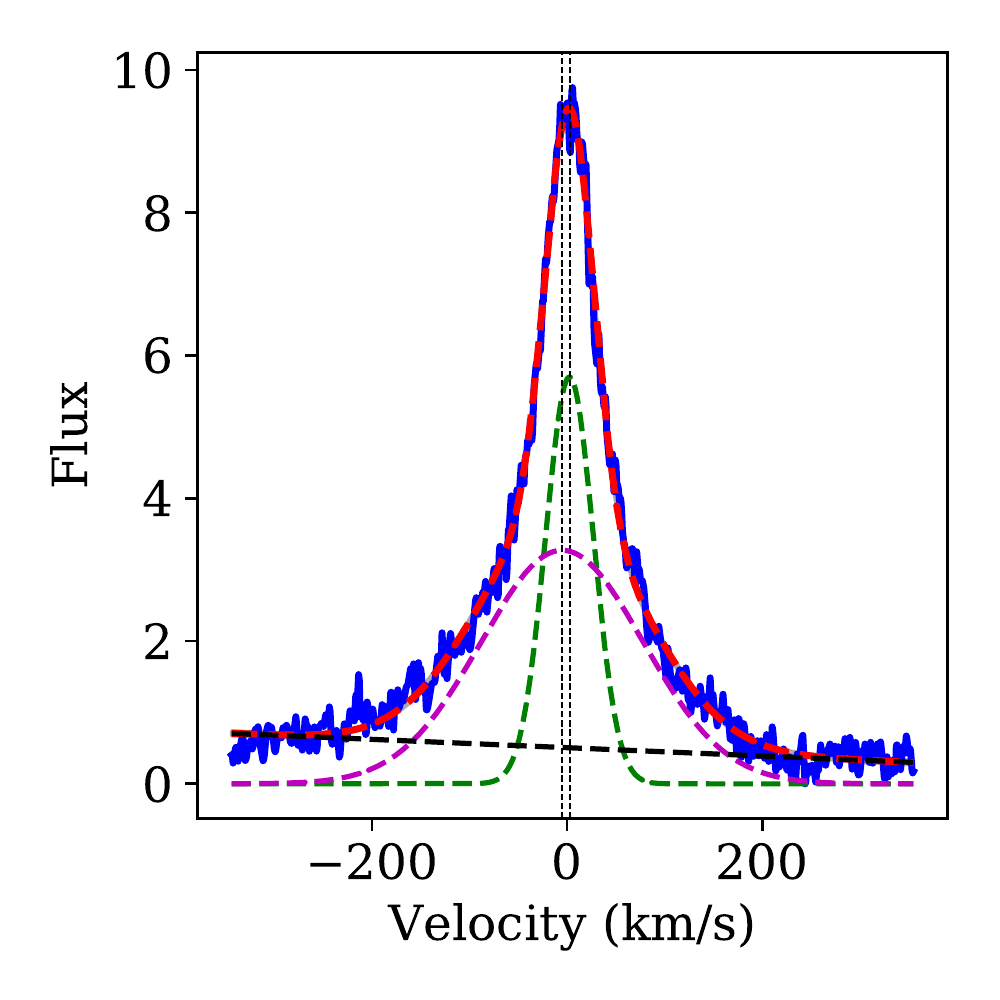} &
	\includegraphics[width=0.23\textwidth]{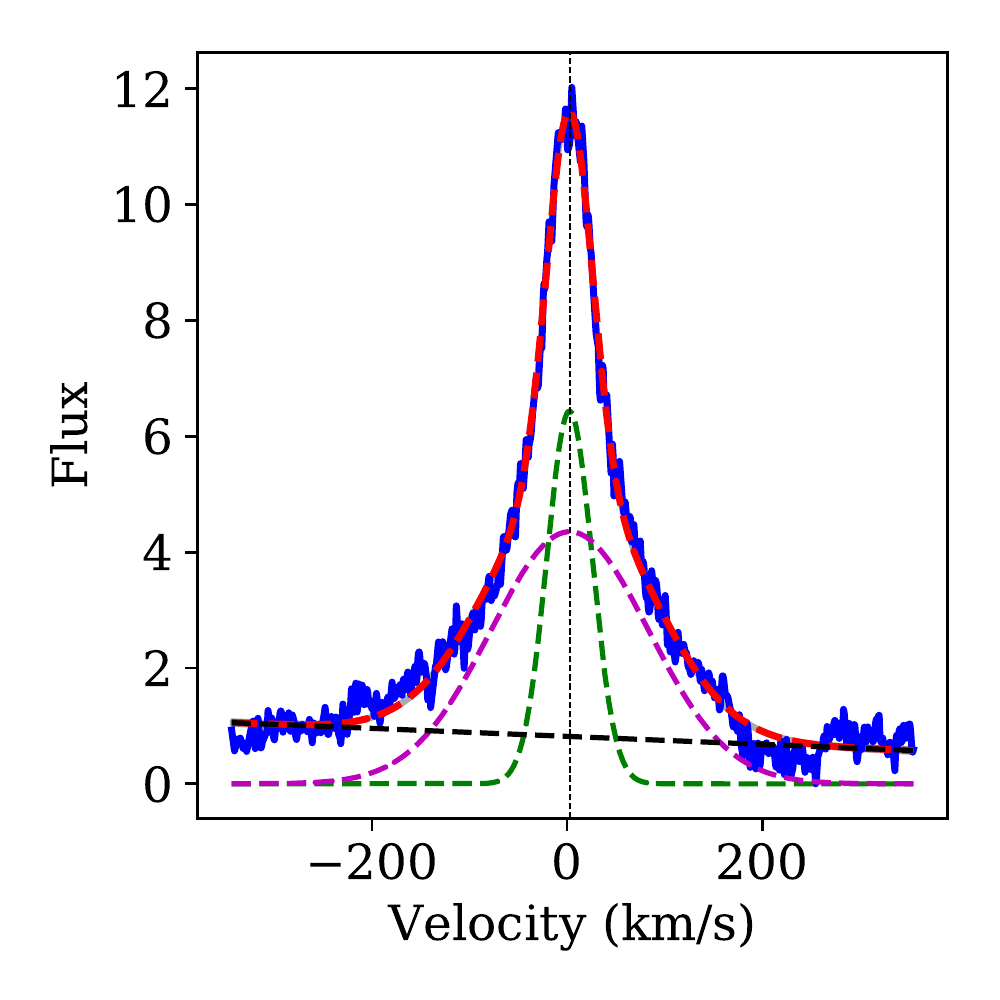} &
	\includegraphics[width=0.23\textwidth]{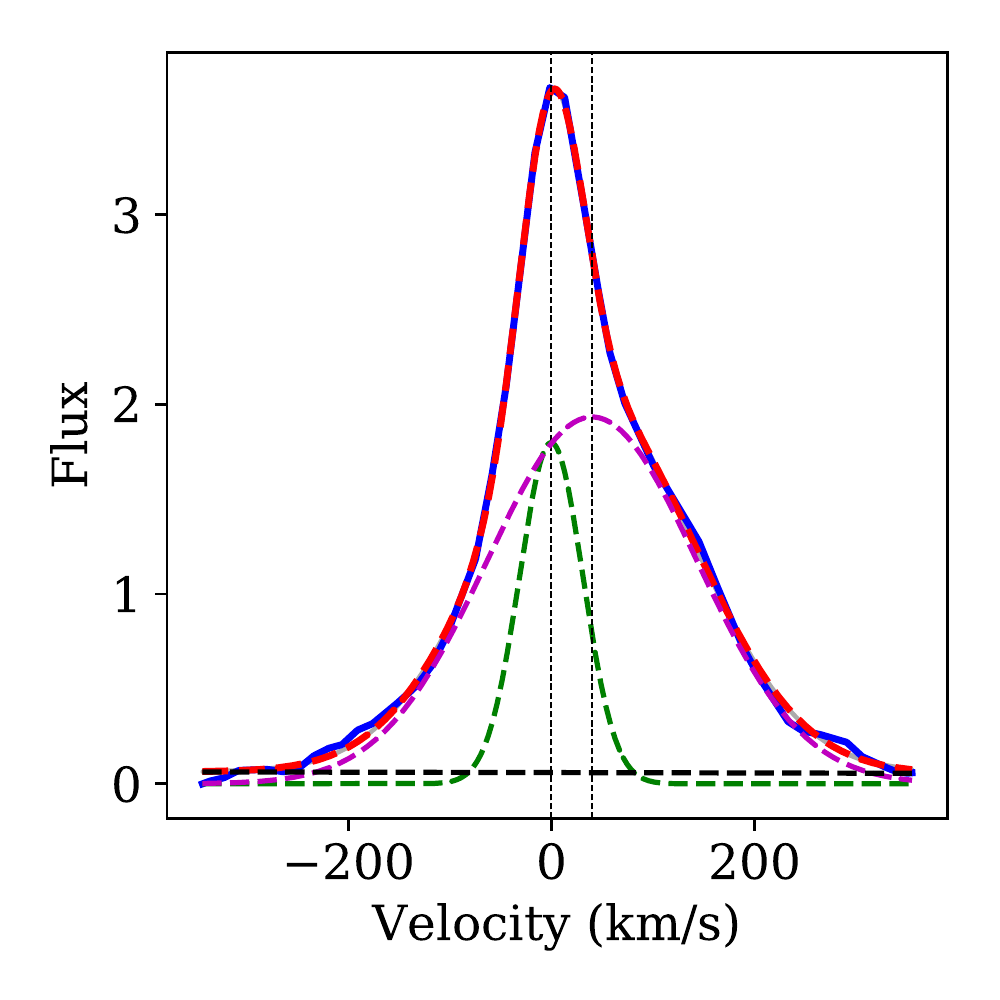} &
	\includegraphics[width=0.23\textwidth]{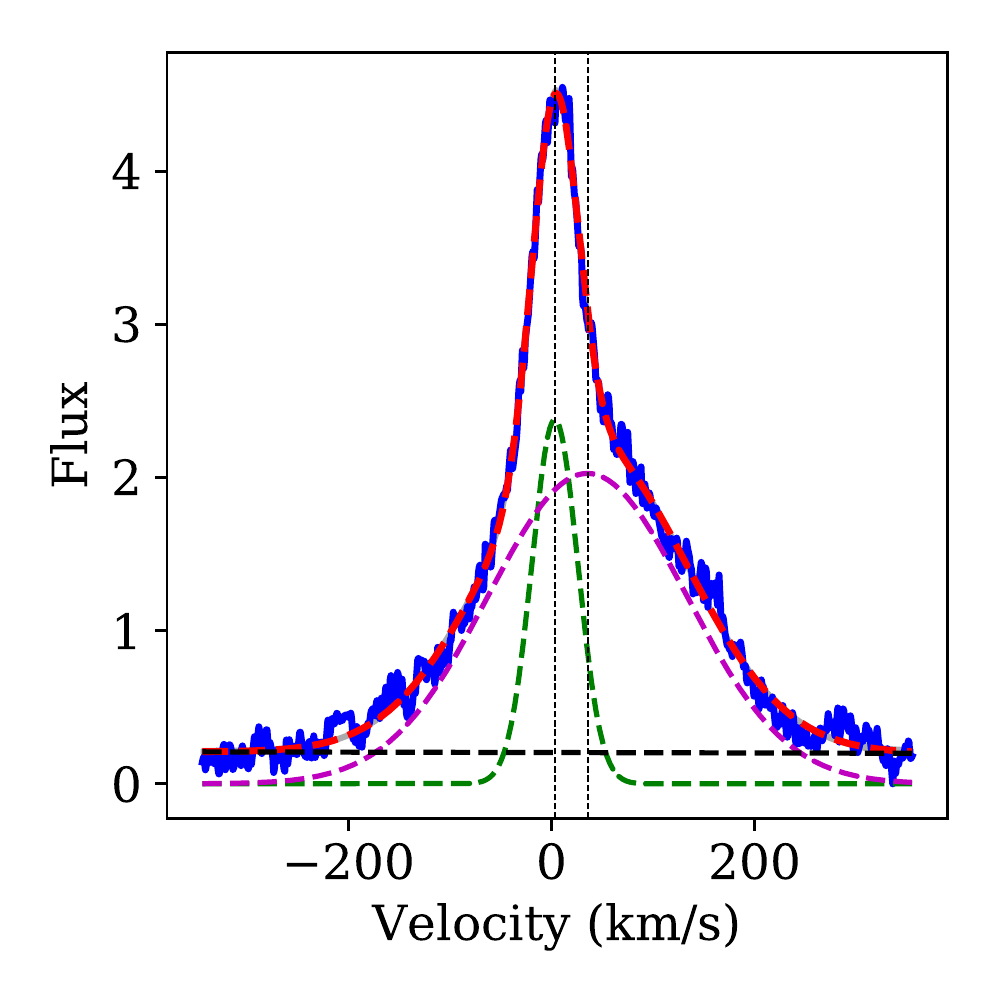}	 
	\end{tabular}

	\caption{Model fits of the CVSO109 lines, input shown in blue, best fit shown in red. Top: Ca II K 3934\,\AA. Bottom: H I 4102\,\AA, each across the four epochs. In each instance, the line is fitted well by a broad and a narrow Gaussian components, indicated by the dashed lines. The redshifted emission wing can be clearly seen in the final two epochs.}
	\label{fig:CVSO_H_fit}
\end{figure*}

\begin{figure}
	\includegraphics[width=1\columnwidth]{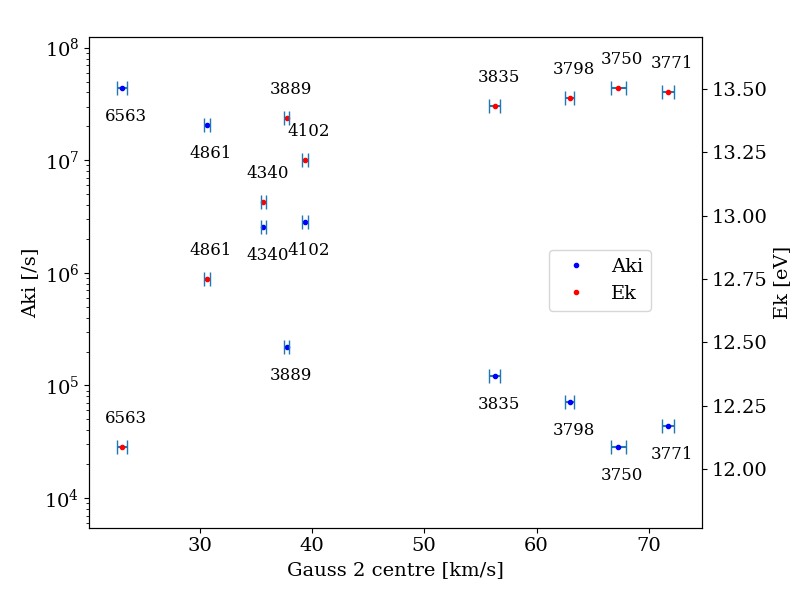}
	\caption{CVSO109 redshifted emission wing component central Gaussian velocity vs Ek and Aki for the well-fitted H lines in the XSHOOTER data. Error bars indicate the standard error in the Gaussian component centre value. Spearman's Rank-Order Correlation coefficients of 0.96 (p=2e-5) and -0.93 (p=2e-4) are calculated for position and the Ek and Aki, respectively.}
	\label{fig:gauss_vs_aki}
\end{figure}

We used the Saha equation to investigate the temperature and density ranges for the emitting regions. Figure\,\ref{fig:GQLup_saha} shows the results for four of the He lines identified in the GQ Lupi spectra: He I 4026, 4471, 5876\,\AA\ and He II 4686\,\AA. The longer wavelength He lines are not included due to potential extinction affects and significantly different line profiles from these four lines. The median XSHOOTER spectra were used to determine the integrated flux ratios for the calculations. These results show similar densities and higher temperatures required for the He production than those shown for EX Lupi. As before, such high ionisation levels suggest non-thermal effects. The degenerate best-fit area is better constrained than for the EX Lupi He lines, this may be because the line profile of the He II is more similar to the He I for GQ Lupi. The best-fit solution is dominated by this single He II line, while part of the He I emission may originate in a different location.

To summarise, for GQ Lupi A, \SM\ has allowed us to systematically extract detailed emission line results across many years of archival data. These analyses show a highly variable accreting system, with periodic and quasi-periodic rotational modulations in the emission lines consistent and semi-consistent with the observed photometric period. This is suggestive of spots on either side of the star, as well as non-constant magnetically channeled accretion along the line of sight, as shown by the inverse PCygni profiles; even in a system not so highly inclined. In this case, although the radial velocity signatures are inconclusive, the line profiles allow us to put an upper limit to the size of the magnetosphere, which is smaller than in the case of EX Lupi, and the line variability indicates that the accretion-related spots are very irregular in distribution and rapidly variable. Both findings show a stark contrast when compared to EX Lupi, although both stars have similar mass and age. In addition, the fact that the star has a low inclination while the disc and the accretion column seem more highly inclined suggests that there could be a star-disc, in addition to a star-magnetic field, misalignment.

\subsection{CVSO109 \label{cvso-sect}}

CVSO109 (V* V462 Ori) is one of the ten \textit{HST}/ULLYSES targets \citep{2020RNAAS...4..205R} located in the Orion OB1 association that was also observed by the ESO PENELLOPE large program \citep{2021arXiv210312446M}. PENELLOPE is a contemporaneous survey to the \textit{HST}/ULLYSES program and it is part of the ODYSSEUS\footnote{\url{https://sites.bu.edu/odysseus/}} collaboration (Outflows and Disks around Young Stars: Synergies for the Exploration of ULLYSES Spectra, Espaillat et al. in prep.), which will analyse the \textit{HST}/ULLYSES YSO data and the connected ground-based data. In this work, we analyse the variability observed in CVSO109 from the four ESO VLT epochs to show what can be achieved with \SM\ across a shorter temporal coverage.

\textit{HST}/ULLYSES observations revealed that CVSO109 is a binary that is unresolved in the ground based data \citep{2021RNAAS...5...36P}. However, the secondary star has little to no excess emission (Espaillat et al. in prep.). Therefore, the emission lines we analyse here are all assumed to have originated from the primary, which is an accreting M0 CTT star (\citealt{2014ApJ...790...47I}, Espaillat et al. in prep.).  The object has also been observed with \textit{TESS} in sectors 6 and 32. \textit{TESS} data, extracted using the principal component analysis (PCA) photometry tool to remove the systematics from nearby stars using the Python package \textsc{eleanor} \citep{feinstein19,brasseur19}, reveals a potential period around 6.853\,d (see Fig.\, \ref{fig:CVSO109_tess}). 

PENELLOPE observations were carried out across three nights with the ESO VLT instruments UVES and XSHOOTER (see Appendix\,\ref{app:inst} for instrument descriptions), with the XSHOOTER spectra being taken $\sim$25 minutes before the final UVES epoch\footnote{These final two epochs were near simultaneous to the \textit{HST}/ULLYSES COS and STIS observations}. From the UVES spectra, we measure the projected rotational and radial velocities of the star as \vsini=8.1$\pm$1.6 and RV=16.76$\pm$0.21 km/s. Lines were identified using a first order polynomial and a wavelength filter of 5\,\AA\ for the continuum subtraction. We identify 45 emission lines over a 3$\sigma$ detection threshold. These lines were all identified using the NIST database (see Tab.\,\ref{tab:lines}), and 29 of them had been detected in our previous work  in the spectra of EX Lupi, ASASSN-13db and ZCMa \citep{sicilia-aguilar_optical_2012,sicilia-aguilar_accretion_2015,sicilia-aguilar_2014-2017_2017,sicilia-aguilar_reading_2020}. 

Many of the identified H and Ca II lines display a clear variability across the four epochs. Nine H lines, the Ca-K line and the Ca II IR triplet have a well-defined redshifted emission wing (see Figs.\,\ref{fig:cvso_bump_comp} and \ref{fig:CVSO_H_fit}). The feature is also prominent in the Ca-H line, although blended with the adjacent H7 line, and in further H lines but with lower S/N. Figure\,\ref{fig:cvso_bump_comp} shows variability plots of the Ca II K 3934\,\AA\, He I 5876\,\AA\, H I 4102\,\AA\ and H I 3750\,\AA\ lines. The standard deviation can be used as a variability indicator, showing the redshifted wing feature clearly in the Ca II K line. The He I 5875\,\AA\ line also shows an extended redshifted component. The feature is not detected in other He lines.

This redshifted component is highly variable, being detected in the XSHOOTER and final UVES epoch taken with $\sim$0.5 hr difference, but is absent from the first two UVES epochs, taken 1 and 2 days earlier. However, composite model fits of two Gaussians and a linear continuum to these lines both reveal the wing component structure, and still accurately model the first two UVES epochs as a narrow and broad component, but without the redshifted wing of the broad component. As with GQ Lupi A, we compared these model fits for the XSHOOTER data to the photosphere subtracted XSHOOTER spectra from ROTFIT \citep{2021arXiv210312446M}. There was little change for the Balmer series and Ca II K lines. For these lines, the measured central Gaussian positions of the wing component from the ROTFIT spectra were within two standard errors of the original XSHOOTER spectra positions. The Ca II IR triplet lines from ROTFIT show a larger maximum flux, since the absorption component is accounted for, and the redshifted wing appears more defined. There is a difference of up to 10 km/s between the central Gaussian velocities of the ROTFIT and original XSHOOTER data for these IR lines. We note a difference in the \vsini\ from the value of 3.5$\pm$1.1 km/s, calculated with the best veiling from ROTFIT \citep{2021arXiv210312446M}. Other users of the package should be aware of this for stars with strong accretion. We will investigate line dependent veiling with \SM\ in future work. 

Figure\,\ref{fig:CVSO_H_fit}, top row, shows the model fits from each epoch for the Ca II K line and the bottom row shows the model fits to the H I 4102\,\AA\ line. These model fits have a high accuracy for each of the epochs, with the H I 3750\,\AA\ line being the lowest S/N of the H Balmer lines still accurately fitted (in the XHSOOTER and one of the UVES epochs, the other lines listed are well-fitted across all four epochs). Table\,\ref{tab:CVSO109_wing_properties} shows the central position and FWHM of each Gaussian component, with standard errors from the model fits, for nine H I lines and the Ca II K line. The XSHOOTER fits to the Ca II IR triplet are also shown.

From the XSHOOTER fits, the wing components have centre Gaussian velocity ranges of $\sim$20-70 km/s for the H lines and $\sim$75-120 km/s in the Ca II K + IR triplet lines (see Tab.\,\ref{tab:CVSO109_wing_properties}). The mean FWHM of these Gaussian components for the H and Ca K lines is 263 km/s with a standard deviation of 25 km/s, and for the Ca II IR lines the mean FWHM is 164 km/s with a standard deviation of 2 km/s. This suggests that the H I emission within the broad component is at a different location compared to the Ca II emission. The wing component central velocity from the H and Ca II K lines shows a blueshift between the XSHOOTER and final UVES epoch, across all observations except the H$\beta$ line. There is also an observed redshift of the broad component central positions of these lines between the first two UVES epochs. These measured differences are  clearly larger than the standard error of the Gaussian fits. If this broad component emission is due to rotating material, these results show that the maximum rotational velocity happens between the second UVES epoch and the XSHOOTER epoch, and that the material may have been at maximum blueshift $\sim$one-two days before the first UVES epoch.

Using the typical mass for an M0 star \citep[M$_*$ $\approx$ 0.55M$_\odot$,][]{2000A&A...358..593S}, a stellar radius of 1.9 R$_*$ \citep[from the T and L$_*$,][]{2021arXiv210312446M}, and an inclination of $\sim$35 deg, which we calculate from the orbital period of 6.8 d and the measured \vsini, we calculate upper limits for the radii of the material, assuming it is due to rotation. The mean observed emission wing velocities correspond to material producing emission at radii of  $\sim$ 2 and 11 R$_*$ for the Ca II and H I lines, respectively; the opposite trend to that observed in GQ Lupi A for these species. These are still upper limits to the radii since although we consider the stellar inclination, what we observe during the XSHOOTER epoch is likely not at the maximum, as previously discussed. The co-rotation radius of this system is $\sim$ 6.5 R$_*$, therefore infall contributions, especially from the Ca II, are also likely. Furthermore, using the lower inclination obtained from the ROTFIT \vsini\ also gives smaller radii, which is more suggestive of infall contributions to the overall motion if the material is in the orbital plane of the stellar rotation. Rapidly evolving velocity changes could result from a non-axisymmetric accretion structure or inner disc feature, similar to that identified from the spectra of EX Lupi \citep{sicilia-aguilar_accretion_2015,sicilia-aguilar_optical_2012}. This may explain the dipper-like behaviour of the light curve, whereby the material in the disc is occulting the emission from the star. Without further data, we cannot confirm this, but the extrapolated timescale of this feature (if it is periodic) would also be consistent with the photometric period.

We further investigated the observed velocities in the H Balmer lines with respect to the transition properties of these lines. Figure\,\ref{fig:gauss_vs_aki} shows the the central position of the wing Gaussian component versus the transition probability (Aki) and excitation energy (Ek) for the H lines in the XSHOOTER spectra. There is a clear correlation between each parameter and the central position of the wing component. The material closest to the star has the lower transition probability and higher excitation energy. Spearman's Rank-Order Correlation coefficients of 0.96 (p=2e-5) and -0.93 (p=2e-4) are calculated for position and the Ek and Aki, respectively. This result, together with the range in velocities across the epochs, not only suggests that we are tracing rotating material at different radial locations, but also at different temperatures and densities at the varying distances. Although we do include the H$\alpha$ line because the measured standard error of the fit is relatively low, this line will be originating from many locations. This could be both accretion- and wind-related, and the wings could be strongly suppressed \citep{mendigutia_accretion_2011}. The H 3889\,\AA\ line appears to be a slight outlier in the plot. This line has more of a blueshift across the narrow components for all epochs; if we were to instead plot the broad component position minus the narrow component central position for all lines, this line then correlates better with the overall results. It is not clear why this line in particular shows this behaviour. The other lower velocity H lines still show the good correlation with the overall data between the central velocities and the transition properties.

These results from CVSO109 show that even with short temporal coverages, we are able to extract detailed information using \SM\ pertaining to variability and the geometry of accretion and inner disc structures. This is promising for further application to the rest of the PENELLOPE survey data, as well as other short time-scale spectral observations.

\section{Conclusions \& Future Work}
\label{sec:conc}

This paper introduces the Python package \SM, which we have developed to analyse the stellar spectra of YSOs and apply the emission line tomography analysis. Data were analysed from UVES, XSHOOTER, FEROS, HARPS, and ESPaDoNs. \SM\ is also compatible with many further instruments, including ESPRESSO and SOPHIE. \SM\ can be run interactively within a Jupyter notebook, allowing for easy customisation, selections, and results that are compatible with other widely used Python libraries. The key capabilities of \SM\ were detailed within and include:
\begin{itemize}
	\item Read in spectra data directly from the \texttt{FITS} files, with allowances for homogenization of data (accounting for barycentric corrections and flux scaling). 
	\item Convenience functions of target and data selection, wavelength exclusions, and line plotting for both wavelength and velocity scales. 
	\item RV and \vsini\ calculation using template stellar spectra (we plan to include synthetic model spectra for comparisons in a future version). 
	\item Automatic emission line identification and matching to the NIST database of lines from the continuum subtracted spectra. 
	\item Line fitting using Gaussian models. Customisable constraints may be applied to the centre/width of the Gaussian(s). $\chi^2$ goodness-of-fit criteria applied. 
	\item Periodicity analysis using periodograms and phase folded variability plots. 
	\item Physical property approximations using Saha/Boltzmann equations and the Sobolev large velocity gradient approximation. 
\end{itemize}
In addition to detailing the \SM\ package functionality, we also presented its application to three PMS stars, EX Lupi, GQ Lupi A and CVSO109. 

The EX Lupi quiescence data from 2007-2019 were used to confirm that \SM\ produces the same results that we found in our previous work on this star \citep{kospal_radial_2014,sicilia-aguilar_accretion_2015}. All previously identified strong lines were recovered and well-fitted. As were the periodicities and offset in phase between the highly energetic lines, such as the He II and Fe lines and the lower energy lines such as the Mg I. Line emission from He I and He II is consistent with very high temperature (over 20,000K) and density,
indicative of a small spot size and non-thermal ionisation that will require a more detailed treatment in the future. The temperature derived from Fe I/Fe II lines is significantly lower.  

GQ Lupi A displays inverse PCygni YY-Ori type profiles in many of the H, He and Ca II lines. Accurate model fitting to these data reveal the maximum position of the infall velocity is anti-correlated with the excitation energy, suggesting that the highest velocities are achieved by lines produced in the lowest temperature regions. There is also suggestion of a misalignment between the stellar and disc rotation axes. The infall velocities allow us to put an upper limit on the size of the magnetosphere, which is smaller than the one observed in EX Lupi.

CVSO109 is the first of the PENELLOPE survey \citep{2021arXiv210312446M} targets to be analysed by \SM. Clear variability is observed across the four epochs by a redshifted emission wing in the H and Ca II lines. The central position of this emission wing correlates with the transition probability and upper-energy level of the lines. Such a feature is indicative of rotating material, which we show may span a range of radial locations with different temperatures. 

By using as examples stars with different amounts of data and time sampling, we also reveal that important constraints to the accretion properties and associated structures can be placed even with incomplete time coverage, if the multiple emission lines observed are accurately extracted and fitted.
Although the three stars examined here are similar in age and stellar mass, we also find a variety of behaviours regarding variability and stability of accretion structures as well as sizes of the magnetospheres (upper limits of $\sim$ 9, 5, and 6 R$_*$ for EX Lupi, GQ Lupi A and CVSO109, respectively). This may indicate differences in the stellar structure and/or in the properties of the inner disc, including differences in the star-disc inclination and in the  magnetic field topology \citep{johnstone14}. The analysis of the line variability reveals that EX Lupi is accreting in a very stable mode, while GQ Lupi has unstable accretion \citep{kurosawa_spectral_2013} and may display a misalignment between the inner and outer disc and/or the disc and accretion channels. Further data are required for confirmation of the nature of the variability in CVSO109, but our results show good evidence for periodic rotating material with infall contributions, which may be occulting the star. This also suggests that material at high latitude, crossing the line-of-sight may be more common than previously thought, being observed in stars that are not highly inclined.

\SM\ automates many aspects of our analysis procedure, allowing for systematic results to be obtained quickly. However, line identification remains a complex process, requiring careful examination of the results obtained. This paper has shown that we can effectively use \SM\ to facilitate our analysis of time-resolved spectra of PMS stars; with many further applications for emission line analysis. For example, we are currently using \SM\ to identify and analyse emission lines from hyper-velocity impact flash spectra \citep{2021LPI....52.1625S}, as well as time- and spatially-resolved coronal mass ejections from solar data.

\SM\ is available to the community on \href{https://github.com/justyncw/STAR_MELT}{GitHub} and the example notebook can be run online via the \href{https://mybinder.org/v2/gh/justyncw/STAR_MELT/HEAD?filepath=STAR_MELT_example_notebook.ipynb}{binder platform}.

\section*{Acknowledgements}

We thank the anonymous referee for their useful comments and
suggestions that have improved this work. JCW, ASA and SM are supported by the  STFC grant number ST/S000399/1 ("The Planet-Disk Connection: Accretion, Disk Structure, and Planet Formation").
This work benefited from discussions with the ODYSSEUS team (HST AR-16129), \url{https://sites.bu.edu/odysseus/}. This project has received funding from the European Union's Horizon 2020 research and innovation programme under the Marie Sk{\l}odowska-Curie grant agreement No 823823 (DUSTBUSTERS). 
This research received financial support from the project PRIN-INAF 2019 "Spectroscopically Tracing the Disk Dispersal Evolution". We also thank the ESO Data Archive Helpdesk and, in particular, John Pritchard, for their valuable help.

\section*{Data availability}
All spectral data used in this work are publicly available. Data for EX Lupi and GQ Lupi A are accessed from the ESO archive (\url{http://archive.eso.org/}). Data for CVSO109 from the ESO PENELLOPE survey are available within the ODYSSEUS repository at \url{https://doi.org/10.5281/zenodo.4477091} for the XSHOOTER data and \url{https://doi.org/10.5281/zenodo.4478360} for the UVES data. The \SM\ package is available at \url{https://github.com/justyncw/STAR_MELT}. The authors also welcome collaborative projects using the package. 




\bibliographystyle{mnras}
\bibliography{papers.bib} 


\appendix
\section{Standard stars for RV and \vsini\ calculation}

\begin{table}
  \caption{List of standard stars used for the RV and \vsini\ calculations.}
  \begin{tabular}{lclc}
    \hline
    Star & Sp Type & RV [km/s] & Reference \\
    \hline
	* tau Sco &	B0.2V &	2.0	$\pm$ 0.9	& \citet{evans_revision_1967} \\
	* alf Leo &	B8IVn &	5.9	$\pm$ 2.0 & \citet{evans_revision_1967} \\
     HD 18331 & A1V & -10.7 $\pm$ 0.9 & \citet{gontcharov_pulkovo_2006} \\
    * alf PsA &	A4V &	6.5	$\pm$ 0.5	& \citet{gontcharov_pulkovo_2006} \\
	* alf CMi &	F5IV &	-15.9	$\pm$ 0.1 & {\citet{buder_galah_2019}} \\	
	* alf Cen A & G2V &	-21.4	$\pm$ 0.8 & {\citet{valenti_spectroscopic_2005}} \\
	* tau Cet &	G8V &	-16.6		& {\citet{soubiran_gaia_2018}} \\
	* 70 Oph &	K0V &	 -5.8		& {\citet{soubiran_gaia_2018}} \\
	HD 114386 &	K3V	 &33.4		& {\citet{soubiran_gaia_2018}} \\
	GJ488 &	M0V &	5.0     	 &	 {\citet{2019A&A...629A..80H}}\\
	* alf Cet &	M1.5IIIa &	-26.1	 &	 {\citet{famaey_local_2005}}\\
	* psi Phe &	M4III &	2.9	$\pm$ 1.5 & \citet{gontcharov_pulkovo_2006} \\
    \hline
  \end{tabular}
  \label{tab:standard_stars}
\end{table}

\section{Instrument Descriptions}
\label{app:inst}
This section contains technical descriptions of the instruments used to obtain the data for this work.

\subsection{FEROS and HARPS}
The ESO La Silla instruments FEROS \citep[Fiber-fed Extended Range Optical Spectrograph;][]{1999Msngr..95....8K} has a resolution of R = 48000, and a wavelength coverage between 3500-9200\AA, containing in total 39 echelle orders. HARPS \citep[High Accuracy Radial velocity Planet Searcher;][]{2003Msngr.114...20M} has a higher resolution of R = 115000, and a lower spectral coverage of 3780-6910\AA. Data were taken from the ESO Science Archive and had been reduced by the standard pipelines. Since the majority of the 1-dimensional spectral data on the ESO Science Archive now follows the ESO Science Data Products (SDP) standard, \SM\, was designed to be fully conformative with this standard. We incorporate an adapted version of the ESO python script \textsc{1dspectrum.py}, version: 2017-08-07 available at \\ \url{https://archive.eso.org/cms/eso-data/help/1dspectra.html\#Python}, to extract the 1-D spectrum data from the \texttt{FITS} files. Further modifications also allow for other ESO and non-ESO instruments to be read in directly. 

\subsection{UVES and XSHOOTER}
The ESO VLT instrument UVES \citep[Ultraviolet and Visual Echelle Spectrograph][]{2000SPIE.4008..534D} uses a single long slit, dividing the spectra in two arms, Red (480 nm) and Blue (450 nm). The 0.6'' wide slits in both arms were used for these observations, leading to a spectral resolution R$\sim$70,000. The wavelength ranges of 330-450nm, 480-575 nm, and 585-680 nm were simultaneously covered. For the data in this work, the exposure times allowed for S/N at 630 nm of at least 50 in each observation. 

The ESO VLT instrument XSHOOTER \citep{2011A&A...536A.105V} is a slit-fed (11'' long) spectrograph, providing simultaneous coverage of the wavelength region between $\sim$300 nm and $\sim$2500 nm, divided into three arms, UVB (300$\lesssim \lambda \lesssim$ 500 nm), VIS (500$\lesssim \lambda \lesssim$ 1000 nm), and NIR (1000$\lesssim \lambda \lesssim$ 2500 nm). Exposures with a set of a 1.0\arcsec-0.4\arcsec-0.4\arcsec ~ wide slits for the UVB, VIS, and NIR arms, respectively, allows for high S/N spectra with $R\sim$5400, 18400, and 11600 in the three arms, respectively. The exposure times were such that the S/N at 400 nm is about $\gtrsim$3-5 and in the VIS and NIR arms $\gtrsim$100.

Data reduction for the data acquired with each instrument is done using the ESO Reflex workflow v2.8.5 \citep{2013A&A...559A..96F}, using the pipeline for each instruments: the UVES v6.1.3 pipeline \citep{2000Msngr.101...31B}, the XSHOOTER pipeline v3.5.0 \citep{2010SPIE.7737E..28M}, as described by \citet{2021arXiv210312446M}. The pipelines apply the standard flat, bias, and dark correction, wavelength calibration, spectral rectification and extraction of the 1D spectrum.

\subsection{ESPaDOnS}
The CFHT instrument ESPaDOnS: an Echelle SpectroPolarimetric Device
for the Observation of Stars is a fibre-fed echelle spectropolarimeter with a spectral resolution R = 65000, and a wavelength coverage between 3700-10500\AA, over 40 echelle orders. This work only features the spectral intensity (Stokes \textit{I}) data, which was obtained from the CHFT archive and has been reduced with the standard pipeline.

\section{Observation log}

\begin{table}  
\begin{center}  
\footnotesize  
\caption{\label{tab:obs_log} Summary of observations used.}  
\begin{tabular}{c|c| c |c  }    
\hline 
   Star/    &  &  &  \\

   Programme ID    & Instrument  & Obs. Start Date & \# Obs. \\
\hline 
\hline
\textbf{EX Lupi} & 		&  &  \\
	079.A-9017 & 	FEROS	& 2007-07-29T03:45:21 & 3 \\
   082.C-0390 &	HARPS	& 2009-02-12T08:06:05 & 5\\
 	082.C-0427 &	HARPS &	2009-03-01T07:12:24 & 4\\
  083.A-9011 &	FEROS &	2009-08-13T00:55:13 & 3 \\
  083.A-9017 &	FEROS &	2009-09-02T00:51:49 & 1 \\
  084.A-9011 &	FEROS &	2010-03-02T07:39:24 & 4 \\
  085.A-9027 &	FEROS &	2010-04-23T09:27:19 & 15 \\
  086.A-9006 &	FEROS &	2011-01-25T08:54:32 & 3 \\
  086.A-9012 &	FEROS &	2011-03-10T07:33:59 & 3 \\
  087.A-9013 &	FEROS &	2011-04-17T04:33:53 & 8 \\
 	089.A-9007 &	FEROS &	2012-07-03T00:19:10 & 10\\
 	091.A-9013 &	FEROS &	2013-07-15T23:16:04& 14\\
 	092.A-9007 &	FEROS &	2014-02-16T06:47:50 & 17\\
 	16AF03 &	 ESPaDOnS &	2016-06-09T09:00:00 &  50 \\ 
 	099.A-9010 &	FEROS &	2017-06-07T03:15:27 & 4\\
 	19AF50 &	 ESPaDOnS &	2019-05-31T09:56:09 &  24 \\ 

 \hline
\textbf{GQ Lupi} &	 & &\\
	074.C-0037 &	HARPS &	2005-02-06T08:43:33 & 7\\
 	075.C-0710 &	HARPS &	2005-05-05T01:10:18 & 9\\
 	079.A-9009 &	FEROS &	2007-05-05T05:41:06 & 3\\
 	081.A-9005 &	FEROS &	2008-04-18T03:28:44 & 2\\
 	081.A-9023 &	FEROS &	2008-06-16T06:23:08 & 3 \\
 	082.C-0390 &	HARPS &	2009-02-12T08:20:46 & 3\\
 	083.A-9004 &	FEROS &	2009-04-22T07:37:29 & 14\\
 	083.A-9011 &	FEROS &	2009-08-11T00:18:23& 1\\
 	083.A-9017 &	FEROS &	2009-06-09T05:55:37& 2\\
 	085.C-0764 &	XSHOOTER	 & 2010-05-05T06:16:33 & 3\\
 	085.A-9027 &	FEROS &	2010-05-24T05:08:24& 14\\
 	097.C-0669 &	XSHOOTER	 & 2016-05-03T05:15:55& 15\\
 \hline
\textbf{CVSO 109} & & & \\
 106.20Z8.009 & UVES & 2020-11-26T03:29:45 & 3 \\
  106.20Z8.002 & XSHOOTER & 2020-11-28T03:36:03 & 1 \\
\hline 
\end{tabular} 
\end{center} 
\end{table}



\onecolumn

\section{List of identified lines}
Included in the supplementary material of the article.

\begin{longtable}{llrlrrll}
\caption{List of lines identified in EX Lupi, CVSO 109, GQ Lupi A. Previous ID column corresponds to lines identified in our previous work. o: outburst, q: quiescence. E12,E15: EX Lupi \citep{sicilia-aguilar_optical_2012,sicilia-aguilar_accretion_2015}, A17: ASASSN-13db \citep{sicilia-aguilar_2014-2017_2017}, Z20: ZCMa \citep{sicilia-aguilar_reading_2020}. Wavelengths marked by * indicate possible blends or multiple matches to the identified line. Acc denotes the accuracy for each transition strength from the NIST database, ranging from $\leq$0.3\% for AAA to > 50\% for E.} \label{tab:lines}\\
\toprule
$\lambda$ &  Element &     Aki &  Acc &    Ei &    Ek &                           Previous ID &     Star \\
(\r{A}) &  & [/s]  &   &  [eV] & [eV] &   &   \\

\midrule
\endhead
\midrule
\multicolumn{8}{r}{{Continued on next page}} \\
\midrule
\endfoot

\bottomrule
\endlastfoot
   3119.49 &   Fe I & 8.3e+06 &    B &  2.43 &  6.41 &                             &                CVSO 109 \\
   3192.06* &  Fe II & 5.0e+05 &    D &  3.81 &  7.70 &                             &                CVSO 109 \\
   3192.27* &  Ti II & 2.2e+06 &   B' &  1.08 &  4.96 &                             &                CVSO 109 \\
   3227.74* &  Fe II & 8.9e+06 &    C &  1.67 &  5.51 &                             &         CVSO 109/GQ Lupi \\
   3227.80* &   Fe I & 5.0e+07 &    B &  2.43 &  6.27 &                             &                CVSO 109/GQ Lupi \\
   3255.89 &  Fe II & 2.8e+05 &    B &  0.99 &  4.79 &                             &         CVSO 109/GQ Lupi \\
   3277.35 &  Fe II & 3.3e+05 &    B &  0.99 &  4.77 &                             &         CVSO 109/GQ Lupi \\
   3278.92 &  Ti II & 8.8e+07 &   B+ &  1.08 &  4.86 &                             &                CVSO 109 \\
   3387.83 &  Ti II & 2.8e+07 &   B+ &  0.03 &  3.69 &                             &                CVSO 109 \\
   3483.01 &   Fe I & 1.4e+05 &    B &  0.91 &  4.47 &                             &                CVSO 109 \\
   3691.56 &    H I & 3.7e+03 &  AAA & 10.20 & 13.56 &                             &                 GQ Lupi \\
   3697.16 &    H I & 4.9e+03 &  AAA & 10.20 & 13.55 &                             &                 GQ Lupi \\
   3703.86 &    H I & 6.7e+03 &  AAA & 10.20 & 13.55 &                             &                 GQ Lupi \\
   3711.98 &    H I & 9.2e+03 &  AAA & 10.20 & 13.54 &                           Z20o &                 GQ Lupi \\
   3721.94 &    H I & 1.3e+04 &  AAA & 10.20 & 13.53 &                             &         CVSO 109/GQ Lupi \\
   3734.37 &    H I & 1.9e+04 &  AAA & 10.20 & 13.52 &                             &         CVSO 109/GQ Lupi \\
   3736.90 &  Ca II & 1.7e+08 &    C &  3.15 &  6.47 &                           Z20o &                CVSO 109 \\
   3750.15 &    H I & 2.8e+04 &  AAA & 10.20 & 13.50 &                           Z20o &         CVSO 109/GQ Lupi \\
   3761.32* &  Ti II & 1.2e+08 &   A' &  0.57 &  3.87 &                           Z20o &                CVSO 109 \\
   3761.41* &   Fe I & 1.0e+06 &    B &  2.59 &  5.88 &                           Z20o &                CVSO 109 \\
   3770.63 &    H I & 4.4e+04 &  AAA & 10.20 & 13.49 &                           Z20o &         CVSO 109/GQ Lupi \\
   3797.91 &    H I & 7.1e+04 &  AAA & 10.20 & 13.46 &                           Z20o &         CVSO 109/GQ Lupi \\
   3825.88 &   Fe I & 6.0e+07 &    A &  0.91 &  4.15 &                             &                CVSO 109 \\
   3835.40 &    H I & 1.2e+05 &  AAA & 10.20 & 13.43 &                           Z20o/E12q &         CVSO 109/GQ Lupi \\
   3889.06 &    H I & 2.2e+05 &  AAA & 10.20 & 13.39 &                      Z20o/E12q &         CVSO 109/GQ Lupi \\
   3933.66 &  Ca II & 1.5e+08 &    C &  0.00 &  3.15 &                 Z20q/Z20o/E12q &  CVSO 109/GQ Lupi/EX Lupi \\
   3968.47 &  Ca II & 1.4e+08 &    C &  0.00 &  3.12 &                      Z20o/E12q &  CVSO 109/GQ Lupi/EX Lupi \\
   3970.04 &    H I & 4.4e+05 &  AAA & 10.20 & 13.32 &                           E12q &  CVSO 109/GQ Lupi/EX Lupi \\
   4026.19 &   He I & 2.9e+06 &  AAA & 20.96 & 24.04 &                      E15q/E12q &                 CVSO 109/GQ Lupi \\
   4100.74 &   Fe I & 2.9e+04 &    A &  0.86 &  3.88 &                             &                 GQ Lupi \\
   4101.71 &    H I & 2.9e+06 &  AAA & 10.20 & 13.22 &                 Z20q/Z20o/E12q &  CVSO 109/GQ Lupi/EX Lupi \\
   4178.86 &  Fe II & 1.7e+05 &    C &  2.58 &  5.55 &            Z20q/E15q/Z20o/E12q &                CVSO 109 \\
   4226.34* &   Fe I & 4.3e+06 &   D+ &  2.85 &  5.78 &                           A17o &                 EX Lupi \\
   4226.73* &   Ca I & 2.2e+08 &   B+ &  0.00 &  2.93 &                 Z20q/E15q/E12q &                 EX Lupi \\
   4233.17 &  Fe II & 7.2e+05 &   C+ &  2.58 &  5.51 &  E12o/Z20q/E15q/Z20o/E12q/A17o &                 EX Lupi \\
   4340.47 &    H I & 2.5e+06 &  AAA & 10.20 & 13.05 &                 Z20q/Z20o/E12q &         CVSO 109/GQ Lupi/EX Lupi \\
   4416.83 &  Fe II & 2.1e+05 &    D &  2.78 &  5.58 &                           E15q &                 EX Lupi \\
   4471.48 &   He I & 6.8e+05 &  AAA & 20.96 & 23.74 &                           E15q &  CVSO 109/GQ Lupi/EX Lupi \\
     4481.13 &  Mg II & 2.3e+08 &    A &  8.86 & 11.63 &                             &                 EX Lupi \\
   4508.29 &  Fe II & 7.0e+05 &    D &  2.86 &  5.60 &            Z20q/E15q/Z20o/E12q &                 EX Lupi \\
   4549.47 &  Fe II & 1.0e+06 &   C+ &  2.83 &  5.55 &  E12o/Z20q/E15q/Z20o/E12q/A17o &                 EX Lupi \\
   4583.83 &  Fe II & 7.2e+05 &   C+ &  2.81 &  5.51 &                      Z20q/Z20o/E12q &                 EX Lupi \\
   4685.80 &  He II & 2.2e+08 &   AA & 48.37 & 51.02 &                           E12q &          GQ Lupi/EX Lupi \\
   4861.28 &    H I & 1.7e+07 &  AAA & 10.20 & 12.75 &       E12o/E12q/A17o/Z20q/Z20o &         CVSO 109/GQ Lupi/EX Lupi \\
   4921.93 &   He I & 2.0e+07 &  AAA & 21.22 & 23.74 &                      E15q/E12q &                 EX Lupi \\
   4923.92 &  Fe II & 4.3e+06 &    C &  2.89 &  5.41 &       Z20q/E15q/Z20o/E12q/A17o &          GQ Lupi/EX Lupi \\
   5015.68* &   He I & 1.3e+07 &  AAA & 20.62 & 23.09 &                      E15q/E12q &          GQ Lupi/EX Lupi \\
   5015.75* &  Fe II & 6.2e+07 &   B+ & 10.35 & 12.82 &                             &          GQ Lupi/EX Lupi \\
   5016.61 &   V II & 2.6e+06 &    C &  5.54 &  8.01 &                             &                 EX Lupi \\
   5018.43 &  Fe II & 2.0e+06 &   D+ &  2.89 &  5.36 &  E12o/Z20q/E15q/Z20o/E12q/A17o &  CVSO 109/GQ Lupi/EX Lupi \\
   5167.32* &   Mg I & 1.1e+07 &   B+ &  2.71 &  5.11 &                      E12q/A17o &                 EX Lupi \\
   5167.48* &   Fe I & 2.7e+06 &   B+ &  1.48 &  3.88 &                           E12q &                 EX Lupi \\
   5169.03 &  Fe II & 4.2e+06 &    C &  2.89 &  5.29 &                 E15q/Z20o/E12q &  CVSO 109/GQ Lupi/EX Lupi \\
   5172.68 &   Mg I & 3.3e+07 &    B &  2.71 &  5.11 &       Z20q/E15q/Z20o/E12q/A17o &                 EX Lupi \\
   5183.49* &  Fe II & 1.8e+06 &   D+ & 10.43 & 12.82 &                             &                 EX Lupi \\
   5183.60* &   Mg I & 5.6e+07 &    B &  2.72 &  5.11 &       Z20q/E15q/Z20o/E12q/A17o &                 EX Lupi \\
   5185.91 &  Ti II & 1.0e+06 &   B+ &  1.89 &  4.28 &                      E15q/E12q &                 EX Lupi \\
   5197.58 &  Fe II & 5.5e+05 &    C &  3.23 &  5.62 &            Z20q/E15q/Z20o/E12q &                 EX Lupi \\
   5234.62 &  Fe II & 2.5e+05 &   D+ &  3.22 &  5.59 &            E12o/Z20q/E15q/Z20o &                 EX Lupi \\
   5269.54 &   Fe I & 1.3e+06 &    A &  0.86 &  3.21 &            Z20q/E15q/Z20o/E12q &                 EX Lupi \\
   5276.00 &  Fe II & 3.8e+05 &    C &  3.20 &  5.55 &            E12o/Z20q/Z20o/E12q &                 EX Lupi \\
   5284.11 &  Fe II & 1.9e+04 &    E &  2.89 &  5.24 &            E12q/E12o/Z20q/E15q/Z20o &                 EX Lupi \\
   5316.62 &  Fe II & 3.9e+05 &    C &  3.15 &  5.48 &       E12o/Z20q/E15q/Z20o/E12q/A17o &                 EX Lupi \\
   5362.86 &  Fe II & 1.5e+07 &   B+ & 10.50 & 12.81 &                           E12q &                 EX Lupi \\
   5534.85 &  Fe II & 3.0e+04 &    E &  3.24 &  5.48 &  E12o/Z20q/E15q/Z20o/E12q/A17o &                 EX Lupi \\
   5577.34 &    O I & 1.3e+00 &   B+ &  1.97 &  4.19 &                      Z20q/Z20o &                 GQ Lupi \\
   5875.62 &   He I & 1.8e+07 &  AAA & 20.96 & 23.07 &                 E12o/E12q/Z20q/E15q/Z20o &  CVSO 109/GQ Lupi/EX Lupi \\
   5889.95 &   Na I & 6.2e+07 &   AA &  0.00 &  2.10 &            Z20q/Z20o/E12q/A17o &                 EX Lupi \\
   6147.73 &  Fe II & 1.3e+05 &    E &  3.89 &  5.90 &                           E15q &                 EX Lupi \\
   6247.56 &  Fe II & 1.6e+05 &    D &  3.89 &  5.88 &  E12o/Z20q/E15q/Z20o/E12q/A17o &                 EX Lupi \\
   6300.30 &    O I & 5.6e-03 &   C+ &  0.00 &  1.97 &                      Z20q/Z20o &  CVSO 109/GQ Lupi/EX Lupi \\
   6347.10 &  Si II & 5.8e+07 &   B+ &  8.12 & 10.07 &       E12o/E15q/Z20o/E12q/A17o &                 EX Lupi \\
   6363.78 &    O I & 1.6e-03 &   C+ &  0.02 &  1.97 &                      Z20q/Z20o &                 EX Lupi \\
   6371.36 &  Si II & 6.8e+07 &   C+ &  8.12 & 10.07 &            E12o/E15q/E12q/A17o &                 EX Lupi \\
   6456.38 &  Fe II & 1.7e+05 &    D &  3.90 &  5.82 &  E12o/Z20q/E15q/Z20o/E12q/A17o &                 EX Lupi \\
   6562.71 &    H I & 5.4e+07 &  AAA & 10.20 & 12.09 &                             &  CVSO 109/GQ Lupi/EX Lupi \\
   6677.98 &   Fe I & 6.3e+05 &    B &  2.69 &  4.55 &                             &                 GQ Lupi \\
   6678.15 &   He I & 6.4e+07 &  AAA & 21.22 & 23.07 &       E12o/Z20q/E15q/Z20o/E12q &  CVSO 109/GQ Lupi/EX Lupi \\
   7065.19 &   He I & 1.5e+07 &  AAA & 20.96 & 22.72 &                 E12o/E15q/E12q &          GQ Lupi/EX Lupi \\
   7771.94 &    O I & 3.7e+07 &    A &  9.15 & 10.74 &                 E15q/Z20o/E12q &                 EX Lupi \\
   8446.25 &    O I & 3.2e+07 &    B &  9.52 & 10.99 &            E15q/Z20o/E12q/A17o &          GQ Lupi/EX Lupi \\
   8498.02 &  Ca II & 1.1e+06 &    C &  1.69 &  3.15 &            E12o/Z20o/E12q/A17o &  CVSO 109/GQ Lupi/EX Lupi \\
   8499.62 &  Fe II & 2.3e+07 &   B+ &  9.76 & 11.22 &                             &                CVSO 109 \\
   8542.09 &  Ca II & 9.9e+06 &    C &  1.70 &  3.15 &                 Z20o/E12q/A17o &  CVSO 109/GQ Lupi/EX Lupi \\
   8545.38 &    H I & 6.5e+03 &  AAA & 12.09 & 13.54 &                             &                CVSO 109 \\
   8662.14 &  Ca II & 1.1e+07 &    C &  1.69 &  3.12 &            E12o/Z20o/E12q/A17o &  CVSO 109/GQ Lupi/EX Lupi \\
   8665.02 &    H I & 1.3e+04 &  AAA & 12.09 & 13.52 &                             &                CVSO 109 \\
   8862.78 &    H I & 3.2e+04 &  AAA & 12.09 & 13.49 &                           Z20o &                 GQ Lupi \\
   8886.18 &  Fe II & 9.0e+05 &   D+ & 11.26 & 12.65 &                             &                 GQ Lupi \\
   9014.91 &    H I & 5.2e+04 &  AAA & 12.09 & 13.46 &                           A17o &                 GQ Lupi \\
   9015.30 &    H I & 5.2e+04 &  AAA & 12.09 & 13.46 &                           Z20o &                 GQ Lupi \\
   9229.01 &    H I & 8.9e+04 &  AAA & 12.09 & 13.43 &                             &                 GQ Lupi \\
   9440.44 &  Fe II & 3.4e+05 &    E & 12.38 & 13.69 &                             &                CVSO 109 \\
  10049.37 &    H I & 3.4e+05 &  AAA & 12.09 & 13.32 &                             &         CVSO 109/GQ Lupi \\
  10826.48 &  Fe II & 7.2e+05 &   C+ &  9.70 & 10.85 &                             &                 GQ Lupi \\
  10830.25 &   He I & 1.0e+07 &  AAA & 19.82 & 20.96 &                             &         CVSO 109/GQ Lupi \\
  10937.99 &    H I & 9.6e+05 &  AAA & 12.09 & 13.22 &                             &         CVSO 109/GQ Lupi \\
  12818.07 &    H I & 2.2e+06 &  AAA & 12.09 & 13.05 &                             &         CVSO 109/GQ Lupi \\
\end{longtable}

\section{Radial velocity vs MJD fits for EX Lupi}
Included in the supplementary material of the article. 

\noindent The tables sown here contain the results of fitting the radial velocity vs MJD with the rotational modulation equation (eq. \ref{rota-eq}) used to derive the position of the line-emitting region for various lines observed in the HARPS, FEROS, and ESPaDOnS spectra of EX Lupi. 

\begin{table}
\centering
  \caption{Examples of best fit to the line radial velocity. As period we list the maximum of the GLS periodogram excluding aliases, \textcolor{black}{ and we also provide the false-alarm-probability (FAP, assuming a white noise model).  $^*$ The phase offset is calculated using the period for the He II line and the zero point MJD 54309.1147. The amplitude is obtained  by fitting Equation \ref{rota-eq} to the line velocities. Objects marked as having quasi-periods tend to show multiple low-significance lines around the quasi-period value. Note that wavelengths $>$6700\AA\ are not covered by HARPS.} 
  } \label{exlup-rvline}
  \begin{tabular}{lccccccl}
    \hline
    Species/$\lambda$  & Amplitude & Offset & Phase$^*$ & Period & FAP & \# of & Notes\\
    (\AA) & (km/s) & (km/s) & (deg) & (d) &  & epochs & \\
    \hline
HeII 4686 & 2.5$\pm$0.3 &  4.5$\pm$0.2 &  -38$\pm$8 & 7.41684 & 2e-8  & 126 & \\
HeI 4471  & 1.1$\pm$0.2 &  4.2$\pm$0.1 &  6$\pm$10 & 7.41710 & 3e-4 & 112 &  \\
HeI 4713  & 1.8$\pm$0.3 &  4.9$\pm$0.2 &  20$\pm$20 & 6.45204 & 0.002  & 65 &  Very noisy, quasiperiods 6-8 d\\
HeI 4922  & 0.8$\pm$2 &  4.0$\pm$0.1 &  -20$\pm$20 & 8.59905  & 0.008 & 63 & Very noisy line\\
HeI 5016  & 1.0$\pm$0.1 &  0.37$\pm$0.08 &  -6$\pm$6 & 7.41481 & 5e-12 & 120  & \\
HeI 7065  & 2.0$\pm$0.2 &  4.2$\pm$0.2 &  33$\pm$8 & 7.41299 & 3e-6 & 72 & Near tellurics \\
FeII 4179 &  1.5$\pm$0.2 &  -0.8$\pm$0.1 &  7$\pm$8 & 6.98646 & 6e-7  & 90 & Quasiperiod\\
FeII 4233 &  1.0$\pm$0.1 &  -0.77$\pm$0.09 &  -13$\pm$7 & 7.41527 & 6e-9 & 103 & \\
FeII 4523 &  1.1$\pm$0.2 &  -0.8$\pm$0.1 &  20$\pm$10 & 6.99165 & 2e-5 & 89 & Quasiperiods 7-8d \\
FeII 4549 &  1.3$\pm$0.2 &  1.4$\pm$0.1 &  -41$\pm$9 & 7.41811 & 3e-7 & 117 &  \\
FeII 4584 &  1.1$\pm$0.2 &  -0.1$\pm$0.1 &  11$\pm$9 & 7.41373 & 1e-4 & 88 & \\
FeII 4923  &  1.0$\pm$0.1 &  0.7$\pm$0.1 &  -4$\pm$9 & 6.75582 & 2e-8 & 132 & Near HeI line \\
FeII 5018 &  1.0$\pm$0.1 &  0.94$\pm$0.09 &  10$\pm$10 & 6.11926  & 3e-7  & 129 & Quasiperiod 6-8d \\
FeII 5198 &  1.1$\pm$0.1 &  -0.6$\pm$0.1 &  11$\pm$8 & 7.41787 & 5e-7 & 82 & \\
FeII 5235 & 1.6$\pm$0.2 &  0.4$\pm$0.1 &  1$\pm$6 & 7.41507  & 1e-12 & 121 & \\ 
FeII 5317  &  1.5$\pm$0.2 &  1.4$\pm$0.1 &  -18$\pm$6 & 7.41633 & 7e-13  & 113 & \\
FeII 5362 & 1.1$\pm$0.2 &  0.8$\pm$0.1 &  30$\pm$10 & 7.59673 & 2e-4 & 63 & \\ 
FeII 6247 &  0.8$\pm$0.2 &  -1.0$\pm$0.1 &  -50$\pm$20 & 9.91940 & 0.04 & 32  &  \\
FeII 6456 &  1.3$\pm$0.2 &  -0.4$\pm$0.2 &  -30$\pm$20 & 7.59270 & 0.004  & 47 &  \\
FeI 5269 &  1.6$\pm$0.2 &  -0.4$\pm$0.1 &  -6$\pm$6 & 7.41557  & 3e-13  & 114 & \\
SiII 6347 &  1.3$\pm$0.1 &  -0.3$\pm$0.1 &  14$\pm$6 & 7.41586 & 2e-9 & 96 & \\
MgI 5167 &  1.5$\pm$0.2 &  2.8$\pm$0.1 &  -24$\pm$8 & 7.41583 & 7e-9 & 117 & \\
MgI 5172 &  1.3$\pm$0.2 &  0.2$\pm$0.1 &  -11$\pm$7 & 7.41380 & 6e-10 & 128 & \\
MgI 5184 &  1.4$\pm$0.1 &  0.01$\pm$0.09 &  4$\pm$5 & 7.41557 & 5e-15 & 125 & \\
MgII 4481 &  2.8$\pm$0.6 &  2.4$\pm$0.4 &  -10$\pm$20 & 6.65702  & 0.04  & 59  & Quasiperiod 6-8d, little data \\
CaII 8662 &  0.37$\pm$0.06 &  0.99$\pm$0.04 &  -110$\pm$20 & 3.10937  & 2e-4 & 109 & Double spot, half a rotational period?\\
OI 8446 &  2.5$\pm$0.4 &  5.3$\pm$0.3 &  40$\pm$20 & 5.940127 & 6e-4 & 68 & Nearby tellurics \\
\hline 
\end{tabular} 
\end{table}  

\begin{longtable}{lccccc}
\caption{Results of fitting the radial velocity of strong lines in EX Lupi, following Eq. \ref{rota-eq}. } \label{tab:exlupi-polar}\\
\toprule
Species/$\lambda$ &     MJD interval &  Amplitude &    V$_0$ &    Phase &    \# of points \\
(\r{A}) & [d] & [km/s]  & [km/s] & [deg] & \\
\midrule
\endhead
\midrule
\multicolumn{6}{r}{{Continued on next page}} \\
\midrule
\endfoot

\bottomrule
\endlastfoot
CaII 8662 &  56489.0 - 56550.1 &   0.8$\pm$0.1 &   1.2$\pm$0.1 &   -90$\pm$7 &  11 \\
FeI 5269 &  54874.3 - 55058.0 &   1.8$\pm$0.5 &   -0.2$\pm$0.4 &   12$\pm$16 &  11 \\
" &  55257.3 - 55407.1 &   0.5$\pm$0.2 &   -0.5$\pm$0.1 &   2$\pm$17 &  18 \\
" &  55586.4 - 55674.3 &   0.9$\pm$0.3 &   -0.1$\pm$0.2 &   -52$\pm$21 &  14 \\
" &  56704.3 - 56709.4 &   3.0$\pm$0.3 &   -0.5$\pm$0.2 &   5$\pm$4 &  17 \\
" &  56489.0 - 56709.4 &   2.1$\pm$0.3 &   -0.4$\pm$0.3 &   3$\pm$9 &  26 \\
" &  57548.4 - 57563.4 &   1.8$\pm$0.3 &   -0.5$\pm$0.3 &   -32$\pm$14 &  28 \\
FeII 4233 &  55257.3 - 55407.1 &   0.4$\pm$0.2 &   -0.6$\pm$0.1 &   -13$\pm$20 &  16 \\
" &  55586.4 - 55674.3 &   1.0$\pm$0.3 &   -0.2$\pm$0.2 &   -73$\pm$20 &  13 \\
" &  56704.3 - 56709.4 &   3$\pm$1 &   -1.1$\pm$0.7 &   -12$\pm$15 &  15 \\
" &  56489.0 - 56709.4 &   2.0$\pm$0.8 &   -1.0$\pm$0.5 &   -16$\pm$21 &  20 \\
" &  57548.4 - 57563.4 &   1.1$\pm$0.2 &   -1.0$\pm$0.2 &   -47$\pm$15 &  31 \\
FeII 4549 &  54874.3 - 55058.0 &   0.6$\pm$0.2 &   1.3$\pm$0.2 &   -35$\pm$23 &  10 \\
" &  55257.3 - 55407.1 &   0.2$\pm$0.3 &   2.2$\pm$0.2 &   -72$\pm$73 &  19 \\
" &  55586.4 - 55674.3 &   1.6$\pm$0.3 &   2.6$\pm$0.2 &   -63$\pm$12 &  14 \\
" &  56111.0 - 56121.1 &   1.9$\pm$0.6 &   1.4$\pm$0.5 &   -30$\pm$19 &  10 \\
" &  56704.3 - 56709.4 &   5$\pm$1 &   0.6$\pm$0.8 &   -40$\pm$12 &  15 \\
" &  56489.0 - 56709.4 &   3.5$\pm$0.9 &   0.9$\pm$0.6 &   -56$\pm$14 &  24 \\
" &  57548.4 - 57563.4 &   1.7$\pm$0.2 &   0.6$\pm$0.2 &   -35$\pm$12 &  31 \\
FeII 4924 &  54874.3 - 55058.0 &   1.2$\pm$0.4 &   0.2$\pm$0.3 &   25$\pm$18 &  11 \\
" &  55257.3 - 55407.1 &   0.1$\pm$0.2 &   0.5$\pm$0.1 &   -165$\pm$118 &  18 \\
" &  55587.4 - 55674.3 &   0.9$\pm$0.3 &   0.4$\pm$0.2 &   -60$\pm$19 &  13 \\
" &  56111.0 - 56121.1 &   1.9$\pm$0.5 &   0.5$\pm$0.4 &   27$\pm$17 &  10 \\
" &  56489.0 - 56550.1 &   2.1$\pm$0.7 &   0.6$\pm$0.4 &   -65$\pm$16 &  11 \\
" &  56704.3 - 56709.4 &   2.2$\pm$0.2 &   0.3$\pm$0.2 &   19$\pm$6 &  17 \\
" &  56489.0 - 56709.4 &   1.6$\pm$0.3 &   0.3$\pm$0.2 &   5$\pm$12 &  28 \\
" &  57548.4 - 57563.4 &   1.2$\pm$0.2 &   1.2$\pm$0.2 &   -16$\pm$11 &  42 \\
" &  58634.4 - 58643.5 &   5$\pm$2 &   3$\pm$5 &   90$\pm$60 &  12 \\
FeII 5018 &  54874.3 - 55058.0 &   0.6$\pm$0.3 &   0.5$\pm$0.2 &   88$\pm$31 &  11 \\
" &  55257.3 - 55407.1 &   0.4$\pm$0.2 &   0.4$\pm$0.1 &   -114$\pm$32 &  19 \\
" &  55586.4 - 55674.3 &   0.9$\pm$0.3 &   0.6$\pm$0.2 &   -76$\pm$22 &  14 \\
" &  56111.0 - 56121.1 &   2.0$\pm$0.8 &   0.3$\pm$0.6 &   1$\pm$26 &  10 \\
" &  56489.0 - 56550.1 &   1.9$\pm$0.7 &   0.8$\pm$0.5 &   -60$\pm$19 &  11 \\
" &  56704.3 - 56709.4 &   2.1$\pm$0.2 &   1.3$\pm$0.2 &   24$\pm$7 &  15 \\
" &  56489.0 - 56709.4 &   1.7$\pm$0.4 &   1.0$\pm$0.3 &   7$\pm$14 &  26 \\
" &  57548.4 - 57563.4 &   1.2$\pm$0.3 &   1.6$\pm$0.2 &   -16$\pm$14 &  43 \\
" &  58634.4 - 58643.5 &   8.7$\pm$4.8 &   0.4$\pm$5.4 &   119$\pm$28 &  10 \\
FeII 5235 &  55257.3 - 55407.1 &   0.4$\pm$0.2 &   -0.1$\pm$0.1 &   16$\pm$25 &  18 \\
" &  55586.4 - 55674.3 &   0.8$\pm$0.4 &   0.1$\pm$0.3 &   -41$\pm$27 &  14 \\
" &  56489.0 - 56550.1 &   1.2$\pm$0.4 &   0.6$\pm$0.4 &   -9$\pm$31 &  10 \\
" &  56704.3 - 56709.4 &   2.4$\pm$0.3 &   -0.1$\pm$0.2 &   21$\pm$7 &  16 \\
" &  56489.0 - 56709.4 &   1.9$\pm$0.3 &   0.2$\pm$0.2 &   22$\pm$9 &  26 \\
" &  57548.4 - 57563.4 &   1.7$\pm$0.3 &   1.0$\pm$0.2 &   -17$\pm$11 &  35 \\
FeII 5317 &  55257.3 - 55407.1 &   0.5$\pm$0.3 &   2.0$\pm$0.2 &   -7$\pm$30 &  19 \\
" &  55586.4 - 55674.3 &   1.2$\pm$0.2 &   2.2$\pm$0.2 &   -49$\pm$11 &  14 \\
" &  56704.3 - 56709.4 &   2.8$\pm$0.3 &   1.3$\pm$0.2 &   -3$\pm$6 &  16 \\
" &  56489.2 - 56709.4 &   1.8$\pm$0.5 &   1.2$\pm$0.3 &   -14$\pm$14 &  23 \\
" &  57548.4 - 57563.4 &   1.9$\pm$0.1 &   0.9$\pm$0.1 &   -35$\pm$6 &  34 \\
HeI 4471 &  54874.3 - 55058.0 &   1.8$\pm$0.4 &   4.0$\pm$0.2 &   -12$\pm$10 &  10 \\
" &  55257.3 - 55407.1 &   0.4$\pm$0.4 &   5.1$\pm$0.3 &   71$\pm$61 &  17 \\
" &  55586.4 - 55674.3 &   1.0$\pm$0.3 &   5.9$\pm$0.2 &   -36$\pm$15 &  14 \\
" &  56111.0 - 56121.1 &   2.5$\pm$0.7 &   4.0$\pm$0.6 &   -7$\pm$19 &  10 \\
" &  56704.3 - 56709.4 &   2.8$\pm$0.3 &   4.6$\pm$0.2 &   -3$\pm$6 &  14 \\
" &  56489.0 - 56709.4 &   1.7$\pm$0.4 &   4.4$\pm$0.3 &   6$\pm$16 &  21 \\
" &  57548.4 - 57563.4 &   1.3$\pm$0.2 &   3.0$\pm$0.1 &   -9$\pm$9 &  29 \\
HeI 5016 &  55257.3 - 55407.1 &   0.5$\pm$0.2 &   0.4$\pm$0.2 &   2$\pm$23 &  17 \\
" &  55586.4 - 55674.3 &   0.9$\pm$0.2 &   0.6$\pm$0.2 &   -28$\pm$14 &  13 \\
" &  56111.0 - 56121.1 &   2.0$\pm$0.2 &   -0.2$\pm$0.2 &   9$\pm$7 &  10 \\
" &  56704.3 - 56709.4 &   2.2$\pm$0.2 &   -0.2$\pm$0.1 &   -0$\pm$4 &  16 \\
" &  56489.0 - 56709.4 &   1.7$\pm$0.3 &   -0.3$\pm$0.2 &   3$\pm$8 &  19 \\
" &  57548.4 - 57563.4 &   0.8$\pm$0.1 &   0.7$\pm$0.1 &   -20$\pm$7 &  41 \\
HeII 4686  &  55257.3 - 55407.1 &   1.7$\pm$0.6 &   5.6$\pm$0.5 &   -18$\pm$21 &  15 \\
" &  55586.4 - 55674.3 &   3.1$\pm$0.5 &   7.3$\pm$0.4 &   -45$\pm$9 &  13 \\
" &  56489.0 - 56550.1 &   0.9$\pm$1.0 &   5.4$\pm$0.6 &   -68$\pm$57 &  10 \\
" &  56704.3 - 56709.4 &   4.4$\pm$0.6 &   4.5$\pm$0.4 &   -36$\pm$6 &  14 \\
" &  56489.0 - 56709.4 &   2.3$\pm$0.6 &   4.9$\pm$0.4 &   -41$\pm$15 &  24 \\
" &  57548.4 - 57563.4 &   2.9$\pm$0.3 &   2.7$\pm$0.2 &   -17$\pm$6 &  38 \\
" &  58634.4 - 58643.5 &   1.5$\pm$1.1 &   2.3$\pm$2.5 &   -91$\pm$110 &  10 \\
MgI 5167  &  55257.3 - 55407.1 &   0.2$\pm$0.2 &   3.0$\pm$0.2 &   -103$\pm$75 &  18 \\
" &  55586.4 - 55674.3 &   0.9$\pm$0.3 &   3.3$\pm$0.2 &   -59$\pm$18 &  14 \\
" &  56489.0 - 56709.4 &   1.6$\pm$0.4 &   2.2$\pm$0.4 &   -63$\pm$19 &  17 \\
" &  57548.4 - 57563.4 &   2.2$\pm$0.2 &   1.8$\pm$0.2 &   -63$\pm$8 &  28 \\
MgI 5173 &  54874.3 - 55058.0 &   0.8$\pm$0.3 &   -0.4$\pm$0.3 &   89$\pm$27 &  10 \\
" &  55257.3 - 55407.1 &   0.9$\pm$0.4 &   -0.3$\pm$0.3 &   15$\pm$25 &  19 \\
" &  55586.4 - 55674.3 &   0.8$\pm$0.3 &   0.4$\pm$0.2 &   -44$\pm$21 &  14 \\
" &  56489.0 - 56550.1 &   2.3$\pm$0.7 &   -0.0$\pm$0.4 &   -70$\pm$13 &  11 \\
" &  56704.3 - 56709.4 &   2.5$\pm$0.4 &   -0.1$\pm$0.2 &   3$\pm$8 &  16 \\
" &  56489.0 - 56709.4 &   1.8$\pm$0.4 &   -0.2$\pm$0.3 &   -10$\pm$13 &  27 \\
" &  57548.4 - 57563.4 &   1.7$\pm$0.3 &   -0.6$\pm$0.2 &   -39$\pm$11 &  38 \\
MgI 5184 &  54874.3 - 55058.0 &   1.4$\pm$0.3 &   0.0$\pm$0.2 &   47$\pm$14 &  10 \\
" &  55257.3 - 55407.1 &   0.4$\pm$0.2 &   0.1$\pm$0.1 &   9$\pm$25 &  19 \\
" &  55586.4 - 55674.3 &   1.0$\pm$0.3 &   0.4$\pm$0.2 &   -31$\pm$15 &  14 \\
" &  56111.0 - 56121.1 &   2.0$\pm$0.6 &   0.1$\pm$0.4 &   2$\pm$19 &  10 \\
" &  56489.0 - 56550.0 &   2.4$\pm$1.0 &   0.3$\pm$0.6 &   -61$\pm$21 &  10 \\
" &  56704.3 - 56709.4 &   2.5$\pm$0.2 &   0.2$\pm$0.1 &   12$\pm$4 &  16 \\
" &  56489.0 - 56709.4 &   1.9$\pm$0.4 &   0.2$\pm$0.3 &   2$\pm$12 &  26 \\
" &  57548.4 - 57563.4 &   1.6$\pm$0.2 &   -0.4$\pm$0.2 &   -14$\pm$9 &  36 \\
SiII 6347 &  55257.3 - 55407.1 &   0.7$\pm$0.2 &   -0.1$\pm$0.1 &   18$\pm$16 &  19 \\
" &  55586.4 - 55674.3 &   1.1$\pm$0.3 &   0.1$\pm$0.2 &   -26$\pm$16 &  14 \\
" &  56111.0 - 56121.1 &   1.6$\pm$0.4 &   0.1$\pm$0.3 &   35$\pm$17 &  10 \\
" &  56704.3 - 56709.4 &   2.7$\pm$0.3 &   -0.8$\pm$0.2 &   14$\pm$7 &  17 \\
" &  56489.0 - 56709.4 &   1.9$\pm$0.5 &   -0.7$\pm$0.3 &   6$\pm$15 &  26 \\
" &  57548.4 - 57563.4 &   2.1$\pm$0.5 &   -0.7$\pm$0.3 &   14$\pm$12 &  14 \\
\end{longtable}

\section{CVSO109 Gaussian fit properties}
Included in the supplementary material of the article. 

\begin{table*}
\caption{CVSO109 summary of the composite Gaussian fit to the lines possessing a clearly defined emission wing, which could be accurately fitted across the four epochs; and the Ca II IR triplet fitted in the XSHOOTER data. Standard errors for the Gaussian components are indicated.}
\label{tab:CVSO109_wing_properties}
\begin{tabular}{llrrrrrrr}
\toprule
UTC observation date (Instrument) & Species & $\lambda$ (air) & g1\_cen & g2\_cen & g1\_fwhm & g2\_fwhm &     Aki &    Ek \\
 &  & [\AA] & [km/s] & [km/s] & [km/s] & [km/s] &     [/s] &    [eV] \\
\midrule
	2020-11-26T03:29:49.172\_UVES	&	H I &	3750.2	&	2.1	$\pm$	1.2	&	-17.1	$\pm$	3.8	&	45.9	$\pm$	3.7	&	197.5	$\pm$	11.5	&	2.8E+04	&	13.5	\\
	2020-11-28T03:36:03.569\_XSHOOTER	&	H I &	3750.2	&	-6.1	$\pm$	0.5	&	67.3	$\pm$	0.7	&	59.1	$\pm$	1.4	&	286.3	$\pm$	1.7	&	2.8E+04	&	13.5	\\
	\hline
	2020-11-26T03:29:49.172\_UVES	&	H I &	3770.6	&	1.9	$\pm$	0.5	&	-8.8	$\pm$	1.5	&	38.4	$\pm$	1.5	&	189.3	$\pm$	4.9	&	4.4E+04	&	13.5	\\
	2020-11-27T03:02:55.805\_UVES	&	H I &	3770.6	&	1.8	$\pm$	0.5	&	-1.1	$\pm$	1.6	&	49.7	$\pm$	1.6	&	230	$\pm$	5.7	&	4.4E+04	&	13.5	\\
	2020-11-28T03:36:03.569\_XSHOOTER	&	H I &	3770.6	&	-7.7	$\pm$	0.3	&	71.7	$\pm$	0.6	&	68.8	$\pm$	0.9	&	293.7	$\pm$	1.3	&	4.4E+04	&	13.5	\\
	2020-11-28T03:59:39.410\_UVES	&	H I &	3770.6	&	1	$\pm$	0.5	&	60	$\pm$	1.3	&	45.6	$\pm$	1.3	&	277	$\pm$	3.5	&	4.4E+04	&	13.5	\\
	\hline
	2020-11-26T03:29:49.172\_UVES	&	H I &	3797.9	&	-0.8	$\pm$	0.4	&	-13.6	$\pm$	1.5	&	56.7	$\pm$	1.3	&	232.3	$\pm$	5.4	&	7.1E+04	&	13.5	\\
	2020-11-27T03:02:55.805\_UVES	&	H I &	3797.9	&	0.5	$\pm$	0.4	&	-7.8	$\pm$	1.5	&	60.5	$\pm$	1.5	&	228	$\pm$	5.6	&	7.1E+04	&	13.5	\\
	2020-11-28T03:36:03.569\_XSHOOTER	&	H I &	3797.9	&	-5.1	$\pm$	0.2	&	63	$\pm$	0.4	&	68.2	$\pm$	0.5	&	295.7	$\pm$	1	&	7.1E+04	&	13.5	\\
	2020-11-28T03:59:39.410\_UVES	&	H I &	3797.9	&	-0.7	$\pm$	0.4	&	57.5	$\pm$	1	&	50.5	$\pm$	1	&	267.5	$\pm$	2.6	&	7.1E+04	&	13.5	\\
	\hline
	2020-11-26T03:29:49.172\_UVES	&	H I &	3835.3	&	2.6	$\pm$	0.5	&	-2.4	$\pm$	1.6	&	46.8	$\pm$	1.4	&	201.4	$\pm$	5.6	&	1.2E+05	&	13.4	\\
	2020-11-27T03:02:55.805\_UVES	&	H I &	3835.3	&	2.8	$\pm$	0.4	&	1.2	$\pm$	1.4	&	50.2	$\pm$	1.2	&	209.4	$\pm$	5.1	&	1.2E+05	&	13.4	\\
	2020-11-28T03:36:03.569\_XSHOOTER	&	H I &	3835.3	&	-2.2	$\pm$	0.2	&	56.3	$\pm$	0.5	&	71.3	$\pm$	0.7	&	279.5	$\pm$	1.3	&	1.2E+05	&	13.4	\\
	2020-11-28T03:59:39.410\_UVES	&	H I &	3835.3	&	3.3	$\pm$	0.4	&	52.4	$\pm$	1.2	&	50	$\pm$	1.1	&	264.5	$\pm$	3.3	&	1.2E+05	&	13.4	\\
	\hline
	2020-11-26T03:29:49.172\_UVES	&	H I &	3889.1	&	-7	$\pm$	0.2	&	-15.4	$\pm$	1.2	&	68.4	$\pm$	0.8	&	248.1	$\pm$	4.9	&	2.2E+05	&	13.4	\\
	2020-11-27T03:02:55.805\_UVES	&	H I &	3889.1	&	-5.6	$\pm$	0.3	&	-11.2	$\pm$	1	&	66.2	$\pm$	1	&	218	$\pm$	4.1	&	2.2E+05	&	13.4	\\
	2020-11-28T03:36:03.569\_XSHOOTER	&	H I &	3889.1	&	-9.4	$\pm$	0.1	&	37.8	$\pm$	0.2	&	78	$\pm$	0.2	&	276.1	$\pm$	0.6	&	2.2E+05	&	13.4	\\
	2020-11-28T03:59:39.410\_UVES	&	H I &	3889.1	&	-5.9	$\pm$	0.3	&	33.8	$\pm$	0.9	&	64.1	$\pm$	0.8	&	260.9	$\pm$	2.6	&	2.2E+05	&	13.4	\\
	\hline
	2020-11-26T03:29:49.172\_UVES	&	H I &	4101.7	&	2.6	$\pm$	0.2	&	-7.7	$\pm$	0.7	&	59.2	$\pm$	0.6	&	205.7	$\pm$	2.5	&	2.9E+06	&	13.2	\\
	2020-11-27T03:02:55.805\_UVES	&	H I &	4101.7	&	2.6	$\pm$	0.2	&	0.9	$\pm$	0.6	&	55.2	$\pm$	0.7	&	194.8	$\pm$	2.4	&	2.9E+06	&	13.2	\\
	2020-11-28T03:36:03.569\_XSHOOTER	&	H I &	4101.7	&	0	$\pm$	0.1	&	39.4	$\pm$	0.3	&	73.7	$\pm$	0.4	&	248.3	$\pm$	0.7	&	2.9E+06	&	13.2	\\
	2020-11-28T03:59:39.410\_UVES	&	H I &	4101.7	&	3.4	$\pm$	0.2	&	34.9	$\pm$	0.5	&	53.3	$\pm$	0.5	&	234.2	$\pm$	1.5	&	2.9E+06	&	13.2	\\
	\hline
	2020-11-26T03:29:49.172\_UVES	&	H I &	4340.5	&	0.2	$\pm$	0.2	&	-7.6	$\pm$	0.8	&	63.9	$\pm$	0.6	&	217	$\pm$	3	&	2.5E+06	&	13.1	\\
	2020-11-27T03:02:55.805\_UVES	&	H I &	4340.5	&	0.8	$\pm$	0.2	&	0.7	$\pm$	0.6	&	59	$\pm$	0.7	&	198.8	$\pm$	2.5	&	2.5E+06	&	13.1	\\
	2020-11-28T03:36:03.569\_XSHOOTER	&	H I &	4340.5	&	-3.2	$\pm$	0.1	&	35.7	$\pm$	0.2	&	73	$\pm$	0.3	&	232.9	$\pm$	0.6	&	2.5E+06	&	13.1	\\
	2020-11-28T03:59:39.410\_UVES	&	H I &	4340.5	&	0.8	$\pm$	0.2	&	34.1	$\pm$	0.5	&	57.6	$\pm$	0.5	&	230	$\pm$	1.4	&	2.5E+06	&	13.1	\\
	\hline
	2020-11-26T03:29:45.199\_UVES	&	H I &	4861.4	&	-1.2	$\pm$	0.2	&	-15	$\pm$	1.1	&	71.3	$\pm$	0.6	&	256.6	$\pm$	3.9	&	2.1E+07	&	12.7	\\
	2020-11-27T03:02:45.220\_UVES	&	H I &	4861.4	&	-0.1	$\pm$	0.2	&	-0.8	$\pm$	0.7	&	64.8	$\pm$	0.8	&	200.1	$\pm$	3.1	&	2.1E+07	&	12.7	\\
	2020-11-28T03:36:03.569\_XSHOOTER	&	H I &	4861.4	&	-5.4	$\pm$	0.1	&	30.7	$\pm$	0.3	&	83.8	$\pm$	0.4	&	244.6	$\pm$	0.7	&	2.1E+07	&	12.7	\\
	2020-11-28T03:59:35.437\_UVES	&	H I &	4861.4	&	-1.7	$\pm$	0.2	&	33.4	$\pm$	0.5	&	69.1	$\pm$	0.6	&	237.2	$\pm$	1.3	&	2.1E+07	&	12.7	\\
	\hline
	2020-11-26T03:29:45.199\_UVES	&	H I &	6562.8	&	-3.4	$\pm$	0.2	&	2.7	$\pm$	0.9	&	96.2	$\pm$	1	&	220.5	$\pm$	4.8	&	4.4E+07	&	12.1	\\
	2020-11-27T03:02:45.220\_UVES	&	H I &	6562.8	&	-0.6	$\pm$	0.6	&	0	$\pm$	1.9	&	70	$\pm$	2.6	&	175.3	$\pm$	3.5	&	4.4E+07	&	12.1	\\
	2020-11-28T03:36:08.759\_XSHOOTER	&	H I &	6562.8	&	-3.5	$\pm$	0.1	&	23.1	$\pm$	0.4	&	122.1	$\pm$	0.5	&	252.3	$\pm$	1.4	&	4.4E+07	&	12.1	\\
	2020-11-28T03:59:35.437\_UVES	&	H I &	6562.8	&	-6.2	$\pm$	0.8	&	12.9	$\pm$	2	&	61.9	$\pm$	2.6	&	242.2	$\pm$	7	&	4.4E+07	&	12.1	\\
	\hline
	2020-11-26T03:29:49.172\_UVES	&	Ca II &	3933.7	&	-1.9	$\pm$	0.1	&	1.1	$\pm$	0.2	&	16.2	$\pm$	0.3	&	45.5	$\pm$	0.8	&	1.5E+08	&	3.2	\\
	2020-11-27T03:02:55.805\_UVES	&	Ca II &	3933.7	&	-1.3	$\pm$	0.1	&	2.1	$\pm$	0.3	&	16.5	$\pm$	0.3	&	48.5	$\pm$	0.9	&	1.5E+08	&	3.2	\\
	2020-11-28T03:36:03.569\_XSHOOTER	&	Ca II &	3933.7	&	-3.3	$\pm$	0.1	&	76.7	$\pm$	0.2	&	62.1	$\pm$	0.2	&	223.5	$\pm$	0.4	&	1.5E+08	&	3.2	\\
	2020-11-28T03:59:39.410\_UVES	&	Ca II &	3933.7	&	-0.5	$\pm$	0.1	&	65.6	$\pm$	0.8	&	26.6	$\pm$	0.3	&	217	$\pm$	2.1	&	1.5E+08	&	3.2	\\
	\hline
	2020-11-28T03:36:08.759\_XSHOOTER	&	Ca II &	8498.0	&	3	$\pm$	0.2	&	113.9	$\pm$	1.2	&	23.4	$\pm$	0.4	&	164.2	$\pm$	3.4	&	1.1E+06	&	3.2	\\	
		\hline
	2020-11-28T03:36:08.759\_XSHOOTER	&	Ca II &	8542.1	&	3	$\pm$	0.2	&	110.9	$\pm$	1.2	&	28.6	$\pm$	0.5	&	162	$\pm$	3.5	&	9.9E+06	&	3.2	\\
	\hline
	2020-11-28T03:36:08.759\_XSHOOTER	&	Ca II &	8662.1	&	3.4	$\pm$	0.2	&	118.9	$\pm$	1.1	&	23.5	$\pm$	0.4	&	166.7	$\pm$	3.1	&	1.1E+07	&	3.1	\\
\bottomrule
\end{tabular}
\end{table*}

\bsp	
\label{lastpage}
\end{document}